\documentclass{article}

\usepackage{amsfonts}

\title{Quantization of affine bodies. Theory and applications in mechanics of structured media.}

\author{Jan J. S\l awianowski\\
Institute of Fundamental Technological Research,\\
Polish Academy of Sciences,\\
21 \'{S}wi\c{e}tokrzyska str., 00-049 Warsaw, Poland\\
e-mail: jslawian@ippt.gov.pl}

\begin{document}

\maketitle

\begin{abstract}

Discussed is kinematics and dynamics of bodies with affine degrees of freedom, i.e., homogeneously deformable "gyroscopes". The special stress is laid on the status and physical justification of affine dynamical invariance. On the basis of classical Hamiltonian formalism the Schroedinger quantization procedure is performed. Some methods of the partial separation of variables, analytical treatment and search of rigorous solutions are developed. The possiblity of applications in theory of structured media, nanophysics, and molecular physics is discussed.

\end{abstract}


\section*{Introduction}

The mechanics of affine bodies was a subject of many papers
\cite{Bog85,Cap00,Chan69,Chev04,Coh81,Coh-Mac89,Coh-Mac91,Coh-Mac92,Coh-Mac94,Coh-Mac96,Coh-Man89,CPS74,Coh-Sun88,Coh-Sun91,Dys68,Erin68,HLN77,Mart02,Mart03,Mart04_1,Mart04_2,ORe96,ORe-Var98,Pap01,RWL02,Ros-Tr98,Ros-Tr99,Rub85,Rub86,AKS89,AKS-JJS91,JJS73_2,JJS75_2,JJS75_4,JJS76,JJS91_5,JJS02_1,JJS02_2,JJS03,JJS04,JJS04S,JJS-VK02,JJS-VK03,JJS-VK04,JJS-VK04S,all-book04,all04,all05,JJS-AKS93,Sol-Pap99,Sol-Pap00,Dias94,Trz-JJS90,West67,Wul-Rob02}.
It has been a field of intensive studies in our group at the Institute of Fundamental
Technological Research in Warsaw. Up to our knowledge, for the first time the idea of objects
with affine degrees of freedom  in mechanics appeared in papers of Eringen
\cite{Erin62,Erin68,Erin75_1,Erin75_2} devoted to structured continua, to be more precise in
his theory of micromorphic media. Micromorphic continuum is an affine extension of the
micropolar Cosserat continuum. Roughly speaking, the Cosserat medium is a deformable continuum
of infinitesimal gyroscopes. Similarly, the micromorphic body is a deformable continuum of
infinitesimal homogeneously deformable gyroscopes. Affine model of collective degrees of
freedom was also used in the theory of collective phenomena in atomic nuclei
\cite{Bohr-Mot75}. The idea of affine body is interesting in itself from the point of view of
analytical mechanics and theory of dynamical systems. It is an instructive example of systems
with degrees of freedom ruled by Lie groups. In mechanics of non-constrained continua the
configuration space may be identified with the group of all diffeomorphisms of the physical
space (volume-preserving diffeomorphisms in mechanics of ideal incompressible fluids). It is
rather difficult to be rigorous with such infinite-dimensional groups. Affine model is placed
between rigid-body mechanics and the general theory of deformable continua, i.e., it involves
deformations but at the same time one deals there with a finite number of degrees of freedom.
The Lie-group background of the geometry of the configuration space offers the possibility of
the effective use of powerful analytic techniques. One can realize certain finite-dimensional
generalizations when the configuration space geometry is ruled, e.g., by the projective or
conformal group. Also other finite-dimensional discretized approaches are useful but of course
the models based on geometric transformation groups are particularly interesting and
efficient.

The range of applications of affine model of collective and internal degrees of freedom is
very wide and has to do with various scales of physical phenomena:
\begin{itemize}
\item macroscopic elastic problems when the length of excited waves is comparable with the
linear size of the body.

\item purely computational and engineering problems connected with the finite elements
methods. A mixture of analytic and numerical procedures.

\item structured bodies, e.g., micromorphic continua and molecular crystals.

\item vibrations of astrophysical objects (stars, concentrations of the cosmic dust), theory
of the shape of the Earth.

\item molecular vibrations.

\item nuclear dynamics.
\end{itemize}

Obviously, the last two subjects must be based on the quantized version of the theory. Quantum
description is also necessary in various problems concerning the nanoscale phenomena,
fullerens, etc. It is a new fascinating subject where one deals with the very intriguing
convolution of the classical and quantum levels, perhaps also with some yet non-solved
paradoxes from the realm of quantum-mechanical foundations like decoherence, etc.

Quantization as a purely mathematical procedure is connected with certain ambiguities which
may be solved only a posteriori, on the bases of experimental data. There are some well-known
problems with the ordering of operators. In models with a firm group-theoretic background
there are some canonical procedures, usually confirmed by experiments. Because of this, the
extensive geometric introduction presented below, almost a treatise as a matter of fact, is a
constitutive element of the theory, motivated by deeper reasons than the purely mathematical
curiosity or artificial sophistication.

Affine models of degrees of freedom of the structured elements is very natural. When one deals
with fullerens, macromolecules, microdefects, affine modes of motion are certainly the most
relevant ones. There are molecules, e.g., $P_{4}$, which have no other degrees of freedom;
there are also such ones for which non-affine behaviour is a merely small correction. As
mentioned, affine modes have also to do with finite elements, when the body is described as an
aggregate of small affine objects.

Perhaps the quantization of such an approach might be a procedure alternative to the phonon
description based on the quantized plane elastic waves.

And finally, one of the most important things. The group-theoretical description of internal
and collective modes is really effective when the dynamics is invariant (or in some sense
almost invariant) with respect to the group underlying kinematics of the problem. And this is
not the case in all models of affine bodies met in literature. Kinematics is there affine but
the group of dynamical symmetries is broken to the isometry group. Because of this, there is
no full use and the full profit of Lie-group techniques. Unlike this, we formulate here
affinely-invariant dynamics, where elastic interactions may be encoded in appropriate kinetic
energy models without (or "almost" without) any use of potential energy terms. This procedure
is similar to that following from the Maupertuis variational principle. There are indications
that just such models may be useful in condensed matter theory, where the structural elements
are more sensitive to the geometry of a surrounding piece of the body, e.g., to the Cauchy
deformation tensor than to the "true" metric tensor of the physical space. This is something
similar to the effective mass tensors of electrons in crystals.

There is an interesting link between our models and theories of integrable lattices like
Calogero-Moser, Sutherland, and others \cite{Cal-Mar74,Mos75_1,Mos75_2}. On the quantum level
the deformation invariants behave like indistinguishable, exotically parastatistical
one-dimensional "particles".

Obviously, the real world, the arena of mechanical phenomena, is three-dimensional. However,
certain important invariance and other problems are explained in a more lucid way when
described with non-physical generality, i.e., in $n$ dimensions. By the way, two-dimensional
problems are also interesting not only in "Flatland" \cite{Abb84} but also in some realistic
physical problems. At the same time, they are computationally simple due to some exceptional,
so to speak pathological, feature of GL$(2,\mathbb{R})$ among all GL$(n,\mathbb{R})$
(SO$(2,\mathbb{R})$ is Abelian, whereas SO$(n,\mathbb{R})$ for $n>2$ are semisimple).

\section{Classical preliminaries}

Let us briefly describe various models of the configuration space of affinely rigid body. It
depends on the particular problem under consideration which of them is more convenient. The
possibility and usefulness of many choices of geometric structures underlying physically the
same degrees of freedom was pointed out by Capriz
\cite{Cap89,Cap00,Cap-Mar(eds)03,Cap-Mar03,Mar00,Mar01,Mar03}. Various descriptions differ in
assuming some auxiliary geometric objects.

We begin with some elementary concepts of affine geometry, just to fix the language and
notation. Affine space is given by a triple $(X,E,\rightarrow)$, where $X$ is a point set,
just the "space itself", $E$ is a linear space of translations in $X$, and the arrow
$\rightarrow$ denotes a mapping from the Cartesian product $X\times X$ onto $E$; the vector
assigned to $(p,q)\in X\times X$ is denoted by $\overrightarrow{pq}$. The arrow operation
satisfies some axioms, namely,
\begin{itemize}
\item[(i)] $\overrightarrow{pq}+\overrightarrow{qr}+\overrightarrow{rp}=0$ for any $p,q,r\in
X$,

\item[(ii)] for any $p\in X$ and $v\in E$ there exists exactly one $q\in X$ such that
$\overrightarrow{pq}=v$; we write $q=t_{v}(p)$.
\end{itemize}
For any $v\in E$, $t_{v}:X\rightarrow X$ is a one-to-one mapping of $X$ onto $X$, the
translation by $v$. And obviously,
\[
t_{v}\circ t_{u}=t_{u}\circ t_{v}=t_{v+u},\qquad t_{o}={\rm id}_{X},\qquad t_{u}^{-1}=t_{-u}.
\]
In this way, $E$ considered as an additive-rule Abelian group acts freely and transitively on
$X$.

Any linear space $E$ may be considered as an affine space $(E,E,-)$, i.e.,
$\overrightarrow{uv}=v-u$.

The axiom $(i)$ implies that $\overrightarrow{pp}=0$,
$\overrightarrow{pq}=-\overrightarrow{qp}$ for any $p,q\in X$.

Let $\Omega$ be an arbitrary set, in general, structure-less one. The set of all mappings from
$\Omega$ in $X$, denoted by $X^{\Omega}$, is simply the $\Omega$-indexed Cartesian product of
$X$. For any mapping $f:\Omega\rightarrow X$, the image $f(\omega)\in X$ is interpreted as an
$\omega$-th component of $f$. When $\Omega$ is a finite $N$-element set, e.g.,
$\Omega=\{1,2,\ldots,N\}$, this is just the familiar finite Cartesian product $X^{N}$.

The set $X^{\Omega}$ is in a natural way an affine space. Its translation space is identical
with $E^{\Omega}$, the set of all mappings from $\Omega$ into $X$. If $F$, $G$ are mappings
from $\Omega$ into $X$, then the corresponding translation vector $\overrightarrow{FG}\in
E^{\Omega}$ is simply given by
\begin{equation}\label{a2}
\left(\overrightarrow{FG}\right)(\omega):=\overrightarrow{F(\omega)G(\omega)},
\end{equation}
for any $\omega\in\Omega$. One can easily show that all axioms of affine geometry are
satisfied then.

Affine mappings, by definition, preserve all affine relationships between figures and points.
So, if $(N,U,\rightarrow)$, $(M,V,\rightarrow)$ are affine spaces, then we say that
$\Phi:N\rightarrow M$ is affine if there exists such a linear mapping $L[\Phi]:U\rightarrow V$
denoted also by $D\varphi$ that for any $p,q\in N$ the following holds:
\[
\overrightarrow{\Phi(p)\Phi(q)}=L[\Phi]\overrightarrow{pq}.
\]
The mapping $L[\Phi]:U\rightarrow V$ is referred to as a linear part of $\Phi$. The set of all
affine mappings from $N$ to $M$ will be denoted by Af$(N,M)$; similarly, L$(U,V)$ denotes the
set of linear mappings. If $\Phi_{1}\in$ Af$(P,M)$ and $\Phi_{2}\in$ Af$(N,P)$, then
$\Phi_{1}\circ\Phi_{2}\in$ Af$(N,M)$ and $L[\Phi_{1}\circ\Phi_{2}]=L[\Phi_{1}]L[\Phi_{2}]$.

Dimension of the translation space $E$ is referred to as the dimension of $X$ itself. Any
fixed point $p\in M$ establishes the bijection of $M$ onto $V$ given by $M\ni q\mapsto
\overrightarrow{pq}\in V$. Such $V$-valued charts establish in $M$ the structure of analytical
differential manifold just of dimension dim $V$. The manifold of affine injections from $N$
into $M$ will be denoted by AfI$(N,M)$, and the corresponding set of linear injections from
$U$ into $V$ by LI$(U,V)$. They are open submanifolds of Af$(N,M)$, L$(U,V)$, respectively.
Obviously, they are non-empty if and only if dim $M\geq$ dim $N$. If dim $M=$ dim $N$, they
become respectively the manifolds of affine and linear isomorphisms.

If $N=M$ and $U=V$, i.e., when we work within some fixed affine space $(M,V,\rightarrow)$,
then some simplified notation is used, namely,
\[
{\rm L}(V,V),\qquad {\rm Af}(M,M),\qquad {\rm LI}(V,V),\qquad {\rm AfI}(M,M)
\]
are denoted respectively by
\[
{\rm L}(V),\qquad {\rm Af}(M),\qquad {\rm GL}(V),\qquad {\rm GAf}(M).
\]
Obviously, the last two sets are groups, respectively, the general linear and affine groups in
$V$, $M$. Translations are affine isomorphisms; their set $T[V]=\{t_{v}:v\in V\}$ is a normal
subgroup of GAf$(M)$. This subgroup is the kernel of the group epimorphism:
\[
{\rm GAf}(M)\ni \varphi\mapsto L[\varphi]\in{\rm GL}(V).
\]
The quotient group GAf$(M)/T(V)$ is isomorphic with GL$(V)$ but in a non-canonical way; any
choice of centre $o\in M$ gives rise to some isomorphism.

The set of affine mappings from $(N,U,\rightarrow)$ to $(M,V,\rightarrow)$, i.e., Af$(N,M)$,
is an affine subspace of $M^{N}$ in the sense of (\ref{a2}); the translation space is
identified with $V^{N}$.

If in the space $N$ some origin point $\mathcal{O}\in N$ is chosen then the manifold Af$(N,M)$
may be simplified to the Cartesian product $M\times$ L$(U,V)$. Namely, with any $\Phi\in$
Af$(N,M)$ we associate a pair $(x,\varphi)\in M\times$ L$(U,V)$ in such a way that
$x=\Phi(\mathcal{O})$ and $\overrightarrow{\Phi(\mathcal{O})\Phi(a)}=\varphi\cdot
\overrightarrow{\mathcal{O}a}$. When we restrict ourselves to the open submanifold of affine
isomorphisms AfI$(N,M)\subset$ Af$(N,M)$, then $\varphi$ in the above expression runs over the
open submanifold LI$(U,V)\subset$ L$(U,V)$.

And finally, let us fix some linear frame, i.e., an ordered basis in $U$,
$E=(E_{1},\ldots,E_{A},\ldots,E_{n})$, $n=$ dim $U=$ dim $V$. When it is kept fixed, any
linear mapping $\varphi\in$ L$(U,V)$ may be identified with the system
$e=(e_{1},\ldots,e_{A},\ldots,e_{n})$, where $e_{A}=\varphi E_{A}$, $A=\overline{1,n}$. When
$\Phi\in$ AfI$(N,M)$, i.e., $\varphi\in$ LI$(U,V)$, then $e$ is a linear frame in $V$. In this
way LI$(U,V)$ is identified with F$(V)$, the manifold of linear frames in $V$. And AfI$(N,M)$
is identified with $M\times$ F$(V)$, the manifold of affine frames in $M$ (the pairs
consisting of points in $M$ and ordered bases in $V$).

Fixing an affine frame $(\mathcal{O},E)$ in $N$ we turn it into the arithmetic space
$\mathbb{R}^{n}$. Linear isomorphisms of $U$ onto $V$ become then linear frames in $V$; their
inverse isomorphisms are identified with the dual co-frames:
$\widetilde{e}=(e^{1},\ldots,e^{A},\ldots,e^{n})$, $\langle
e^{A},e_{B}\rangle=\delta^{A}{}_{B}$

As frames and dual co-frames mutually determine each other, AfI$(N,M)$ may be as well
identified with $M\times$ F$(V^{\ast})=M\times$ F$(V)^{\ast}$; here F$(V^{\ast})$ is the
manifold of frames in the dual space $V^{\ast}$ denoted also as F$(V)^{\ast}$.

If in addition some affine frame
$(o,\mathcal{E})=(o;\mathcal{E}_{1},\ldots,\mathcal{E}_{A},\ldots,\mathcal{E}_{n})$ in $M$ is
fixed, then also $V$ becomes identified with $\mathbb{R}^{n}$. The manifold LI$(U,V)$ is then
analytically identified with the general linear group GL$(n,\mathbb{R})$, and AfI$(N,M)$ may
be identified with the semi-direct product GAf$(n,\mathbb{R})\simeq$
GL$(n,\mathbb{R})\times_{s}\mathbb{R}^{n}$.

The manifold AfI$(N,M)$ is a homogeneous space of affine groups GAf$(M)$, GAf$(N)$ acting
respectively on the left and on the right:
\begin{eqnarray}
A\in{\rm GAf}(M)&:&{\rm AfI}(N,M)\ni\Phi\mapsto A\circ\Phi,\label{b6}\\
B\in{\rm GAf}(N)&:&{\rm AfI}(N,M)\ni\Phi\mapsto \Phi\circ B.\label{a6}
\end{eqnarray}
Similarly, linear groups GL$(V)$, GL$(U)$ act transitively on LI$(U,V)$:
\begin{eqnarray}
\alpha\in{\rm GL}(V)&:&{\rm LI}(U,V)\ni\varphi\mapsto \alpha\varphi,\label{d6}\\
\beta\in{\rm GL}(U)&:&{\rm LI}(U,V)\ni\varphi\mapsto \varphi\beta.\label{c6}
\end{eqnarray}

Let us observe that although GL$(V)$, GL$(U)$ are logically distinct disjoint sets, the
corresponding transformation groups intersect nontrivially. Namely, dilatations belong to both
of them, the left and right actions of $\alpha=\lambda$ Id$_{V}$, $\beta=\lambda$ Id$_{U}$
result in multiplying $\varphi$ by $\lambda$, i.e., $\varphi\mapsto\lambda\varphi$.

When some origin $\mathcal{O}\in N$ is fixed and AfI$(N,M)$ is identified with $M\times$
LI$(U,V)$, the left-acting transformation groups GAf$(M)$, GL$(V)$ may be represented as
follows:
\begin{eqnarray}
A\in{\rm GAf}(M)&:&M\times{\rm LI}(U,V)\ni(x,\varphi)\mapsto(A(x),L(A)\varphi),\label{a7}\\
\alpha\in{\rm GL}(V)&:&M\times{\rm LI}(U,V)\ni(x,\varphi)\mapsto(x,\alpha\varphi),\label{b7}
\end{eqnarray}
The origin $\mathcal{O}$ enables one to identify GAf$(N)$ with the semi-direct product
GL$(U)\times_{s}U$. Namely, $B\in$ GAf$(N)$ is represented by the pair
$(L(B),\overrightarrow{\mathcal{O}B(\mathcal{O})})$. And conversely the pair $(\beta,u)\in$
GL$(U)\times_{s}U$ gives rise to the mapping $B\in$ GAf$(N)$ such that for any $a\in N$:
\[
\overrightarrow{\mathcal{O}B(a)}=\beta\cdot\overrightarrow{\mathcal{O}a}+u.
\]
The right action (\ref{a6}) of $B$ on AfI$(N,M)$ is represented in $M\times$ LI$(U,V)$ as
follows:
\begin{eqnarray}\label{c7}
(\beta,u)\in{\rm GL}(U)\times_{s}U&:&(x,\varphi)\mapsto(t_{\varphi u})(x),\varphi\beta).
\end{eqnarray}
If we put $u=0$, then the group GL$(U)$ itself acts only on the second component:
\begin{eqnarray}\label{d7}
\beta\in{\rm GL}(U)&:&(x,\varphi)\mapsto(x,\varphi\beta).
\end{eqnarray}

The standard language of continuum mechanics is based on the use of two affine spaces: the
physical and material ones. We denote them respectively by $(M,V,\rightarrow)$ and
$(N,U,\rightarrow)$. If we deal with the infinite continuum medium filling up the whole
physical space, configurations are described by diffeomorphisms of $N$ onto $M$. The
smoothness class of these diffeomorphisms depends on peculiarities of the considered problem.
The manifold $N$ is interpreted as the set of material points. In configuration given by
$\Phi:N\rightarrow M$, the material point $a\in N$ occupies the spatial position $\Phi(a)\in
M$. Diffeomorphism groups Diff$(M)$ and Diff$(N)$ give rise to transformation groups acting on
the configuration space Diff$(N,M)$, i.e.,
\begin{eqnarray}
A\in{\rm Diff}(M)&:&{\rm Diff}(N,M)\ni\Phi\mapsto A\circ\Phi,\\
B\in{\rm Diff}(N)&:&{\rm Diff}(N,M)\ni\Phi\mapsto \Phi\circ B.\label{b8}
\end{eqnarray}
They are referred to respectively as spatial and material transformations. Obviously, spatial
and material transformations mutually commute. In continuum mechanics they have to do with
symmetries of space and material itself.

Obviously, when one deals with realistic bounded bodies, this description should be modified,
e.g., manifolds with boundary become a better model of the material space. Another possibility
is to use a smooth smeared-out model of the boundary, i.e., to describe the bounded body as a
non-bounded, one however, with the mass density quickly vanishing outside the real object.

Deeper modifications are necessary when describing continua with degenerate dimension like
membranes, strings, infinitesimally thin shells, rods, etc. And obviously, for discrete
systems the description based on the affine space $N$ as a material body is not applicable in
the literal sense, unless some tricks like smeared out density functions and so on are used.
It would be a good thing, especially when dealing with affine systems in microscopic
applications (molecular, microstructural, etc.) to start from some general formulation
applicable both to discrete and continuous systems of various kinds. There is also some more
subtle point to that, namely, the material space is primarily the abstract set of material
points or their labels, so-to-say "identification cards". A priori this set is structure-less;
it is a kind of "powder" of material points. Let us denote it by $\Omega$. Configurations are
mappings from $\Omega$ to $M$, i.e., elements of $M^{\Omega}$ (the usual finite Cartesian
product $M^{N}$ when one deals with an $N$-particle system). More precisely, $M^{\Omega}$ is
the set of singular configurations, i.e., ones admitting coincidences of different material
points at the same spatial point. To avoid such a "catastrophe" one must decide that the
"true" configuration space is the set of injections from $\Omega$ into $M$, i.e.,
Inj$(\Omega,M)$. As far as $\Omega$ is structure-less the only well-defined set of material
transformations is Bij$(\Omega)$, the set of bijections of $\Omega$ onto $\Omega$. They act on
configurations according to the following rule:
\begin{eqnarray}
B\in{\rm Bij}(\Omega)&:&{\rm Inj}(\Omega,M)\ni\Phi\mapsto\Phi\circ B.\label{b10}
\end{eqnarray}
These "permutations" of material points are just the only admissible material transformations
on this yet amorphous stage. Transformations of $M$ onto itself, in particular diffeomorphisms
of $M$ onto itself of an appropriate class, act on Inj$(\Omega,M)$ according to the following
rule:
\begin{eqnarray}
A\in{\rm Bij}(M)&:&{\rm Inj}(\Omega,M)\ni\Phi\mapsto A\circ\Phi.\label{a10}
\end{eqnarray}
In general, this action is not transitive and splits into orbits, i.e., transitively classes.
Any fixed class carries over geometric structures from $M$ to $\Omega$. For example, if
$\Omega$ has the continuum cardinal number and only the bijections of $\Omega$ onto $M$ are
admitted as configurations, then any fixed orbit of the left-acting diffeomorphism group
Diff$^{r}(M)$ induces in $\Omega$ some structure of $C^{r}$-class differentiable manifold. The
powder of material points becomes the continuous body and its configuration space is
identified with Diff$^{r}(\Omega,M)$, i.e., the set of $C^{r}$-class diffeomorphisms of
$\Omega$ onto $M$. If some orbit of the left-acting affine group GAf$(M)$ is fixed, $\Omega$
becomes endowed with the induced structure of affine space. And the one can sensibly tell
about affine mappings from $\Omega$ onto $M$ and about affine relationships between material
points in $\Omega$.

Different orbits induce structures in $\Omega$, which are literally different, although
usually isomorphic.

In general, when dealing with constrained systems of material points, it is not the total
group Bij$(M)$ or the diffeomorphism group Diff$^{r}(M)$, but some rather peculiar proper
subgroup $G\subset$ Bij$(M)$ that rules geometry of degrees of freedom and perhaps also the
dynamics. Configuration spaces are constructed as orbits of such groups. Let us assume that
some orbit $Q$, i.e., some particular model of degrees of freedom is fixed. By the very
definition, $G$ acts transitively on $Q$, i.e., $Q$ is a homogeneous space of the action
(\ref{a10}). The point is how to define some right-hand-side action analogous to (\ref{b8}) or
(\ref{a6}). In general, the transformation group Bij$(\Omega)$ acting through (\ref{b10}) is
too poor. For example, when one deals with a finite system of material points, Bij$(\Omega)$
is the permutation group of $\Omega$ and nothing like continuous groups of material
transformations, e.g., (\ref{b8}), (\ref{a6}) can be constructed on the basis of
Bij$(\Omega)$. One feels intuitively that there is something non-satisfactory here. And
indeed, it is possible to define some rich, in general continuous, group of material
transformations acting on the right on $Q$.

The construction is more lucid when one forgets for a moment about details and considers an
abstract homogeneous space with the underlying point set $Q$ and the group $G$ acting
transitively on $Q$ on the left. For simplicity, we denote the action of $G$ by
\begin{eqnarray}
g\in G&:& Q\ni q\mapsto gq\in Q.\nonumber
\end{eqnarray}
This graphical convention is well-suited to the left-hand-side nature of this action:
\[
(g_{1}g_{2})q=g_{1}(g_{2}q).
\]
The action is assumed to be effective, i.e., the group identity $e\in G$ (its neutral element)
is the only element of $G$ satisfying the condition $eq=q$ for any $q\in Q$. Roughly speaking,
$G$ is a "proper" transformation group, not something homomorphically (with a non-trivial
kernel) mapped into the group Bij$(Q)$. This is exactly the case in problems described
previously. Let $q_{0}\in Q$ denote some arbitrarily fixed points, and $H(q_{0})\subset G$
denote its isotropy group, i.e., the set of elements which do not move $q$:
\[
H(q_{0}):=\left\{g\in G:gq_{0}=q_{0}\right\}.
\]

It is well-known that $Q$ may be identified in a one-to-one way with the set of left cosets,
i.e., with the quotient space $G/H(q_{0})$. The action of $G$ on $G/H(q_{0})$ is represented
by left translations, namely, the coset $xH(q_{0}):=\left\{xh:h\in H(q_{0})\right\}$ is
transformed by $g\in G$ into $gxH(q_{0})$; obviously, the result does not depend on the
particular choice of $x$ within its coset, i.e., on the replacement $x\mapsto xh$, $h\in H$.

The question is now whether there exist some right translations of representants, $x\mapsto
xg$, admitting an interpretation in terms of transformations acting on $G/H(q_{0})$. It is
easy to see that the answer is affirmative.

Namely, let $N\left(H(q_{0})\right)\subset G$ denote the maximal subgroup of $G$ for which
$H(q_{0})\subset N\left(H(q_{0})\right)$ is a normal subgroup. It is easy to see that for
every $n\in N\left(H(q_{0})\right)$ the corresponding right regular translation $G\ni x\mapsto
xn\in G$ is projectable to the manifold of left cosets $G/H(q_{0})$. Indeed, cosets are
transformed onto cosets:
\[
\left(xH(q_{0})\right)n=xH(q_{0})n=x\left(nH(q_{0})n^{-1}\right)n=xnH(q_{0})=(xn)H(q_{0}).
\]
In this way transformations may be performed on representants, $G\ni x\mapsto xn\in G$.
Obviously, the choice of representants does not matter because
\[
(xh)nH(q_{0})=(xhn)H(q_{0})=xnn^{-1}hnH(q_{0}),
\]
and for any $h\in H(q_{0})$, $n^{-1}hn\in H(q_{0})$, thus $n^{-1}hnH(q_{0})=H(q_{0})$, and
finally $((xh)n)H(q_{0})=(xn)H(q_{0})$. The non-effectiveness kernel of the right action of
$N\left(H(q_{0})\right)$ on the coset manifold coincides with the group $H(q_{0})$ itself,
thus the true group of right-acting transformations is given by
\[
N\left(H(q_{0})\right)/H(q_{0})=H(q_{0})\setminus N\left(H(q_{0})\right).
\]
The above construction of right-acting transformations pre-assumes some choice of the
reference point $q_{0}\in Q$. The question arises as to what extent does the presented
prescription depend on $q_{0}$. It turns out that the constructed transformation group itself
is well-defined and the particular choice of $q_{0}$ influence only the "parameterization",
so-to-say identification labels of the group elements. Let $q_{1},q_{2}\in Q$ be two
arbitrarily chosen reference point. The subset of $G$ consisting of elements $k$ transforming
$q_{1}$ into $q_{2}$, i.e., $kq_{1}=q_{2}$, will be denoted by $H(q_{1},q_{2})$. Obviously,
$H(q_{1},q_{2})$ is simultaneously the left and right coset of the subgroups $H(q_{1})$,
$H(q_{2})$, respectively; if $k$ is an element of $H(q_{1},q_{2})$, then so is $h_{2}kh_{1}$
for any $h_{1}\in H(q_{1})$, $h_{2}\in H(q_{2})$. (Incidently, let us notice: perhaps it would
be convenient to write $H(q_{1},q_{1})$, $H(q_{2},q_{2})$ instead of $H(q_{1})$, $H(q_{2})$.)
Obviously, for any $k\in H(q_{1},q_{2})$ we have
\[
H(q_{2})=kH(q_{2})k^{-1},\qquad N(q_{2})=kN(q_{1})k^{-1}.
\]
Any choice of $k\in H(q_{1},q_{2})$ fixes some isomorphisms of $H(q_{1})$, $N(q_{1})$
respectively onto $H(q_{2})$, $N(q_{2})$.

Let us take some $g_{1}\in G$ and the point $q\in Q$ produced by it from $q_{1}\in Q$, i.e.,
\[
q=g_{1}q_{1}.
\]
When $q$ is fixed, $g_{1}$ is defined up to the gauging $g_{1}\mapsto g_{1}h$, $h\in
H(q_{1})$. And now we transform $q$ by the right action of $n_{1}\in N\left(H(q_{1})\right)$:
\[
q\mapsto q^{\prime}=g_{1}n_{1}q_{1}.
\]
The result does not depend on the gaugings $g_{1}\mapsto g_{1}h$, $n_{1}\mapsto
\chi_{1}n_{1}\chi_{1}$, where $h,\chi_{1},\chi_{2}\in H(q_{1})$. Let us now express this
action in terms of the reference point $q_{2}=kq_{1}$
\[
q^{\prime}=g_{1}n_{1}q_{1}=g_{1}n_{1}k^{-1}q_{2}=\left(g_{1}k^{-1}\right)
\left(kn_{1}k^{-1}\right)q_{2}.
\]
Now $q$ is produced from the reference point $q_{2}$ by $g_{2}=g_{1}k^{-1}$; $q=g_{2}q_{2}$.
And its representing group elements $g_{2}$ is affected on the right by $n_{2}=kn_{1}k^{-1}\in
N\left(H(q_{2})\right)$ the $k$-conjugation of $n_{1}\in N\left(H(q_{1})\right)$. In this way
different right-hand-side actions in $G$, i.e.,
\[
G\ni x\mapsto xn_{1},\qquad G\ni x\mapsto xn_{2},
\]
describe the same transformation in $Q$. They are different "labels" of this transformation
corresponding to various choices of reference points $q_{1},q_{2}\in Q$ and various choices of
$k\in H(q_{1},q_{2})$.

We are particularly interested in situations when the action of $G$ on $Q$ is free, i.e., when
the isotropy groups are trivial, $H(q)=\{e\}$ for any $q\in Q$. Then $H(q_{1},q_{2})$ are
one-element sets, $H(q_{1},q_{2})=\{k\}$, i.e., the above $k$-element is unique. The
"labelling" of the right-acting transformation group by elements of $G$ depends only on the
choice of the reference point. It is clear that for any $q_{0}\in Q$,
$N\left(H(q_{0})\right)=G$, i.e., the right-acting transformation group is isomorphic with $G$
itself. Choosing some reference point $q_{0}$ we automatically fix one of these isomorphisms.

The extended affinely rigid body is defined as a system of material points constrained in such
a way that all affine relationships between constituents remain frozen during any admissible
motion. Summarizing the above remarks we can formulate a few geometric models of its
configuration space.
\begin{enumerate}
\item If we use the standard terms of continuum mechanics based on the affine physical and
material spaces $(M,V,\rightarrow)$, $(N,U,\rightarrow)$, then the configuration space is
given by AfI$(N,M)$, i.e., the manifold of affine isomorphisms of $N$ onto $M$. Affine groups
GAf$(M)$, GAf$(N)$ act on AfI$(N,M)$ according to the rules (\ref{b6}), (\ref{a6}) and
describe respectively spatial and material transformations (kinematical symmetries). If we
formally admitted singular configurations with degenerate dimension, the configuration space
would be given by Af$(N,M)$, i.e., the set of all affine mappings of $N$ into $M$, in general
non-invertible ones. By the way, Af$(N,M)$ is also an affine space with Af$(N,V)$ as the
translation space; this is just the special case of (\ref{a2}).

\item If some material origin $\mathcal{O}\in N$ is fixed,
the configuration space may be identified with
\[
Q=Q_{\rm tr}\times Q_{\rm int}=M\times{\rm LI}(U,V);
\]
the first and second factors refer respectively to translational and internal (relative)
motion. And again, when singular configurations are admitted, LI$(U,V)$ is replaced by its
linear shell L$(U,V)$. The natural groups of affine symmetries act on $Q$ according to
(\ref{a7}), (\ref{b7}), (\ref{c7}), (\ref{d7}). When translational motion is neglected, the
configuration space reduces to $Q_{\rm int}=$ LI$(U,V)$, or simply to L$(U,V)$ when singular
internal configurations are admitted.

\item When in addition to $\mathcal{O}\in N$ some linear basis
$E=(E_{1},\ldots,E_{A},\ldots,E_{n})$ is chosen, i.e., when an affine frames $(\mathcal{O},E)$
is fixed in $N$, the configuration space becomes identified with
\[
Q=Q_{\rm tr}\times Q_{\rm int}=M\times F(V);
\]
$F(V)$ denotes as previously the manifold of linear frames in $V$ ($n=$ dim $V$). When we are
not interested in translational motion, simply $Q_{\rm int}=F(V)$ is used as the configuration
space. If, for any reason, singular configurations are admitted, we extend $F(V)$ to
$V^{n}=V\times\cdots\times V$ ($n$ Cartesian factors).

Just as in the model $Q=M\times$ LI$(U,V)$ transformation groups act essentially according to
the rules (\ref{a7}), (\ref{b7}), (\ref{c7}), (\ref{d7}), the linear space $U$ being replaced
by $\mathbb{R}^{n}$ (any choice of $E\in F(U)$ identifies $U$ with $\mathbb{R}^{n}$). More
precisely, spatial transformations are given by
\[
A\in{\rm GAf}(M):M\times F(V)\ni(x;\ldots,e_{K},\ldots) \mapsto(A(x);\ldots,L(A)e_{K},\ldots),
\]
\[
\alpha\in{\rm GL}(V):M\times F(V)\ni(x;\ldots,e_{K},\ldots)\mapsto(x;\ldots,\alpha
e_{K},\ldots).
\]
The frame $(\mathcal{O},E)$ identifies $N$ with $\mathbb{R}^{n}$; namely, the point $a\in N$
is identified with its coordinates $a^{K}(a)$ with respect to $(\mathcal{O},E)$:
\[
\overrightarrow{\mathcal{O}a}=a^{K}E_{K}.
\]
Therefore, GAf$(N)$, GL$(U)$ are identified respectively with
GL$(n,\mathbb{R})\times_{s}\mathbb{R}^{n}$, GL$(n,\mathbb{R})$. Their right-hand-side actions
on $Q=M\times F(V)$ are respectively described as follows:
\[
(\beta,u)\in{\rm GL}(n,\mathbb{R})\times_{s}\mathbb{R}^{n}:(x;\ldots,e_{K},\ldots)
\mapsto(t_{eu}(x);\ldots,e_{L}\beta^{L}{}_{K},\ldots),
\]
\[
\beta\in{\rm
GL}(n,\mathbb{R}):(x;\ldots,e_{K},\ldots)\mapsto(x;\ldots,e_{L}\beta^{L}{}_{K},\ldots),
\]
where $eu\in V$ denotes the vector the coordinates of which with respect to the basis $e$
coincide with $u^{K}$, $k=\overline{1,n}$,
\[
eu=u^{K}e_{K}.
\]

\item Model with the structure-less material space. This is just the model based on orbits and
homogeneous spaces, described in some details above. So, $\Omega$ is the (structure-less) set
of material points and Inj$(\Omega,M)$ denotes the set of injections of $\Omega$ into $M$. The
spatial affine group GAf$(M)$ acts on Inj$(\Omega,M)$ through (\ref{a10}),
\begin{eqnarray}\label{a18}
A\in{\rm GAf}(M)&:&{\rm Inj}(\Omega,M)\ni\Phi\mapsto A\circ\Phi.
\end{eqnarray}
Any orbit of this action may be chosen as the configuration space of affinely rigid body.
Different orbits are related to each other by non-affine transformations. More precisely, we
usually concentrate on such orbits $Q$ that for any $\Phi\in Q$ $\Phi(\Omega)\subset M$ is not
contained in any proper affine subspace of $M$. Therefore, the body is essentially
$n$-dimensional ($n=$ dim $M$), although, obviously, it need not be so in the rigorous
topological sense (e.g., when $\Omega$ is finite the body is topologically zero-dimensional).
Let us mention, however that there are interesting applications of the model of singular
affine body.
%\cite{Rz}.
The configuration space $Q$ is then such an orbit of (\ref{a18}) that
for any $\Phi\in Q$ the subset $\Phi(\Omega)\subset M$ is contained in an affine subspace of
$M$ of dimension $k<n$. The right-acting partner of (\ref{a18}) is then constructed as
described above for the general homogeneous space $(Q,G)$. Now $G=$ GAf$(M)$ and $Q$ is an
orbit of (\ref{a18}) consisting of injections $\Phi$ with $n$-dimensional affine shells of
$\Phi(\Omega)$ ($n=$ dim $M$).

There is another way of fixing the configurations of an extended affine body with the
structure-less material space. Let us assume that the body is non-degenerate. There exists
then an $(n+1)$-element subset $B\subset \Omega$ of material points such that for any $\Phi\in
Q$ $\Phi(B)\subset M$ is not contained in any proper affine subspace, i.e., its affine shell
coincides with $M$. Let us take the elements of $B$ in some peculiar order
$(\omega_{1},\ldots,\omega_{n+1})$. Every configuration $\Phi\in Q$ is uniquely fixed by
position $(\Phi(\omega_{1}),\ldots,\Phi(\omega_{n+1}))$ of the ordered system
$(\omega_{1},\ldots,\omega_{n+1})$. The current positions $\Phi(\omega)$ of all other material
points $\omega\in\Omega$ are uniquely determined by
$(\Phi(\omega_{1}),\ldots,\Phi(\omega_{n+1}))$. The reason is that all affine relationships
between positions of material points, i.e., all linear equations satisfied by vectors
$\overrightarrow{\Phi(\omega)\Phi(\omega^{\prime})}$ ($\omega$, $\omega^{\prime}$ being
arbitrary elements of $\Omega$) are invariant during any admissible (affinely constrained)
motion. In other words, they depend only on $\omega$, $\omega^{\prime}$ but are independent of
$\Phi\in Q$.

In this way, configurations are identified with elements of the Cartesian product
$M^{n+1}=M\times\cdots\times M$ ($(n+1)$ copies of $M$). When the body is non-singular, it is
not the total $M^{n+1}$ that is admitted but its open subset consisting of such $(n+1)$-tuples
$(y_{1},\ldots,y_{A},\ldots,y_{n+1})$, which are not contained in any proper affine subspace
of $M$ (therefore, the affine shell of $y_{A}$, $A=\overline{1,(n+1)}$), coincides with $M$.
This means that the vectors $\overrightarrow{y_{1}y_{A}}$, $A=\overline{2,(n+2)}$, are
linearly independent.

For $k$-dimensional degenerate body ($k\leq n=$ dim $M$) the configuration space $Q$ (an orbit
of GAf$(M)$) may be identified with $M^{k+1}=M\times\cdots\times M$ ($(k+1)$ Cartesian
factors) or rather with its open subset consisting of $(k+1)$-tuple
$(y_{1},\ldots,y_{A},\ldots,y_{k+1})$ with $k$-dimensional affine shells.

Non-degenerate ordered $(n+1)$-tuples $y\in M^{n+1}$ may be interpreted as affine bases
(affine frames) in $M$. For any pair of such bases $y=(y_{1},\ldots,y_{A},\ldots,y_{n+1})$,
$y^{\prime}=(y^{\prime}_{1},\ldots,y^{\prime}_{A},\ldots,y^{\prime}_{n+1})$ there exists
exactly one affine transformation $\Phi\in$ GAf$(M)$ such that $y^{\prime}_{A}=\Phi(y_{A})$,
$A=\overline{1,(n+1)}$. Any such basis may be interpreted as a reference configuration.

\item Finally, if some affine frames $(\mathcal{O},E)$, $(o,e)$ are fixed both in $N$ and $M$,
these affine spaces become identified with $\mathbb{R}$, and the numerical affine group
GAf$(n,\mathbb{R})\simeq$ GL$(n,\mathbb{R})\times_{s}\mathbb{R}^{n}$ may be used as the
configuration space. Spatial and material transformations become then respectively left and
right regular translations.
\end{enumerate}

Obviously, all the above models are mutually equivalent and their formal utility and practical
usefulness depend on the kind of considered problems. For example, when describing discrete
systems we shall use the model (4) with the structure-less material space. In certain
microstructural applications and in fundamental physics one uses internal degrees of freedom
which are not interpreted in terms of composed multi-particle systems and perhaps by some
principal reasons do not admit such interpretation at all (cf. the concept of spin of
elementary particles). Obviously, then the model (3) based on the manifold of linear frames
$FM$ is the most adequate one. When one deals with the motion of structured bodies in
non-Euclidean spaces, this is practically the only adequate approach if one wishes to remain
within the framework of finite-dimensional analytical mechanics
\cite{JJS75_4,JJS76,JJS85,JJS91_3,JJS91_5,JJS02_1,all-book04,all04,all05}.

There are some delicate points concerning the connectedness of the configuration space.
Obviously, within the standard continuum treatment the singular situations are forbidden,
$\Phi$ and $\varphi$ must be bijections, and $e$ is a linear frame, not an arbitrary $n$-tuple
of vectors. Therefore, the genuine configuration space is then one of the two connected
components of AfI$(U,V)$, LI$(U,V)$, F$(V)$. Otherwise one would have to pass through the
forbidden "singular configurations". Only the connected components of group unity, i.e.,
orientation preserving subgroups GAf$^{+}(M)$, GL$^{+}(V)$, GAf$^{+}(N)$, GL$^{+}(U)$,
GL$^{+}(n,\mathbb{R})$ are admitted as transformation groups. However, it is not the case when
one deals with discrete affine bodies. Then there is nothing catastrophic in passing through
"singular" situations when at some instant of time $\Phi(\Omega)$ is contained in a proper
affine subspace of $M$. And there is nothing bad in mirror-reflected configurations forbidden
in continuum mechanics. Obviously, the above description of configuration spaces must be
modified then, e.g., $Q$ is not any longer the homogeneous space of GAf$(M)$; rather, it may
be a union of various transitivity orbits corresponding to all possible dimensions $k\leq n$
of affine shells of $\Phi(\Omega)$.

Analytical formulas will be usually expressed in terms of rectilinear Cartesian coordinates
$x^{i}$, $a^{K}$ respectively in $M$ and $N$. They are fixed by affine frames $(o,e)$,
$(\mathcal{O},E)$
\[
\overrightarrow{ox}=x^{i}(x)e_{i},\qquad \overrightarrow{\mathcal{O}a}=a^{K}(a)E_{K}.
\]
Coordinates $x^{i}$, $a^{K}$ induce parameterization of configurations. Eulerian and
Lagrangian coordinates (spatial and material variables) are related to each other by the
formula:
\[
y^{i}=\Phi^{i}(a)=x^{i}+\varphi^{i}{}_{K}a^{K},
\]
where $y^{i}$ are coordinates of the spatial position of the $a$-th material point, and
$x^{i}$ are coordinates of the spatial position $\Phi(\mathcal{O})$ of the fixed reference
point $\mathcal{O}\in N$. The quantities $(x^{i},\varphi^{i}{}_{K})$ are labels of $\Phi$ and
may be used as generalized coordinates $q^{\alpha}$, $\alpha=\overline{1,n(n+1)}$ on the
configuration space $Q$.

The reference point $\mathcal{O}\in N$, i.e., origin of Lagrange coordinates
($a^{K}(\mathcal{O})=0$) was chosen here in a completely arbitrary way. In practical problems
the choice of $\mathcal{O}$ is, as a rule, physically motivated. If there exist additional
constraints due to which some material point is immovable (i.e., the body is pinned at it),
then the material origin $\mathcal{O}$ usually is chosen just at this point. There is no
translational motion then. If translations are non-constrained, the centre of mass is usually
chosen as the material reference point. Let us remember that in situations other than
continuous medium filling up the whole space, centre of mass may happen to be placed "in
vacuum". And even if it is not the case, the centre of mass is something else than the
material point coinciding with it.

Let the reference mass distribution be described by some positive regular measure $\mu$ on
$N$; this means that the mass of the sub-body $B\subset N$ is given by
\[
\mu(B)=\int_{B}d\mu.
\]
Centre of mass $C(\mu)$ in $N$ is the only point satisfying
\[
\int \overrightarrow{C(\mu)a}d\mu(a)=0;
\]
the dipole moment of $\mu$ with respect to $C(\mu)$ vanishes. Any configuration $\Phi$ gives
rise to the $\Phi$-transported measure $\mu_{\Phi}$ on $M$,
\[
\mu_{\Phi}(\Phi(B))=\mu(B),\qquad \mu_{\Phi}(A)=\mu(\Phi^{-1}(A)).
\]
The measure $\mu_{\Phi}$ describes the Eulerian mass distribution on $M$ (current mass
distribution). The current centre of the mass distribution $C(\mu_{\Phi})\in M$ is given by
the formula:
\[
\int \overrightarrow{C(\mu_{\Phi})y}d\mu_{\Phi}(y)=0.
\]
It is well-known that the centre of mass is an invariant of affine transformations; by the
way, affine transformations may be defined just as those preserving centres of mass.
%\cite{B}.
Therefore,
\[
\Phi(C(\mu))=C(\mu_{\Phi}).
\]
And besides, for any affine transformation $A\in$ GAf$(M)$,
\[
A(C(\mu_{\Phi}))=C(\mu_{A\circ\Phi}).
\]
This is not true for non-affine configurations and transformations.

If the material reference point $\mathcal{O}$ is chosen as $C(\mu)$, i.e., Lagrangian centre
of mass, then generalized coordinates $(x^{i},\varphi^{i}{}_{K})$ on $Q$ are especially
convenient because $x^{i}$ are spatial (Eulerian) coordinates of the instantaneous position of
the centre of mass in $M$. The variables $\varphi^{i}{}_{K}$ refer to the purely relative
(internal) motion.

The physical quantity $\mu$ (Lagrangian mass distribution) is at the same time an auxiliary
geometric object underlying the convenient models $M\times$ LI$(U,V)$, $M\times$ F$(V)$ of the
configuration space of the affine body. Practically all over this treatment we put the origin
of Lagrangian coordinates at the Lagrangian centre of mass.

%eliza

The centre of mass is defined as such a point with respect to
which the dipole moment of the mass distribution vanishes. The
monopole moment equals the total mass of the body,
\[
m= \mu(N)= \int_{N} d\mu.
\]
Higher order multipole moments given an account of inertial
properties of extended bodies. Thus, the Lagrangian second-order
moment $J \in U \otimes U$ is given by
\[
J^{KL}:= \int a^{K}a^{L} d\mu(a).
\]
As mentioned, the origin of $a^{K}$-coordinates is placed at the
Lagrangian centre of mass $C(\mu)$. The above object $J^{KL}$ is
algebraically equivalent to the usual co-moving tensor of inertia
known from the rigid body mechanics. One can $\Phi$-transport it
to the physical space $M$. Obviously, the result $J(\varphi) \in V
\otimes V$ is non-constant; it depends explicitly on the
configuration $\Phi$ but only through its internal part $\varphi$,
\begin{equation}\label{a25}
    J(\varphi)^{ij}= \varphi^{i}{}_{K}\varphi^{j}{}_{L}J^{KL}.
\end{equation}
It is clear that
\begin{equation}\label{b25}
    J(\varphi)^{ij}= \int \left(y^{i} - x^{i}\right)\left(y^{j} - x^{j}\right)d \mu_{\Phi}(y).
\end{equation}
Obviously, the tensors $J$, $J(\varphi) = \Phi_{\ast } \cdot J$
are symmetric and positively definite. The most convenient choices
of the material reference frames $E$ are those diagonalizing $J$.

One can as well define higher-order multipoles, i.e.,
\begin{eqnarray}
J^{K_{1} \cdots K_{l}} &=& \int a^{K_{1}} \cdots a^{K_{l}}
d\mu(a),\nonumber\\
J(\varphi)^{i_{1} \cdots i_{l}}&=& \int \left(y^{i_{1}} -
x^{i_{1}}\right)\cdots\left(y^{j_{l}} - x^{j_{l}}\right)d
\mu_{\Phi}(y).\nonumber
\end{eqnarray}
Obviously, in affine motion
\[
J(\varphi)^{i_{1} \cdots i_{l}} = \varphi^{i_{1}}{}_{K_{1}} \cdots
\varphi^{i_{l}}{}_{K_{l}}J^{K_{1} \cdots K_{l}}.
\]
Inertial multipoles of the order $l > 2$ do not occur in mechanics
of affine bodies, nevertheless they are useful in other problems
of continuum mechanics \cite{Chan69}.

As yet we have used above only affine concepts, i.e., we remained
on the ascetic level of Tales geometry. No metric concepts like
distances and angular were used. The admissible configurations
$\Phi \in {\rm AfI}(N,M)$ are homogeneous in the sense that for
any $a \in N$ the placement $D_{a} \Phi \in {\rm LI}(U,V)$ takes
on the same value $\varphi =$ L$[\Phi]$. Nevertheless, it would be
incorrect to say that deformations are homogeneous, because
without metric (Euclidean) geometry there is no deformation
concept at all. The only well-defined "deformation" is then
violation of affine geometry, i.e., non-constancy of the mapping
$a \mapsto D_{a} \Phi$.

Let us now introduce metrical concepts. The spatial and material
metric tensors will be denoted respectively by $g \in V^{\ast }
\otimes V^{\ast }$, $\eta \in U^{\ast } \otimes U^{\ast }$. By
definition they are symmetric and positively definite. Their
contravariant inverses are denoted by $\widetilde{g} \in V \otimes
V$, $\widetilde{\eta} \in U \otimes U$, but in analytical
expressions we use the same kernel symbols; the distinction is
indicated by the use of lower- and upper-case indices,
\[
g_{ij}, \qquad g^{ij}, \qquad  \eta_{AB}, \qquad \eta^{AB}; \qquad
g^{ik} g_{kj}= \delta^{i}{}_{j}, \qquad \eta^{AC} \eta_{CB}=
\delta^{A}{}_{B}.
\]

For any configuration $\Phi \in {\rm GAf}(M)$ we define Green and
Cauchy tensors $G[\Phi] \in U^{\ast } \otimes U^{\ast }$, $C[\Phi]
\in V^{\ast } \otimes V^{\ast }$,
\[
G[\Phi]= \varphi^{\ast }g, \qquad C[\Phi]= \varphi^{-1\ast } \eta,
\]
i.e., analytically:
\[
G[\Phi]_{AB}= g_{ij}\varphi^{i}{}_{A} \varphi^{j}{}_{B}, \qquad
C[\Phi]= \eta_{AB}\varphi^{-1A}{}_{i} \varphi^{-1B}{}_{j}.
\]
In these formulas, as usual, $\varphi$ denotes the linear part of
$\Phi$, $\varphi =$ L$[\Phi]=D \Phi$. For general non-affine
configurations these tensors become fields respectively on $N$,
$M$, namely:
\[
G[\Phi]_{a}= D_{a}\Phi^{\ast } \cdot g, \qquad C[\Phi]_{y}=
D_{y}\Phi^{-1\ast } \cdot \eta.
\]
Analytically:
\[
G_{AB}=g_{ij} \frac{\partial y^{i}}{\partial a^{A}} \frac{\partial
y^{j}}{\partial a^{B}}, \qquad C_{ij}=\eta_{AB} \frac{\partial
a^{A}}{\partial y^{i}} \frac{\partial a^{B}}{\partial y^{j}}.
\]

Obviously, Green and Cauchy tensors are symmetric and positively
definite. $G$ is built of the spatial metric tensor $g$ and is
independent of the material metric $\eta$. And conversely, $C$ is
independent of $g$ and explicitly depends on $\eta$. Therefore,
the traditional term "deformations tensors" is rather non-adequate
here; the deformation concept presumes comparison of two metrics,
whereas $G$ and $C$ are well-defined even if respectively $\eta$
and $g$ are not fixed at all. Let us assume they are both fixed
and denote the corresponding Euclidean space structures by
$(N,U,\rightarrow, \eta)$, $(M,V,\rightarrow,g)$. When some
configuration $\Phi$ is fixed, any of these two spaces is endowed
by two metric-like tensors, respectively, $G, \eta \in \in U^{\ast
} \otimes U^{\ast }$, $C,g \in V^{\ast } \otimes V^{\ast }$. The
Lagrange and Euler deformation tensors are respectively given by
\[
E[\Phi]= \frac{1}{2} \left(G[\Phi]- \eta \right), \qquad
e[\Phi]=\frac{1}{2} \left(g- e[\Phi]\right).
\]
They vanish in the non-deformed configurations, i.e., when $\Phi$
are isometries. Obviously, these are usual (metrically) rigid body
configurations. Their manifold will be denoted by Is$(N, \eta;
M,g) \subset {\rm AfI}(N,M)$. The isometry groups Is$(N, \eta)
\subset {\rm GAf}(N)$, Is$(M, g) \subset {\rm GAf}(M)$ act on
Is$(N, \eta; M,g)$ respectively on the right and on the left in
the sense of (\ref{b6}), (\ref{a6}). Obviously, in realistic
classical mechanics of (metrically) rigid body mirror-reflected
coordinates are excluded and the genuine configuration space is
given by some connected component of Is$(N, \eta; M,g)$. When
orientations are fixed in $N$, $M$, this will be the manifold
Is$^{+}(N, \eta; M,g)$ of orientation-preserving isometries.
Physically admissible symmetries are given by the connected
subgroups Is$^{+}(N, \eta) \subset {\rm GAf}^{+}(N)$, Is$^{+}(M,g)
\subset {\rm GAf}^{+}(M)$ of orientation-preserving
transformations (obviously, these group themselves do not assume
any fixed orientation; they preserve separately both of them).

When translational degrees of freedom are neglected,
configurations of internal (relative) motion are elements of D$(U,
\eta; V,g)$, i.e., the manifold of linear isometries of $(U,
\eta)$ onto $(V,g)$. The corresponding spatial and material
transformations are respectively elements of the subgroup O$(V,g)
\subset {\rm GL}(V)$, O$(U, \eta) \subset {\rm GL}(U)$; $g$- and
$\eta$-orthogonal transformation groups. And again in classical
rigid body mechanics one should restrict ourselves to one of the
two connected components of O$(U, \eta; V,g)$. When orientations
in $U$, $V$ are fixed, this is SO$(U, \eta; V,g) \subset {\rm
LI}^{+}(U, \eta; V,g)$, i.e., the manifold of
orientation-preserving linear isometries. The connected components
are ruled by the proper orthogonal groups SO$(V,g)$, SO$(U, \eta)$
consisting of isometries with positive, thus plus-one-determinants
(no orientations in $V$, $U$ needed for fixing these subgroups).
When using the matrix representations we describe configurations
by elements of the orthogonal group O$(n, \mathbb{R})$, or in
classical problems, by elements of the proper rotation group
SO$(n, \mathbb{R})$.

Let us now fix some metric tensors $\eta$, $g$. The geometric
structure becomes more rich and certain additional object may be
defined. For example, the material reference frame maybe made less
arbitrary and more based on physical concepts. Let us define two
tensor objects built of the inertial tensor $J \in U \otimes U$
namely $\widetilde{J} \in U^{\ast } \otimes U^{\ast }$ and
$\widehat{J} \in U \otimes U^{\ast } \simeq {\rm L}(U)$,
analytically given by the formulas:
\[
{\widetilde{J}}_{AC}J^{CB} = \delta_{A}{}^{B}, \qquad
{\widehat{J}}^{A}{}_{B} = J^{AC} \eta_{CB}.
\]
Obviously, the object $\widetilde{J}$ is non-metrical, but
$\widehat{J}$ depends explicitly on the material metric tensor
$\eta$. Now we have a well-defined eigenproblem in $U$:
\[
\widehat{J}E = l E.
\]
In the generic non-degenerate case there are $n$ mutually distinct
eigenvalues $l_{A}$, $A = 1, \ldots, n$, and $n$ mutually
orthogonal eigendirections. The directions are determined by
vectors $E_{A}$ which may be chosen $\eta$-normalized to unity and
such that the orthonormal frame $E= \left(E_{1}, \ldots, E_{A},
\ldots, E_{n}\right)$ is oriented positively with respect to the
fixed orientation in $U$. When the eigenvalues $l_{A}$ are ordered
by convention in increasing order then $E_{A}$ are unique up to
multiplying some of them by minus-one-factors. And the inertial
tensor $J$ is then represented as follows:
\begin{equation}\label{a30}
    J = \sum^{n}_{A=1} J^{A} E_{A} \otimes E_{A}.
\end{equation}

This is just the best choice of the material reference frame.
Obviously, it is no longer unique when degeneracy occurs, but
these are non-generic situations. The extreme degeneracy
correspond to the $\eta$ -spherical body, when $J$ is proportional
to $\eta$
\[
J^{AB} = \mu \eta^{AB}.
\]
\textbf{Remark}: If $\varphi$ is not an isometry, then obviously
the co-moving vectors $e_{A} = \varphi E_{A}$ are not orthonormal
in the $g$-sense, however, they are orthonormal with respect to
the Cauchy tensor $C$ used as a "metric" in $V$. And then:
\[
J[\varphi] = \sum^{n}_{A=1} J^{A} e_{A} \otimes e_{A}
\]
with the same values $J^{A}$ which occur in (\ref{a30}).

\noindent\textbf{Remark}: For the sake of economy of symbols we
could as well denote ${\widehat{J}}^{A}{}_{B}$ by $J^{A}{}_{B}$,
following the convention used in Euclidean and Riemannian
geometry. But it is not the case with ${\widetilde{J}}_{AB}$.
Writing it as $J_{AB}$ would suggest $\eta_{AC} \eta_{BD}J^{CD}$,
quite incorrectly, this is not the $\eta$-lowering of indices.
Typical notational shorthands may be misleading. Only for the
$\eta$-manipulation of indices and for the contravariant inverse
of $\eta$ ($\eta^{AC} \eta_{CB}= \delta^{A}{}_{B}$) we can safely
use the index manipulation with non-modified kernel symbols.

When metric tensors are fixed, we can discuss the problem of
interaction between rotations and deformations. In any of linear
spaces $U$, $V$ we are given two symmetric positively definite
tensors: $\eta \in U^{\ast } \otimes U^{\ast }$, $G \in U^{\ast }
\otimes U^{\ast }$, $g \in V^{\ast } \otimes V^{\ast}$, $C \in
V^{\ast} \otimes V^{\ast}$. Obviously, $\eta$, $g$ are fixed
whereas $G$, $C$ are configuration-dependent. Just as previously,
we can define the byproduct-objects $\widetilde{G} \in U \otimes
U$, $\widehat{G} \in U \otimes U^{\ast} \simeq {\rm L}(U)$,
$\widetilde{C} \in V \otimes V$, $\widehat{C} \in V \otimes
V^{\ast} \simeq {\rm L}(V)$. Analytically they are given by
\[
\widetilde{G}^{AC}G_{CB} = \delta^{A}{}_{B}, \qquad
\widehat{G}^{A}{}_{B}= \eta^{AC}G_{CB}, \qquad
\widetilde{C}^{ik}C_{jk} = \delta^{i}{}_{j}, \qquad
\widehat{C}^{i}{}_{j}= g^{ik}C_{kj}.
\]
Again the same care must be taken as to the upper- and lower-case
indices.

And now $G[\varphi]$, $C[\varphi]$ may be expressed in terms of
their $\eta$- and $g$-orthonormal bases $\left( \ldots,
F_{a}[\varphi], \ldots\right)$, $\left( \ldots, f_{a}[\varphi],
\ldots\right)$ in $U$, $V$
\begin{eqnarray}
G[\varphi]&=& \sum^{n}_{a=1} \lambda_{a}[\varphi] F^{a}[\varphi] \otimes F^{a}[\varphi],\nonumber\\
C[\varphi]&=& \sum^{n}_{a=1} \frac{1}{\lambda_{a}[\varphi]}
f^{a}[\varphi] \otimes f^{a}[\varphi];\nonumber
\end{eqnarray}
obviously, $\left( \ldots, F^{a}[\varphi], \ldots\right)$, $\left(
\ldots, f^{a}[\varphi], \ldots\right)$ are the dual orthonormal
bases of $U^{\ast}$, $V^{\ast}$. When there is no danger of
misunderstanding, the label $\varphi$ at $\lambda_{a}$, $F^{a}$,
$f^{a}$, $F_{a}$, $f_{a}$ may be omitted.

The quantities $\lambda_{a}[\varphi]$ are deformation invariants
in the sense that they do not feel spatial and material linear
isometries,
\[
\lambda_{a}[\alpha \varphi \beta] = \lambda_{a}[\varphi]
\]
for any $\alpha \in {\rm O}(V, g)$, $\beta \in {\rm O}(U, \eta)$.
The more so they are non-sensitive with respect to the spatial and
material affine isometries (because translations evidently do not
affect them). Unlike this, the Green and Cauchy deformation
tensors are non-sensitive only with respect to spatial and
material isometries,
\[
G[\alpha \varphi]= G[\varphi], \qquad C[\varphi \beta] =
C[\varphi]
\]
for any $\alpha \in {\rm O}(V, g)$, $\beta \in {\rm O}(U, \eta)$,
but not conversely. $F^{a}[\varphi] \in U$, $f^{a}[\varphi]\in V$
are normalized eigenvectors respectively for
$\widehat{G}[\varphi]$, $\widehat{C}[\varphi]$:
\[
\widehat{G}[\varphi]F_{a}[\varphi]= \lambda_{a}F_{a}[\varphi],
\qquad \widehat{C}[\varphi]f_{a}[\varphi]=
\frac{1}{\lambda_{a}}f_{a}[\varphi],
\]
\[
\eta \left(F_{a}, F_{b}\right) = \delta_{ab}, \qquad g
\left(f_{a}, f_{b}\right) = \delta_{ab},
\]
and they are essentially unique for the non-degenerate spectra.

The fixed bases $\left( \ldots, F_{a}[\varphi], \ldots\right)$,
$\left( \ldots, f_{a}[\varphi], \ldots\right)$, being orthonormal,
give rise some isometry $U[\varphi] \in {\rm O}(U, \eta;V,g)$,
namely such one that
\[
U[\varphi]F_{a}[\varphi]=f_{a}[\varphi], \qquad a=1, \ldots, n.
\]
The $\varphi$ itself may be written in the form
\begin{equation}\label{a33}
    \varphi = U[\varphi] A[\varphi],
\end{equation}
where $A[\varphi] : U \rightarrow U$ is $\eta$-symmetric and
positively definite, i.e.,
\[
\eta\left(A[\varphi] u, v\right) = \eta\left(u, A[\varphi]
v\right), \qquad \eta\left(A[\varphi]u,u\right)>0,
\]
for arbitrary $u,v \in U$ and arbitrary $u \neq 0$ in the
inequality. Analytically:
\[
\eta_{AC}A[\varphi]^{C}{}_{B}= \eta_{BC}A[\varphi]^{C}{}_{A},
\]
and the matrix $\left[A[\varphi]^{A}{}_{B}\right]$ has positive
eigenvalues, coinciding, by the way, with square roots of
deformation invariants $\lambda_{a}$ (which, obviously, are also
positive). Obviously, (\ref{a33}) is a geometric interpretation of
the polar decomposition, and analytically, when we put $U = V =
{\mathbb{R}}^{n}$ it exactly coincides with the usual polar
decomposition known from the matrix theory. It may be also
alternativelly written in the form:
\begin{equation}\label{b33}
    \varphi = B[\varphi] U[\varphi],
\end{equation}
where $B[\varphi] : V \rightarrow V$ is $g$-symmetric and
positively definite, and obviously:
\[
B[\varphi]=U[\varphi]A[\varphi]U[\varphi]^{-1}.
\]
These are simply the left and right polar decomposition known from
the matrix theory. It is clear that Green and Cauchy tensor
satisfy respectively:
\[
\widehat{G}[\varphi]=A[\varphi]^{2}, \qquad
\widehat{C}[\varphi]=B[\varphi]^{-2}.
\]
Let us mention, there are also other possible choices of
deformation invariants; every system of $n$ functionally
independent functions of $\lambda_{a}$, $a=1, \ldots, n$  may be
used as a basic system of invariants. Let us remind a few
popularly used system e.g.,
\[
{\mathcal{K}}_{a}[\varphi]={\rm
Tr}\left(\widehat{G}[\varphi]^{a}\right), \qquad a=1, \ldots, n.
\]
Obviously,
\[
{\mathcal{K}}_{a}= \sum^{n}_{i=1} \left(\lambda_{i}\right)^{a}.
\]
Another possibility is the system of coefficients of the
characteristic polynomial of $\widehat{G}[\varphi]$,
$I_{p}[\varphi]$:
\[
\det \left[\widehat{G}[\varphi]^{A}{}_{B} - \lambda
\delta^{A}{}_{B}\right] = \sum^{n}_{k=0}(-1)^{k}I_{n-k}[\varphi]
\lambda^{k}.
\]
Obviously, $I_{0}=1$, and for $p=1, \ldots, n$, $I_{p}$ is the sum
of all possible products of $p$ quantities $\lambda_{a}$ with
different (but not necessarily disjoint) sets of labels $a=1,
\ldots, n$, e.g.,
\[
I_{1}= \sum^{n}_{i=1} \lambda_{i}={\rm
Tr}\left(\widehat{G}\right),\qquad I_{n}= \lambda_{1} \cdots
\lambda_{n}= \det \left(\widehat{G}\right).
\]
In the physical three-dimensional case
\[
I_{2}= \lambda_{2} \lambda_{3} + \lambda_{3} \lambda_{1} +
\lambda_{1} \lambda_{2}.
\]
Let us observe that orthonormal bases $\left( \ldots,
f_{a}[\varphi], \ldots\right)$, $\left( \ldots, F_{a}[\varphi],
\ldots\right)$ represent formally configurations of two fictitious
rigid bodies, respectively in $(V,g)$ and $(U, \eta)$. They refer
respectively to the eigenaxes of the Cauchy and Green deformations
tensors, therefore, they tell us how the deformation state is
oriented with respect to $V$, $U$ (what are instantaneous
positions of deformation ellipsoids). Unlike this, deformation
invariants contain only the scalar deformation about the
deformation state (how large are stretchings). The manifold of
scalar deformation states (parametrized by deformation invariants)
may be considered as double-coset-space of LI$(U,V)$ with respect
to the left and right actions of O$(V,g)$, O$(U, \eta)$,
\[
{\rm Inv}= {\rm O}(V,g) \backslash {\rm LI}(U,V) / {\rm O}(U,
\eta),
\]
or, when reflections are excluded,
\[
{\rm Inv}= {\rm SO}(V,g) \backslash {\rm LI}^{+}(U,V) / {\rm
SO}(U, \eta).
\]

Let us observe that, as usual linear frames $F[\varphi]$, $f[\varphi]$ may be naturally
identified with linear isomorphisms: $R[\varphi]: {\mathbb{R}}^{n} \rightarrow U$,
$L[\varphi]: {\mathbb{R}}^{n} \rightarrow V$. As they are orthonormal they are linear
isometries of $(U, \eta)$ onto $({\mathbb{R}}^{n},\delta)$ and of $(V,g)$ onto
$({\mathbb{R}}^{n},\delta)$; $\delta$ denotes here the natural Descartes-Kronecker metric of
${\mathbb{R}}^{n}$. Similarly, the dual co-frames $\widetilde{F}[\varphi]$,
$\widetilde{f}[\varphi]$ may be identified with the inverse mappings $R[\varphi]^{-1}: U
\rightarrow  {\mathbb{R}}^{n}$, $L[\varphi]^{-1}:  V \rightarrow {\mathbb{R}}^{n}$; obviously,
they are also linear isometries. One can show that $\varphi$ may be represented as
 \[
\varphi = L[\varphi]D[\varphi]R[\varphi]^{-1},
 \]
where the linear mapping $D[\varphi]: {\mathbb{R}}^{n} \rightarrow
{\mathbb{R}}^{n}$, i.e., simply a matrix, is diagonal,
 \[
D[\varphi]= \left[
\begin{array}{ccc}
  D_{1}[\varphi] & \ldots & 0 \\
  \vdots & \ddots & \vdots \\
  0 & \ldots & D_{n}[\varphi] \\
\end{array}
\right] = {\rm Diag}\left( D_{1}[\varphi], \ldots,
D_{n}[\varphi]\right).
 \]

If there is no danger of misunderstanding we shall omit the label
$\varphi$ at the quantities $L$, $D$, $R$. So, finally, we write
\[
\varphi = LDR^{-1}
\]
and in this way any internal (relative) configuration $\varphi$ is
formally identified with the configuration of two fictitious
(metrically) rigid bodies and $n$ purely oscillatory degrees of
freedom of the stretching state.

If we put $U=V= {\mathbb{R}}^{n}$, then the above decomposition
(two-polar decomposition, triple decomposition) is formally
obtained from the polar one. Namely, for any $\varphi = {\rm
GL}(n,{\mathbb{R}})$ one starts from the polar decomposition
\[
\varphi = UA, \qquad U \in {\rm O}(n,{\mathbb{R}}), \qquad A \in
{\rm Sym}^{+}(n,{\mathbb{R}})
\]
and then $A$ is orthogonally diagonalized,
\[
A = RDR^{-1}, \qquad D\in {\rm Diag}(n,{\mathbb{R}}), \qquad R \in
{\rm O}(n,{\mathbb{R}}),
\]
so, finally,
\[
\varphi = LDR^{-1}, \qquad L= UR \in {\rm O}(n,{\mathbb{R}}).
\]
Unlike the polar decomposition, the two-polar one is non-unique.
In the non-degenerate case, when all diagonal elements of $D$ are
pairwise distinct, this non-uniqueness is discrete and controlled
by the permutation group ${\rm S}^{(n)}$ interchanging deformation
invariants. When degeneracy occurs the non-uniqueness is more
catastrophic, in a sense continuous, and resembles the singularity
of spherical coordinates at $r=0$ (although, one must say, it is
much more complicated).

Let us finish with some kinematical concepts. Generalized velocity
of an affine body is given by the pair $(v, \xi) \in V \times {\rm
L}(U,V)$ consisting of the translational velocity $v$ and the
internal one $\xi$. On a given classical motion ${\mathbb{R}} \ni
t \mapsto \left(x(t), \varphi(t)\right)$ it is analytically given
by the system
\[
\left(\ldots, \frac{dx^{i}}{dt}, \ldots; \ldots, \frac{d
\varphi^{i}{}_{A}}{dt}, \ldots\right).
\]
When $U= {\mathbb{R}}^{n}$, i.e., $Q=M \times F(V)$, then
velocities are elements of $V \times V^{n}=V^{n+1}$.

It is convenient to use affine velocities in the spatial and
material (co-moving) representations, $\Omega \in$ L$(V)$,
$\widehat{\Omega} \in$ L$(U)$, namely:
\[
\Omega = \xi \varphi^{-1}= \frac{d \varphi}{dt} \varphi^{-1},
\qquad \widehat{\Omega} = \varphi^{-1} \xi= \varphi^{-1} \frac{d
\varphi}{dt}.
\]
They are interrelated as follows:
\[
\Omega = \varphi \widehat{\Omega} \varphi^{-1}.
\]
Analytically:
\[
\Omega^{i}{}_{j} = \frac{d \varphi^{i}{}_{A}}{dt}
\varphi^{-1A}{}_{j}, \qquad \widehat{\Omega}^{A}{}_{B} =
\varphi^{-1A}{}_{i} \frac{d \varphi^{i}{}_{B}}{dt},\qquad
\Omega^{i}{}_{j} = \varphi^{i}{}_{A} \widehat{\Omega}^{A}{}_{B}
\varphi^{-1B}{}_{j}.
\]
Eringen in his micromorphic theory \cite{Erin68,Erin75_1,Erin75_2}
uses for them the term "gyration". They are Lie-algebraic objects
related respectively to the right-invariant and left-invariant
vector fields and differential forms,
\[
X[E]_{\varphi} = E \varphi, \qquad
X\left[\widehat{E}\right]_{\varphi} = \varphi \widehat{E},
\]
\begin{equation}\label{a37}
    \omega = d \varphi \varphi^{-1}, \qquad \widehat{\omega} = \varphi^{-1}
    d \varphi.
\end{equation}

In very rough, formal terms we would say that $\omega$,
$\widehat{\omega}$ are obtained from $\Omega$, $\widehat{\Omega}$
via the $dt$-multiplying. More rigorously, they are respectively
$L(V)$-valued and $L(U)$-valued differential one-forms on
LI$(U,V)$. Their evaluations on vectors tangent to trajectories
just coincide with $\Omega$, $\widehat{\Omega}$.

The right and left invariance is meant obviously in  the sense of
transformations (\ref{d6}), (\ref{c6}). They become right and left
regular group translations when $U =V = {\mathbb{R}}^{n}$ and
LI$(U,V)$ is identified with GL$(n,{\mathbb{R}})$. In the above
formulas for vector fields $E$ and $\widehat{E}$ are respectively
fixed elements of L$(V)$, L$(U)$.

Affine velocities are non-holonomic in the sense that there no
generalized coordinates for which they would be time derivatives.
This is due to the non-commutativity of the full linear group.

In continuum mechanics $\Omega$ may be interpreted in terms of the
Euler velocity field. Namely, the material point which at a given
instant of time passes the spatial point $y \in M$ has the
velocity:
\[
v(y) = v + \Omega \overrightarrow{xy};
\]
analytically,
\[
v^{i}(y)= v^{i} + \Omega^{i}{}_{j}\left(y^{j} - x^{j}\right).
\]
When the motion is metrically-rigid, i.e., gyroscopic, then affine
velocity becomes skew-symmetric with respect to the appropriate
metric tensor,
\begin{eqnarray}
&&\Omega^{i}{}_{j}=- g^{ia} g_{jb}\Omega^{b}{}_{a} =
-\Omega_{j}{}^{i},\nonumber \\
&&\widehat{\Omega}^{A}{}_{B}=- \eta^{AK}
\eta_{BL}\widehat{\Omega}^{L}{}_{K} =
-\widehat{\Omega}_{B}{}^{A},\nonumber
\end{eqnarray}
These objects are the usual angular velocities, respectively in
the spatial and co-moving representation. In the physical
three-dimensional case they are identified in a standard way the
usual pseudo-vectors of angular velocity; namely, in orthonormal
coordinates
\begin{eqnarray}
\Omega^{i}{}_{j} = - \varepsilon^{i}{}_{jk} \Omega^{k}, &\qquad&
\widehat{\Omega}^{A}{}_{B} = - \varepsilon^{A}{}_{BC}
\widehat{\Omega}^{C},\nonumber\\
\Omega^{i} = - \frac{1}{2} \varepsilon^{i}{}_{j}{}^{k}
\Omega^{j}{}_{k}, &\qquad& \widehat{\Omega}^{A} = - \frac{1}{2}
\varepsilon^{A}{}_{B}{}^{C} \widehat{\Omega}^{B}{}_{C},\qquad
\Omega^{i}= \varphi^{i}{}_{A}\widehat{\Omega}^{A}.\nonumber
\end{eqnarray}
Obviously, $\varepsilon$ denotes here the totally antisymmetric
Ricci symbol and indices are trivially shifted with the use of
Kronecker-delta (orthonormal coordinates).

Obviously, in the two-dimensional case (also physically
interesting), angular velocities are one-dimensional objects,
$\Omega$ and $\widehat{\Omega}$ numerically coincide and in
orthonormal coordinates:
\begin{eqnarray}
&&\left[\Omega^{i}{}_{j}\right]=\left[\widehat{\Omega}^{A}{}_{B}\right]=
\omega \left[
\begin{array}{cc}
  0 & -1 \\
  1 & 0 \\
\end{array}
\right] = \frac{d \theta}{dt}\left[
\begin{array}{cc}
  0 & -1 \\
  1 & 0 \\
\end{array}
\right],\nonumber \\
&&\left[\varphi^{i}{}_{A}\right]=\left[
\begin{array}{cc}
  \cos \theta & - \sin \theta \\
  \sin \theta & \cos \theta \\
\end{array}%
\right].\nonumber
\end{eqnarray}
For the fictitious rigid bodies corresponding to the polar and
two-polar decompositions we also introduce the corresponding
angular velocities:
\begin{eqnarray}
\omega &=& \frac{dU}{dt}U^{-1} \in {\rm SO}(V,g)^{\prime} \subset
{\rm
L}(V), \nonumber \\
\widehat{\omega} &=& U^{-1}\frac{dU}{dt} \in {\rm
SO}(U,\eta)^{\prime} \subset {\rm
L}(U), \nonumber \\
\chi &=& \frac{dL}{dt}L^{-1} \in {\rm SO}(V,g)^{\prime} \subset
{\rm
L}(V), \label{a39} \\
\widehat{\chi} &=& L^{-1} \frac{dL}{dt} \in {\rm
SO}(n,{\mathbb{R}})^{\prime} \subset
{\rm L}(n, {\mathbb{R}}), \nonumber \\
\vartheta &=& \frac{dR}{dt}R^{-1} \in {\rm SO}(U,\eta)^{\prime}
\subset {\rm L}(U), \nonumber \\
\widehat{\vartheta} &=& R^{-1} \frac{dR}{dt} \in {\rm
SO}(n,{\mathbb{R}})^{\prime} \subset {\rm L}(n, {\mathbb{R}}).
\nonumber
\end{eqnarray}
These objects are elements of the of the indicated Lie algebras of
orthogonal groups. So, $\omega$, $\chi$ are $g$-skew-symmetric,
$\widehat{\omega}$, $\vartheta$ are $\eta$-skew-symmetric, and
$\widehat{\chi}$, $\widehat{\vartheta}$ are skew-symmetric in the
usual Kronecker sense.

In certain problems it is convenient to use co-moving
representation of the translational velocity,
\[
{\widehat{v}}^{A}:=\varphi^{-1A}{}_{i}{v}^{i} =
\varphi^{-1A}{}_{i} \frac{dx^{i}}{dt}.
\]

Canonical moments, i.e. linear functional on generalized
velocities are pairs $(p, \pi) \in V^{\ast} \times {\rm L}(V,U)$,
i.e., analytically $\left(\ldots, p_{i}, \ldots; \ldots,
p^{A}{}_{i}, \ldots\right)$. Their evaluations on virtual
velocities are given by
\begin{equation}\label{a40}
    \left\langle(p, \pi), (v, \xi)\right\rangle =
    \left\langle p,v\right\rangle + {\rm Tr}\left(\pi
    \xi\right)=p_{i}v^{i}+ p^{A}{}_{i}\xi^{i}{}_{A}.
\end{equation}
When $U={\mathbb{R}}^{n}$ and $Q = M \times F(V)$, then canonical
momenta are elements of $V^{\ast} \times V^{\ast n}=V^{\ast
(n+1)}$.

The duality between $\pi$ and $\xi$ may be expressed in terms of
the duality between affine spin and affine velocity. More
rigorously, affine spin $\Sigma \in {\rm L}(V)$ and its co-moving
representation $\widehat{\Sigma} \in {\rm L}(U)$ are defined as:
\[
\Sigma= \varphi \pi, \qquad \widehat{\Sigma} = \pi \varphi.
\]

Analytically
\[
\Sigma^{i}{}_{j} = \varphi^{i}{}_{A}p^{A}{}_{j}, \qquad
\widehat{\Sigma}^{A}{}_{B} = p^{A}{}_{i} \varphi^{i}{}_{B}.
\]
One uses also the term "hypermomentum". The quantities $\Sigma$,
$\widehat{\Sigma}$ are respectively Hamiltonian generators of the
transformation groups (\ref{d6}), (\ref{c6}).

Linear momentum $p_{i}$ generates spatial translations. In many
problems it is convenient to use its co-moving representation
\[
\widehat{p}_{A} = p_{i} \varphi^{i}{}_{A}
\]
which has to do with the material translations.

One uses also the orbital affine momentum and the total affine
momentum $\Lambda$, $\mathcal{J}$, given respectively by
\[
\Lambda^{i}{}_{j} = x^{i}p_{j}, \qquad {\mathcal{J}}^{i}{}_{j} =
\Lambda^{i}{}_{j}+ \Sigma^{i}{}_{j}.
\]
Unlike $\Sigma$ the quantities $\Lambda$, $\mathcal{J}$ depend on the choice of the origin $o
\in M$ of affine coordinates in $M$. And $\mathcal{J}$, more precisely $\mathcal{J}(o)$ is a
Hamiltonian generator (momentum mapping) of the centre-affine subgroup GAf$(M,o) \subset {\rm
GAf(M)}$ (affine transformations preserving $o$) acting through (\ref{b6}).

The doubled skew-symmetric parts of hypermomenta,
\begin{eqnarray}
S^{i}{}_{j} &=& \Sigma^{i}{}_{j}- g^{ik}g_{jl}\Sigma^{l}{}_{k},
\nonumber \\
L^{i}{}_{j} &=& x^{i}p_{j}- g^{ik}g_{jl}x^{l}p_{k},
\nonumber \\
\Im^{i}{}_{j} &=& L^{i}{}_{j}+ S^{i}{}_{j}, \nonumber
\end{eqnarray}
are the usual angular momenta: the internal (spin), the orbital,
and the total ones. They are Hamiltonian generators (momentum
mapping) of the corresponding isometry groups. The quantity
\[
V^{A}{}_{B} :=\widehat{\Sigma}^{A}{}_{B} - \eta^{AL}\eta_{BK}
\widehat{\Sigma}^{K}{}_{L},
\]
called by Dyson "vorticity", is the Hamiltonian generator of the
right-acting rotation group (\ref{d7}). When $\varphi$ is not an
isometry, then $V$ is not the co-moving representation of $S$,
\[
S^{i}{}_{j} \neq \varphi^{i}{}_{A}{V}^{A}{}_{B}
\varphi^{-1B}{}_{j}.
\]
$S$ and $V$ generate respectively spatial and material rotations
of internal degrees of freedom. $\Im$ and $p$ together are
Hamiltonian generators of spatial isometries.

In a sense, $\Sigma$ and $\widehat{\Sigma}$ are non-holonomic
canonical momenta; non-holonomic, because their Poisson brackets
do not vanish.

The pairing between internal canonical momenta and velocities may
be now expressed as follows:
\[
\left\langle \Sigma, \Omega \right\rangle = \left\langle
\widehat{\Sigma}, \widehat{\Omega} \right\rangle = {\rm
Tr}\left(\Sigma \Omega\right) = {\rm Tr}\left(\widehat{\Sigma}
\widehat{\Omega}\right) = {\rm Tr}\left(\pi \xi\right).
\]
Analytically:
\[
\left\langle \Sigma, \Omega \right\rangle = \left\langle
\widehat{\Sigma}, \widehat{\Omega} \right\rangle =
\Sigma^{j}{}_{i}\Omega^{i}{}_{j} =
{\widehat{\Sigma}}^{B}{}_{A}{\widehat{\Omega}}^{A}{}_{B}.
\]

Just as $\Omega$, $\widehat{\Omega}$ themselves, $\Sigma$,
$\widehat{\Sigma}$ may be interpreted in terms of right - and
left-invariant vector fields or differential forms.

We use the standard conventions of differential geometry according
to which vector fields with components $Z^{i}$ related to local
coordinates $z^{i}$ are identified with first-order differential
operators
\[
Z = Z^{i} \frac{\partial}{\partial z^{i}}.
\]
Then $\Sigma$, $\widehat{\Sigma}$ as systems of vector fields dual
to systems of Pfaff forms (\ref{a37}) (more precisely, to L$(V)$-
and L$(U)$-valued differential one-forms) are given by
\begin{equation}\label{a43}
    E^{i}{}_{j} =\varphi^{i}{}_{A}\frac{\partial}{\partial
    \varphi^{j}{}_{A}}, \qquad \widehat{E}^{A}{}_{B} =\varphi^{i}{}_{B}\frac{\partial}{\partial
    \varphi^{i}{}_{A}}.
\end{equation}

In other words, at the point $\varphi \in {\rm LI}(U,V)$, the
$\left({}^{k}{}_{A}\right)$-th component of $E^{i}{}_{j}$ equals
$\varphi^{i}{}_{A}\delta^{k}{}_{j}$, and the
$\left({}^{i}{}_{C}\right)$-th component of
$\widehat{E}^{A}{}_{B}$ equals
$\varphi^{i}{}_{B}\delta^{A}{}_{C}$.

Interpreted as invariant forms $\Sigma$- and
$\widehat{\Sigma}$-objects become respectively the following
fields on ${\rm LI}(U,V)$:
\[
Y[F]_{\varphi} = \varphi^{-1}F, \qquad
Y\left[\widehat{F}\right]_{\varphi} = \widehat{F}\varphi^{-1},
\]
where $F$, $\widehat{F}$ are arbitrarilly fixed elements of ${\rm
L}(V)$, ${\rm L}(U)$, and we remember that the dual space ${\rm
L}(U,V)^{\ast}$ is canonically isomorphic with ${\rm L}(V,U)$
through the formula (\ref{a40}).

Let us quote the obvious transformation rules of $\Omega$,
$\widehat{\Omega}$, $\Sigma$, $\widehat{\Sigma}$ under
transformations (\ref{d6}), (\ref{c6}) of internal degrees of
freedom:
\begin{eqnarray}
\alpha \in {\rm GL}(V)&:& \qquad \Omega \mapsto \alpha \Omega
\alpha^{-1}, \qquad \widehat{\Omega} \mapsto \widehat{\Omega}
\nonumber \\
\beta \in {\rm GL}(U)&:& \qquad \Omega \mapsto \Omega, \qquad
\qquad\ \widehat{\Omega} \mapsto \beta^{-1} \widehat{\Omega} \beta
\nonumber \\
\alpha \in {\rm GL}(V)&:& \qquad \Sigma \mapsto \alpha \Sigma
\alpha^{-1}, \qquad \widehat{\Sigma} \mapsto \widehat{\Sigma} \nonumber \\
\beta \in {\rm GL}(U)&:& \qquad \Sigma \mapsto \Sigma, \qquad
\qquad\ \widehat{\Sigma} \mapsto \beta^{-1} \widehat{\Sigma} \beta
\nonumber
\end{eqnarray}
Now let us quote the basic Poisson brackets. The most important of
them, namely those involving the above generators, are determined
by the structure constans of the linear and affine groups.
\begin{eqnarray}
\left\{\Sigma^{i}{}_{j}, \Sigma^{k}{}_{l}\right\} =
\delta^{i}{}_{l}\Sigma^{k}{}_{j} -
\delta^{k}{}_{j}\Sigma^{i}{}_{l}, &\qquad&
\left\{\widehat{\Sigma}^{A}{}_{B},
\widehat{\Sigma}^{C}{}_{D}\right\} =
\delta^{C}{}_{B}\widehat{\Sigma}^{A}{}_{D} -
\delta^{A}{}_{D}\widehat{\Sigma}^{C}{}_{B},\nonumber\\
\left\{\mathcal{J}^{i}{}_{j}, \mathcal{J}^{k}{}_{l}\right\} =
\delta^{i}{}_{l}\mathcal{J}^{k}{}_{j} -
\delta^{k}{}_{j}\mathcal{J}^{i}{}_{l},&\qquad&
\left\{\Sigma^{i}{}_{j},
\widehat{\Sigma}^{A}{}_{B}\right\} = 0, \nonumber\\
\left\{\Lambda^{i}{}_{j}, \Lambda^{k}{}_{l}\right\} =
\delta^{i}{}_{l}\Lambda^{k}{}_{j} -
\delta^{k}{}_{j}\Lambda^{i}{}_{l}&\qquad&
\left\{\widehat{\Sigma}^{A}{}_{B}, \widehat{p}_{C}\right\}
=\delta^{A}{}_{C}\widehat{p}_{B}, \nonumber \\
\left\{\mathcal{J}^{i}{}_{j}, p_{k}\right\} =
\left\{\Lambda^{i}{}_{j}, p_{k}\right\} =\delta^{i}{}_{k}p_{j}.&&
\nonumber
\end{eqnarray}
For any function $F$ depending only on generalized coordinate
$x^{i}$, $\varphi^{i}{}_{A}$ we have:
\begin{eqnarray}
\left\{\Sigma^{i}{}_{j}, F\right\} &=& -E^{i}{}_{j}F
=-\varphi^{i}{}_{A} \frac{\partial F}{\partial \varphi^{j}{}_{A}}, \nonumber \\
\left\{\Lambda^{i}{}_{j}, F\right\} &=& -x^{i} \frac{\partial F}{\partial x^{j}}, \nonumber \\
\left\{\mathcal{J}^{i}{}_{j}, F\right\} &=& -x^{i} \frac{\partial
F}{\partial x^{j}}-\varphi^{i}{}_{A} \frac{\partial F}{\partial \varphi^{j}{}_{A}}, \nonumber \\
\left\{\widehat{\Sigma}^{A}{}_{B}, F\right\} &=&
-\widehat{E}^{A}{}_{B}F =-\varphi^{i}{}_{B} \frac{\partial
F}{\partial \varphi^{i}{}_{A}}. \nonumber
\end{eqnarray}
These Poisson brackets are in principle sufficient for obtaining
equations of motion in the form:
\begin{equation}\label{a44}
    \frac{dF}{dt}= \left\{F,H\right\},
\end{equation}
$H$ donating the Hamilton function.

It is convenient to introduce in addition to vector fields
$E^{i}{}_{j}$, $\widehat{E}^{A}{}_{B}$ and differential one-forms
$\omega^{i}{}_{j}$, $\widehat{\omega}^{A}{}_{B}$ some others
objects, namely, the vector fields  $H_{a}$, $\widehat{H}_{A}$ and
differential one-forms  $\theta^{a}$, $\widehat{\theta}^{A}$, all
defined on the configuration space $Q = M \times {\rm LI}(U,V)$.
In terms of affine coordinates $x^{i}$, $\varphi^{i}{}_{A}$ but
result is coordinate-independent they are given by
\begin{equation}\label{a45}
    \theta^{a}=dx^{a}, \qquad H_{a} = \frac{\partial }{\partial
    x^{a}}, \qquad \widehat{\theta}^{A} = \varphi^{-1A}{}_{i}dx^{i},
    \qquad \widehat{H}_{A} = \varphi^{i}{}_{A} \frac{\partial }{\partial
    x^{i}}.
\end{equation}
The following duality relations hold among them:
\begin{eqnarray}
\left\langle \widehat{\omega}^{A}{}_{B},
\widehat{E}^{C}{}_{D}\right\rangle = \delta^{A}{}_{D}
\delta^{C}{}_{B}, &\qquad& \left\langle
\widehat{\omega}^{A}{}_{B}, \widehat{H}_{C}\right\rangle = 0,
\nonumber \\
\left\langle \widehat{\theta}^{A},
\widehat{E}^{C}{}_{D}\right\rangle = 0, &\qquad& \left\langle
\widehat{\theta}^{A}, \widehat{H}_{B}\right\rangle =
\delta^{A}{}_{B},\nonumber
\end{eqnarray}
and similarly,
\begin{eqnarray}
\left\langle {\omega}^{a}{}_{b}, {E}^{c}{}_{d}\right\rangle =
\delta^{a}{}_{d} \delta^{c}{}_{b}, &\qquad& \left\langle
{\omega}^{a}{}_{b}, {H}_{c}\right\rangle = 0,
\nonumber \\
\left\langle {\theta}^{a}, {E}^{c}{}_{d}\right\rangle = 0,
&\qquad& \left\langle {\theta}^{a}, {H}_{b}\right\rangle =
\delta^{a}{}_{b}. \nonumber
\end{eqnarray}
Therefore, they are mutually dual fields of non-holonomic frames and co-frames on $Q$. As seen
from the structure of Poisson brackets, these fields are geometrically important. In the
theory of principal fibre bundles of linear frames or co-frames they are know as structural
fields and standard horizontal fields \cite{Kob-Nom63,JJS03}.

Additional important Poisson brackets:
\begin{equation}\label{b45}
    \left\{p_{a}, F\right\} = -H_{a}F, \qquad \left\{\widehat{p}_{A}, F\right\} = -\widehat{H}_{A}F
\end{equation}
for any function $F$ depending only on generalized coordinate.

Let us finish the above description of classical geometry of
degrees of freedom with a brief review of symmetry problems
underlying the polar and two-polar decomposition. They are very
important for quantization problems.

First of all we shall modify slightly our notation. We introduce
new generalized coordinates parameterizing deformation invariants.
In many problems it is convenient to denote the diagonal elements
of $D$ by $Q^{a}$, $a = 1, \ldots, n$,
\[
Q^{a}:= D_{aa}, \qquad \lambda_{a} = \left(D_{aa}\right)^{2} =
\left(Q^{a}\right)^{2}.
\]
And many formulas become remarkably simplified when the
logarithmic scale is used for parametrizing deformation
invariants, $q^{a}=\ln Q^{a}$, thus:
\[
Q^{a} = D_{aa} = \exp\left(q^{a}\right), \qquad \lambda_{a} =
\exp\left(2q^{a}\right).
\]

Deformation parameters $q^{a}$ run over the total real range
$\mathbb{R}$. They are fictitious "material points" moving along
the real axis. As such they are essentially identical and
indistinguishable; this has to do with the mentioned
non-uniqueness of the two-polar decomposition. This
non-distinguishability is essentially striking and interesting in
the quantized version of the theory. The volume extension ratio is
given by $\det D = \exp \left(q^{1} + \cdots +q^{n}\right)$, thus
it may measured in a convenient way by the sum $\left(q^{1} +
\cdots +q^{n}\right)$. It is often convenient to split $D$ into
the isochoric (incompressible) and the purely dilatational parts,
\[
D = l \Delta, \qquad \det \Delta = 1, \qquad l \in \mathbb{R}^{+}.
\]
The factor $l$ is the linear size extension ratio; obviously,
\[
l = \sqrt[n]{\det D} = \exp\left(\frac{1}{n}\left(q^{1} + \cdots
+q^{n}\right)\right).
\]
The logarithmic measure of this ratio:
\[
q= \ln l = \frac{1}{n}\left(q^{1} + \cdots +q^{n}\right)
\]
is simply the "centre of mass" of the mentioned "material points".
The isochoric part $\Delta$ depends only on the ratios
$Q^{i}/Q^{j}$, i.e., logarithmically, on the "relative positions"
$q^{i}-q^{j}$.

The splitting of internal configurations $\varphi$ into
dilatational and isochoric parts may be written down as follows:
\[
\varphi= l \Psi = \exp(q)\Psi= \exp(q)L\Delta R^{-1},
\]
where $\Delta$ as preciously is diagonal and isochoric ($\det
\Delta =1$). The term $\Psi$ refers to the shear-rotational
degrees of freedom. It is convenient to use the isochoric affine
velocities
\[
\nu = \frac{d \Psi}{dt} \Psi^{-1}, \qquad \widehat{\nu} =
\Psi^{-1}\frac{d \Psi}{dt}= \Psi^{-1} \nu \Psi.
\]
They are trace-less, i.e, $\nu \in {\rm SL}(V)^{\prime}$,
$\widehat{\nu} \in {\rm SL}(U)^{\prime}$ (elements of the Lie
algebras of ${\rm SL}(V)$, ${\rm SL}(U)$). The total affine
velocities may be expressed as follows:
\begin{equation}\label{a47}
    \Omega = \nu + \frac{d q}{dt} {\rm Id_{V}}, \qquad \widehat{\Omega} = \widehat{\nu} + \frac{d q}{dt} {\rm Id_{U}},
\end{equation}
where ${\rm Id_{V}}$, ${\rm Id_{U}}$ are identity transformations
in $V$, $U$.

Similarly, the affine spin may be decomposed as follows:
\begin{equation}\label{a48}
    \Sigma = \sigma + \frac{p}{n}{\rm Id_{V}}, \qquad \widehat{\Sigma} = \widehat{\sigma} + \frac{p}{n}{\rm Id_{V}},
\end{equation}
where $\sigma \in {\rm SL}(V)^{\prime}$, $\widehat{\sigma} \in
{\rm SL}(U)^{\prime}$ (traceless) and $p$ is the dilatational
canonical momentum,
\begin{equation}\label{b48}
    {\rm Tr}(\Sigma) = {\rm Tr}\left(\widehat{\Sigma}\right) = p.
\end{equation}
This momentum is canonically conjugate to the above $q$-variable
(logarithmic size variable). The pairing between velocities and
momenta may be expressed as follows:
\[
{\rm Tr} \left(\Sigma \Omega \right) = {\rm Tr} \left(
\widehat{\Sigma} \widehat{\Omega} \right) = {\rm Tr} \left(\sigma
\omega \right)+ p \dot{q} = {\rm Tr} \left(\widehat{\sigma}
\widehat{\omega} \right)+ p \dot{q}.
\]
Poisson brackets for the components of $\sigma$ are based on the
structure constants of ${\rm SL}(V)$. The same based for
$\widehat{\sigma}$ with the only precise that the signs are
reverse. The mutual Poisson brackets $\left\{ \sigma,
\widehat{\sigma} \right\}$ vanish. And obviously, $\left\{ q, p
\right\}=1$, and the dilatational phase-space variables have
vanishing Poisson brackets with the shear-rotational quantities
$\Psi$, $\sigma$, $\widehat{\sigma}$.

Dilatational canonical momentum $p$ may be interpreted as the
total linear momentum of the one-dimensional $q^{a}$-particles.

The two-polar decomposition identifies (modulo some
non-uniqueness) internal configuration with the triplets $\left(L;
q^{1},\ldots,q^{n}; R\right)$, where $(L,R)$ is the pair of rigid
bodies (Cauchy and Green deformation tensors principal axes), and
the fictitious one-dimensional material points
$q^{1},\ldots,q^{n}$ are deformation invariants (in logarithmic
scale). The formulas (\ref{a39}) suggest us to make use of two
possible systems of non-holonomic velocities:
\begin{equation}\label{a49}
    \left( \widehat{\chi}; \ldots, \dot{q}^{a}, \ldots;
    \widehat{\vartheta}\right), \qquad \left( \chi; \ldots, \dot{q}^{a}, \ldots;
    \vartheta \right).
\end{equation}
As expected from the (metrically) rigid body mechanics the first
subsystem is more effective in analysis of dynamical models.

Similarly, when the polar decompositions are used, we have at
disposal the following natural systems of non-holonomic
velocities:
\begin{equation}\label{b49}
    \left(\widehat{\omega}, \dot{A}\right), \qquad \left(\widehat{\omega},
    \dot{B}\right),\qquad \left(\omega, \dot{A}\right), \qquad \left(\omega, \dot{B}\right),
\end{equation}
cf. (\ref{a33}), (\ref{b33}). For the qualitative analysis of
practically important dynamical models the
$\left(\widehat{\omega}, \dot{A}\right)$-system is most
convenient.

Non-holonomic canonical momenta conjugate to (\ref{a49}) are respectively denoted by
\[
\left(\widehat{\rho}; \ldots, p_{a}, \ldots;
\widehat{\tau}\right), \qquad \left(\rho; \ldots, p_{a}, \ldots;
\tau\right),
\]
where
\begin{equation}\label{c49}
   \widehat{\rho} \in {\rm SO}(n, \mathbb{R})^{\prime}, \qquad \widehat{\tau} \in {\rm SO}(n, \mathbb{R})^{\prime},
   \qquad g \in {\rm SO}(V, g)^{\prime}, \qquad \tau \in {\rm SO}(U, \eta)^{\prime},
\end{equation}
and $p_{a}$ are canonical momenta conjugate to $q^{a}$.

Canonical spin variables of the Green and Cauchy gyroscope and their dual angular velocities
are considered as elements of the same linear spaces; this is due to the natural isomorphisms
between orthogonal Lie algebras and their duals. So, the corresponding pairings are given by
\begin{eqnarray}
\left\langle \left(\widehat{\rho}, \bar{p}, \widehat{\tau}\right),
\left(\widehat{\chi}, \dot{\bar{q}}, \widehat{\vartheta}\right)
\right\rangle &=& p_{a}\dot{q}^{a} + \frac{1}{2}{\rm
Tr}\left(\widehat{\rho} \widehat{\chi}\right) + \frac{1}{2}{\rm
Tr}\left(\widehat{\tau} \widehat{\vartheta}\right),
\nonumber \\
\left\langle \left(\rho, \bar{p}, \tau \right), \left(\chi,
\dot{\bar{q}}, \vartheta\right) \right\rangle &=& p_{a}\dot{q}^{a}
+ \frac{1}{2}{\rm Tr}\left(\rho \chi\right) + \frac{1}{2}{\rm
Tr}\left(\tau \vartheta\right). \nonumber
\end{eqnarray}
Similarly, the dual objects of (\ref{b49}) will be denoted by
\[
\left(\widehat{\sigma}, \alpha\right), \qquad \left(\widehat{\mu},
\beta\right), \qquad \left(\sigma, \alpha\right), \qquad
\left(\mu, \beta\right),
\]
where
\[
\widehat{\sigma}, \widehat{\mu} \in {\rm SO}(U, \eta)^{\prime};
\qquad \sigma, \mu \in {\rm SO}(V, g)^{\prime},
\]
and $\alpha \in {\rm L}(U)$, $\beta \in {\rm L}(V)$ are
respectively $\eta$- and $g$-symmetric. The corresponding pairings
are given by
\[
\left\langle \left(\widehat{\sigma}, \alpha\right),
\left(\widehat{\omega}, \dot{A}\right) \right\rangle=
\frac{1}{2}{\rm Tr}\left(\widehat{\sigma} \widehat{\omega}\right)
+ {\rm Tr}\left(\alpha \dot{A}\right),
\]
etc., an analogous way for other combinations.

Let us remind that in the physical three-dimensional case the
skew-symmet\-ric tensors are identified with the axial vectors,
e.g.,
\[
\chi^{i}{}_{j}= - \varepsilon^{i}{}_{jk} \chi^{k}, \qquad \chi^{i}
= - \frac{1}{2}\varepsilon^{i}{}_{j}{}^{k} \chi^{j}{}_{k}
\]
in orthonormal coordinates. For the dual angular momentum
quantities we have the reversed-sign-convention, e.g.
\[
\rho^{i}{}_{j}= \varepsilon^{i}{}_{jk} \rho^{k}, \qquad \rho^{i} =
\frac{1}{2}\varepsilon^{i}{}_{j}{}^{k} \rho^{j}{}_{k}.
\]
The shift of indices here is meant in the cosmetic
Kronecker-delta-sense. Then the former formulas are compatible
with the standard ${\mathbb{R}}^{3}$- conventions, e.g.,
\[
\frac{1}{2}{\rm Tr}\left(\rho \chi\right) = \rho_{i} \chi^{i},
\]
and similarly for other angular velocity and angular momentum
quantities.

The quantities $\rho$, $\tau$ coincide respectively with spin $S$
and negative vorticity $-V$. They are Hamiltonian generators of
transformations:
\[
\varphi \mapsto A \varphi, \qquad \varphi \mapsto \varphi C^{-1},
\qquad A \in {\rm SO}(V, g), \qquad C \in {\rm SO}(U, \eta),
\]
i.e., in terms of the two-polar decomposition:
\[
L \mapsto AL, \qquad R \mapsto CR.
\]

The objects $\widehat{\rho}$, $\widehat{\tau}$ are Hamiltonian
generators of transformations
\[
L \mapsto LA, \qquad R \mapsto RC, \qquad A, C \in {\rm SO}(n,
\mathbb{R}).
\]
Similarly, if we use the polar decomposition, $\sigma$ coincides
with spin $S$, because it generates transformations $\varphi
\mapsto W \varphi$, $W \in {\rm SO}(V, g)$, i.e.,
\[
U \mapsto WU, \qquad \varphi = UA \mapsto WUA = W \varphi.
\]
The quantity $\widehat{\sigma}$ is the Hamiltonian generator of
\[
\varphi = UA \mapsto UKA, \qquad K \in {\rm SO}(U, \eta).
\]
Similarly, the objects $\widehat{\mu}$ generates transformations:
\[
C \in {\rm SO}(U, \eta): \qquad \varphi = BU \mapsto BUC = \varphi
C.
\]
Therefore, it coincides with the canonical vorticity V.

And finally, $\mu$ generates the transformation group:
\[
W \in {\rm SO}(V, g): \ \varphi = BU \mapsto BWU.
\]

Everything said above implies that
\begin{eqnarray}
\chi^{i}{}_{j} &=&
L^{i}{}_{a}\widehat{\chi}^{a}{}_{b}L^{-1b}{}_{j},
\nonumber \\
\vartheta^{A}{}_{B} &=&
R^{A}{}_{a}\widehat{\vartheta}^{a}{}_{b}R^{-1b}{}_{B},
\nonumber \\
\omega^{i}{}_{j} &=&
U^{i}{}_{A}\widehat{\omega}^{A}{}_{B}U^{-1B}{}_{j} \nonumber
\end{eqnarray}
and similarly
\begin{eqnarray}
\rho^{i}{}_{j} = L^{i}{}_{a}\widehat{\rho}^{a}{}_{b}L^{-1b}{}_{j},
&\qquad& \tau^{A}{}_{B} =
R^{A}{}_{a}\widehat{\tau}^{a}{}_{b}R^{-1b}{}_{B},
\nonumber \\
\sigma^{i}{}_{j} =
U^{i}{}_{A}\widehat{\sigma}^{A}{}_{B}U^{-1B}{}_{j}, &\qquad&
\mu^{i}{}_{j} = U^{i}{}_{A}\widehat{\mu}^{A}{}_{B}U^{-1B}{}_{j}.
\nonumber
\end{eqnarray}

In situations where the variables $Q^{i}$ are more convenient than
$q^{i} = \ln Q^{i}$, the canonical momenta ${P}_{i}$ conjugate to
$Q^{i}$ will be used instead $p_{i}$ (conjugates of $q^{i}$). The
relationship is as follows:
\[
{P}_{i} = p_{i} \exp \left(-q^{i}\right) = \frac{ p_{i}}{Q^{i}}.
\]

Having in view applications on the fundamental level, including
the atomic and molecular structure, we concentrate here on the
quantization procedure. Because of this, on the classical level we
are interest mainly in Hamiltonian models. Even if dissipative
phenomena are taken into account, they are considered as a
correction to the Hamiltonian background. In any case, the primary
concept is that of the kinetic energy. If we assume that the
mechanism of affine constraint is compatible with the d'Alembert
principle, then the kinetic energy is obtained by restriction of
the primary multiparticle kinetic energy to the tangent bundle of
the constraints manifold. After easy calculations one obtains:
\[
T=T_{\rm tr}+T_{\rm int}= \frac{m}{2}g_{ij} \frac{dx^{i}}{dt}
\frac{dx^{j}}{dt}+ \frac{1}{2}g_{ij} \frac{d
\varphi^{i}{}_{A}}{dt} \frac{d \varphi^{i}{}_{B}}{dt} J^{AB},
\]
where $m$, $J$ denote, as previously, the total mass and the
second-order moment of the mass distribution. They characterize
respectively the translational and internal inertia. It is
instructive to quote two alternative formulas:
\begin{eqnarray}
T &=& \frac{m}{2}G_{AB} \widehat{v}^{A}\widehat{v}^{B}+
\frac{1}{2}G_{AB}\widehat{\Omega}^{A}{}_{K}\widehat{\Omega}^{B}{}_{L}J^{KL},
\nonumber \\
T &=& \frac{m}{2}g_{ij} {v}^{i}{v}^{j}+
\frac{1}{2}g_{ij}{\Omega}^{i}{}_{k}{\Omega}^{j}{}_{l}J[\varphi]^{kl}.
\nonumber
\end{eqnarray}
Legendre transformation may be written in any of the following
equivalent forms:
\begin{eqnarray}
p_{i}= \frac{\partial T}{\partial v^{i}} = mg_{ij}v^{j},&\quad&
{p}^{A}{}_{i}= \frac{\partial T}{\partial \xi^{j}{}_{A}}= g_{ij}
\xi^{j}{}_{B}J^{BA},
\nonumber \\
\widehat{p}_{A}= \frac{\partial T}{\partial \widehat{v}^{A}} = m
G_{AB}\widehat{v}^{B},&\quad& {\widehat{\Sigma}}^{A}{}_{B}=
\frac{\partial T}{\partial \widehat{\Omega}^{B}{}_{A}}= G_{BC}
\widehat{\Omega}^{C}{}_{D}J^{DA},
\nonumber \\
&\quad& {\Sigma}^{i}{}_{j}= \frac{\partial T}{\partial
{\Omega}^{j}{}_{i}}= g_{jk} {\Omega}^{k}{}_{l}J[\varphi]^{li}.
\nonumber
\end{eqnarray}
Obviously, it is assumed here that there is no generalized
potential depending on velocities (e.g., no magnetic forces).
Otherwise we would have to replace the kinetic energy $T$ by the
total Lagrangian. Inverting the above formulas and substituting
them to the kinetic energy expressions we obtain there formulas
for the kinetic Hamiltonian:
\begin{eqnarray}
{\mathcal{T}}&=&{\mathcal{T}}_{\rm tr}+{\mathcal{T}}_{\rm int}=
\frac{1}{2m}g^{ij} {p}_{i}{p}_{j}+
\frac{1}{2}\widetilde{J}_{AB}{p}^{A}{}_{i}{p}^{B}{}_{j}g^{ij},
\label{a54}\\
{\mathcal{T}} &=& \frac{1}{2m}\widetilde{G}^{AB}
\widehat{p}_{A}\widehat{p}_{B}+
\frac{1}{2}\widetilde{J}_{AB}{\widehat{\Sigma}}^{A}{}_{K}{\widehat{\Sigma}}^{B}{}_{L}\widetilde{G}^{KL},
\label{a55} \\
{\mathcal{T}}&=&\frac{1}{2m}g^{ij} {p}_{i}{p}_{j}+
\frac{1}{2}\widetilde{J}[\varphi]_{ij}{\Sigma}^{i}{}_{k}{\Sigma}^{j}{}_{l}g^{kl}, \label{a56}
\end{eqnarray}
where the "tilda" objects are reciprocal tensors.

For Lagrangians $L = T - V(x, \varphi)$, the resulting
Hamiltonians have the form $H = {\mathcal{T}}+ V(x, \varphi)$.

The usual kinetic energy quadratic in velocities is geometrically
equivalent to some Riemann structure on the configuration space,
\begin{eqnarray}
T= \frac{1}{2} \Gamma_{\mu \nu}(q)\frac{dq^{\mu}}{dt}
\frac{dq^{\nu}}{dt}, &\qquad&\Gamma = \Gamma_{\mu \nu}(q) dq^{\mu}
\otimes dq^{\nu},
 \nonumber
\end{eqnarray}
or, in traditional notation using the arc element:
\[
d \sigma^{2} = \Gamma_{\mu \nu}(q) dq^{\mu} dq^{\nu}.
\]
For the kinetic Hamiltonian we have:
\begin{eqnarray}
{\mathcal{T}} =\frac{1}{2}\widetilde{\Gamma}^{\mu \nu}(q) p_{\mu}
p_{\nu}, &\qquad& \widetilde{\Gamma} = \widetilde{\Gamma}^{\mu
\nu}(q) \frac{\partial}{\partial q^{\mu}} \otimes
\frac{\partial}{\partial q^{\nu}}.\nonumber
\end{eqnarray}
Using the previously introduced symbols we can express the metric
tensor underlying our kinetic energy in any of the following
equivalent forms:
\begin{eqnarray}
\Gamma &=& m g_{ij} dx^{i} \otimes dx^{j} + g_{ij}J^{AB} d
\varphi^{i}{}_{A} \otimes d \varphi^{j}{}_{B}, \label{a55a} \\
\Gamma &=& m G_{AB} \widehat{\theta}^{A} \otimes
\widehat{\theta}^{B} + G_{AB}J^{KL} \widehat{\omega}^{A}{}_{K}
\otimes
\widehat{\omega}^{B}{}_{L}, \nonumber \\
\Gamma &=& m g_{ij} {\theta}^{i} \otimes {\theta}^{j} +
g_{ij}J[\varphi]^{kl} {\omega}^{i}{}_{k} \otimes
{\omega}^{j}{}_{l}. \nonumber
\end{eqnarray}
Similarly, for the inverse metric $\widetilde{\Gamma}$ we have the following equivalent
expressions:
\begin{eqnarray}
\widetilde{\Gamma}&=& \frac{1}{m} g^{ij}\frac{\partial}{\partial x^{i}} \otimes
\frac{\partial}{\partial x^{j}} +  g^{ij} \widetilde{J}_{AB} \frac{\partial}{\partial
\varphi^{i}{}_{A}} \otimes
\frac{\partial}{\partial \varphi^{j}{}_{B}}, \nonumber \\
\widetilde{\Gamma}&=& \frac{1}{m}\widetilde{G}^{AB}\widehat{H}_{A}
\otimes \widehat{H}_{B} + \widetilde{G}^{AB}
\widetilde{J}_{KL}\widehat{E}^{K}{}_{A}
\otimes \widehat{E}^{L}{}_{B}, \nonumber \\
\widetilde{\Gamma}&=& \frac{1}{m} g^{ij} H_{i} \otimes H_{j} +
g^{ij}\widetilde{J}[\varphi]_{kl}E^{k}{}_{i} \otimes E^{l}{}_{j}. \nonumber
\end{eqnarray}
It is clear that the above kinetic energies (metric tensors) are
invariant under the group of spatial isometries ${\rm Is}(M,g)$
acting through (\ref{b6}). It is also invariant under O$(U,J)$
acting through (\ref{c6}), i.e., the subgroup of GL$(U)$
preserving $J$. In particular, it is materially isotropic, i.e.,
invariant under O$(U, \eta)$ when the inertial tensor is
spherical, $J = \mu \widetilde{\eta}$. However, there is no total
affine invariance either in the spatial or material sense. This
kinematical symmetry is broken by the tensors $g$, $J$. Therefore,
the traditional d'Alembert model of the kinetic energy does not
belong to the framework of left- or right- (or two-side-)
invariant geodetic systems on Lie groups or their group spaces,
i.e., to the theory developed by Hermann and Arnold on the basis
of rigid body or incompressible ideal fluid dynamics. By the way,
with the above metric tensors the geodetic systems are
non-physical, because they predict the unlimited contraction and
expansion of the body. And when some extra potential is introduced
as a dynamical model of deformative vibrations, then, except some
very special potential shape, none or rather small profit is
gained from the group-theoretical model of degrees of freedom.

So, it is a tempting idea to formulate dynamical models unifying two things: geodetic
description (no potential as far as possible) and affine invariance. It turns out that to some
extent this may be successfully done: not only the inertia but also interactions are encoded
in some affinely-invariant kinetic energy forms (metrics on the configuration space).

The most general and reasonable class of dynamical geodetic models
invariant under the spatial affine group GAf$(M)$ acting through
(\ref{b6}) is given by the metric tensor:
\begin{equation}\label{a57}
    \Gamma = m \eta_{AB} \widehat{\theta}^{A} \otimes \widehat{\theta}^{B}
    + {\mathcal{L}}^{B}{}_{A}{}^{D}{}_{C} \ \widehat{\omega}^{A}{}_{B}
    \otimes  \widehat{\omega}^{C}{}_{D},
\end{equation}
where ${\mathcal{L}}^{B}{}_{A}{}^{D}{}_{C}$ are constant and
symmetric in their bi-indices,
\[
{\mathcal{L}}^{B}{}_{A}{}^{D}{}_{C} =
{\mathcal{L}}^{D}{}_{C}{}^{B}{}_{A}.
\]
The inverse metric is given by
\[
\widetilde{\Gamma}= \frac{1}{m}\eta^{AB}\widehat{H}_{A} \otimes
\widehat{H}_{B} +
{\mathcal{L}}^{B}{}_{A}{}^{D}{}_{C}\widehat{E}^{A}{}_{B} \otimes
\widehat{E}^{C}{}_{D},
\]
where
\[
{\widetilde{\mathcal{L}}}^{A}{}_{B}{}^{K}{}_{L} {\mathcal{L}}^{L}{}_{K}{}^{C}{}_{D} =
\delta^{A}{}_{D}\delta^{C}{}_{B}.
\]
Similarly, the right-invariant metric tensors have the form:
\begin{equation}\label{c58}
    \Gamma = m g_{ij} {\theta}^{i} \otimes {\theta}^{j}
    + {\mathcal{R}}^{j}{}_{i}{}^{l}{}_{k} \ {\omega}^{i}{}_{j}
    \otimes  {\omega}^{k}{}_{l}
\end{equation}
with similar properties of constants ${\mathcal{R}}$:
\[
{\mathcal{R}}^{j}{}_{i}{}^{l}{}_{k}={\mathcal{R}}^{l}{}_{k}{}^{j}{}_{i}.
\]
The inverse contravariant metric (underlying the kinetic
Hamiltonian) is given by
\begin{eqnarray}
\widetilde{\Gamma}= \frac{1}{m} g^{ij} H_{i} \otimes H_{j} +
{\widetilde{\mathcal{R}}}^{j}{}_{i}{}^{l}{}_{k}E^{i}{}_{j} \otimes
E^{l}{}_{k}, &\qquad&
{\widetilde{\mathcal{R}}}^{j}{}_{i}{}^{l}{}_{k}
{\mathcal{R}}^{k}{}_{l}{}^{a}{}_{b} =
\delta^{j}{}_{b}\delta^{a}{}_{i}.\nonumber
\end{eqnarray}
The corresponding explicit expressions for kinetic energies and
kinetic Hamiltonians are:
\begin{eqnarray}
T&=&\frac{m}{2}\eta_{AB} \widehat{v}^{A} \widehat{v}^{B} +
\frac{1}{2}{\mathcal{L}}^{B}{}_{A}{}^{D}{}_{C}
\widehat{\Omega}^{A}{}_{B}\widehat{\Omega}^{C}{}_{D}, \label{a58}\\
{\mathcal{T}}&=&\frac{1}{2m}\eta^{AB} \widehat{p}_{A}
\widehat{p}_{B} +
\frac{1}{2}{\widetilde{\mathcal{L}}}^{B}{}_{A}{}^{D}{}_{C}
\widehat{\Sigma}^{A}{}_{B}\widehat{\Sigma}^{C}{}_{D}, \nonumber \\
T&=&\frac{m}{2}g_{ij} {v}^{i} {v}^{j} +
\frac{1}{2}{\mathcal{R}}^{j}{}_{i}{}^{d}{}_{c}
{\Omega}^{i}{}_{j}{\Omega}^{c}{}_{d}, \label{b58}\\
{\mathcal{T}}&=&\frac{1}{2m}g^{ij} {p}_{i} {p}_{j} +
\frac{1}{2}{\widetilde{\mathcal{R}}}^{j}{}_{i}{}^{d}{}_{c}{\Sigma}^{i}{}_{j}{\Sigma}^{c}{}_{d}.
\nonumber
\end{eqnarray}
In certain problems it is convenient to use another equivalent
expressions for the translational parts. They are respectively
given by
\begin{eqnarray}
T_{\rm tr}=\frac{m}{2}C_{ij} {v}^{i} {v}^{j}, &\qquad&
{\mathcal{T}}_{\rm tr}=\frac{1}{2m}\widetilde{C}^{ij} {p}_{i}
{p}_{j},\label{a59} \\
T_{\rm tr}=\frac{m}{2}G_{AB} \widehat{v}^{A}
\widehat{v}^{B},&\qquad& {\mathcal{T}}_{\rm
tr}=\frac{1}{2m}\widetilde{G}^{AB} \widehat{p}_{A}
\widehat{p}_{B}. \label{b59}
\end{eqnarray}
Let us observe that expressions for GAf$(M)$-invariant may be
interpreted in the following way. No metric tensor $g$ in the
physical space is assumed and even if it exists (as it does in
reality) it does not enter the kinetic energy expression (if it
did, affine symmetry would be broken and restricted to isometric
one). The role of the metric tensor in contraction of tensorial
indices is played by the Cauchy tensor $C$.

There is no kinetic energy model invariant simultaneously under spatial and material affine
transformations. More precisely, any symmetric twice covariant tensor field on $Q=M \times
{\rm LI}(U,V)$ must be degenerate. This is due to the very malicious non-semisimplicity of the
affine group. However, if we neglect the translational motion, then there exist internal
metric on ${\rm LI}(U,V)$ invariant under both spatial and material (homogeneous) affine
transformations (\ref{d6}), (\ref{c6}). They are given by
\begin{eqnarray}
\Gamma^{0}_{\rm int}&=& A\widehat{\omega}^{K}{}_{L} \otimes
\widehat{\omega}^{L}{}_{K} + B\widehat{\omega}^{K}{}_{K} \otimes
\widehat{\omega}^{L}{}_{L}  \label{a60} \\
&=& A{\omega}^{k}{}_{l} \otimes {\omega}^{l}{}_{k} +
B{\omega}^{k}{}_{k} \otimes {\omega}^{l}{}_{l}, \nonumber
\end{eqnarray}
$A$, $B$ denoting constants. Their inverses have the form:
\begin{eqnarray}
\widetilde{\Gamma}^{0}_{\rm int}&=&
\frac{1}{A}\widehat{E}^{K}{}_{L} \otimes \widehat{E}^{L}{}_{K} -
\frac{B}{A(A+nB)}\widehat{E}^{K}{}_{K} \otimes
\widehat{E}^{L}{}_{L}  \nonumber \\
&=& \frac{1}{A}{E}^{k}{}_{l} \otimes {E}^{l}{}_{k} -
\frac{B}{A(A+nB)}{E}^{k}{}_{k} \otimes {E}^{l}{}_{l}. \nonumber
\end{eqnarray}
Such a metric is never positively-definite. The reason is that
SL$(n, \mathbb{R})$ is non-compact and semisimple.
$\Gamma^{0}_{\rm int.}$ becomes the usual Killing metric when
$A=2n$, $B=-2$. But this is the pathological situation, because
$\Gamma^{0}_{\rm int}$ is degenerate for $A/B=-n$ (due to the
dilatational centre in SO$(n, \mathbb{R})$).

The corresponding kinetic energies are given by
\begin{eqnarray}
T^{0}_{\rm int}&=& \frac{A}{2}{\rm
Tr}\left(\widehat{\Omega}^{2}\right)+ \frac{B}{2}\left({\rm
Tr} \ \widehat{\Omega}\right)^{2} \label{a60a}\\
&=& \frac{A}{2}{\rm Tr}\left(\Omega^{2}\right)+
\frac{B}{2}\left({\rm Tr} \ \Omega\right)^{2}, \nonumber \\
{\mathcal{T}}^{0}_{\rm int}&=& \frac{1}{2A}{\rm
Tr}\left(\widehat{\Sigma}^{2}\right)- \frac{B}{2A(A+nB)}\left({\rm
Tr} \ \widehat{\Sigma}\right)^{2} \label{b60}\\
&=& \frac{1}{2A}{\rm Tr}\left(\Sigma^{2}\right)-
\frac{B}{2A(A+nB)}\left({\rm Tr} \ \Sigma\right)^{2}. \nonumber \\
\end{eqnarray}

The $B$-controlled term in $T_{\rm int}$ above is a merely
correction. The main term ($A$-controlled one) has the hyperbolic
signature $\left(n(n+1)/2\ +,n(n-1)/2\ -\right)$, where the "plus"
contribution corresponds to the non-compact dimensions and the
"minus" one to the compact dimensions in GL$(V)$, GL$(U)$. This is
the highest possible symmetry of $T_{\rm int}$, an affine
counterpart of the spherical top. One is rather reluctant to
non-positive "kinetic energies". However, one can show that in the
above model the lock of positive definiteness is not essentially
embarrassing; on the contrary, the negative contributions may
encode the attractive part of the deformation dynamics. By the
way, the same effect may be obtained within the framework of
positive Riemannian structures on $Q$, when we use a slightly
modified version of (\ref{a60a}).

Let us observe that translational kinetic energy (\ref{a58}),
(\ref{a59}) is affinely-invariant in the physical space and
isometry-invariant in the material space. And conversely,
(\ref{b58}), (\ref{b59}) is isometry-invariant (homogeneous and
isotropic) in the physical space and affinely invariant in the
material space. This focuses our attention on Riemannian
structures on $Q=M \times {\rm LI}(U,V)$ invariant under the
spatial affine group GAf$(M)$ and the group of material isometries
${\rm Is}(N,\eta)$; the opposite models are those invariant under
spatial isometries ${\rm Is}(M,g)$ and material affine
transformations GAf$(N)$. The corresponding metric tensors are
respectively given by
\begin{eqnarray}
\Gamma &=& m\eta_{KL}\widehat{\theta}^{K} \otimes
\widehat{\theta}^{L} + I \eta_{KL}\eta^{MN}
\widehat{\omega}^{K}{}_{M}
    \otimes  \widehat{\omega}^{L}{}_{N} + \Gamma^{0}_{\rm int},
    \nonumber \\
\Gamma &=& mg_{ij}{\theta}^{i} \otimes {\theta}^{j} + I
g_{ik}g^{jl} {\omega}^{i}{}_{j}
    \otimes  {\omega}^{k}{}_{l} + \Gamma^{0}_{\rm int},
    \nonumber
\end{eqnarray}
where the constants $I$, $A$, $B$ are generalized moments of
inertia.

The corresponding contravariant inverses have the form
\begin{eqnarray}
\widetilde{\Gamma} &=& \frac{1}{m}\eta^{KL}\widehat{H}_{K} \otimes
\widehat{H}_{L} + \frac{1}{\widetilde{I}} \eta_{KL}\eta^{MN}
\widehat{E}^{K}{}_{M} \otimes  \widehat{E}^{L}{}_{N} \nonumber \\
&+& \frac{1}{\widetilde{A}} \widehat{E}^{K}{}_{L} \otimes
\widehat{E}^{L}{}_{K} + \frac{1}{\widetilde{B}}
\widehat{E}^{K}{}_{K} \otimes  \widehat{E}^{L}{}_{L}, \nonumber \\
\widetilde{\Gamma} &=& \frac{1}{m}g^{ij}{H}_{i} \otimes {H}_{j} + \frac{1}{\widetilde{I}}
g_{ik}g^{jl} {E}^{i}{}_{j} \otimes
{E}^{k}{}_{l} \nonumber \\
&+& \frac{1}{\widetilde{A}}{E}^{i}{}_{j} \otimes {E}^{j}{}_{i} +
\frac{1}{\widetilde{B}}{E}^{k}{}_{k} \otimes {E}^{l}{}_{l},
\nonumber
\end{eqnarray}
where the inertial constants $\widetilde{I}$, $\widetilde{A}$, $\widetilde{B}$ are given by
\[
    \widetilde{I} = \frac{1}{I}\left(I^{2} - A^{2}\right), \qquad
    \widetilde{A} = \frac{1}{A}\left(A^{2} -
    I^{2}\right), \label{c62} \qquad
    \widetilde{B} = -\frac{1}{B}\left(I + A\right)\left(I + A +
    nB\right).
\]
The corresponding kinetic energies are explicitly given by
\begin{eqnarray}
    T= T_{\rm tr}+T_{\rm int} &=& \frac{m}{2}\eta_{AB}\widehat{v}^{A}
    \widehat{v}^{B}+
    \frac{I}{2}\eta_{KL}\widehat{\Omega}^{K}{}_{M}\widehat{\Omega}^{L}{}_{N}\eta^{MN} \label{a62} \\
&+& \frac{A}{2}{\rm Tr}\left(\widehat{\Omega}^{2}\right)+
\frac{B}{2}\left({\rm Tr} \ \widehat{\Omega}\right)^{2}, \nonumber \\
T= T_{\rm tr}+T_{\rm int} &=& \frac{m}{2}g_{ij}{v}^{i}{v}^{j}+
\frac{I}{2}g_{ik} {\Omega}^{i}{}_{j}
{\Omega}^{k}{}_{l}g^{jl} \label{b62} \\
&+& \frac{A}{2}{\rm Tr}\left({\Omega}^{2}\right)+
\frac{B}{2}\left({\rm Tr} \ {\Omega}\right)^{2}. \nonumber
\end{eqnarray}
Obviously, the last two terms in both expressions coincide because
${\rm Tr}\left({\Omega}^{p}\right)={\rm
Tr}\left({\widehat{\Omega}}^{p}\right)$ for any natural $p$.

The corresponding kinetic Hamiltonians have the following form:
\begin{eqnarray}
{\mathcal{T}}= {\mathcal{T}}_{\rm tr}+{\mathcal{T}}_{\rm int} &=&
\frac{1}{2m}\eta^{AB}\widehat{p}_{A} \widehat{p}_{B}+
    \frac{1}{2\widetilde{I}}\eta_{KL}\widehat{\Sigma}^{K}{}_{M}\widehat{\Sigma}^{L}{}_{N}\eta^{MN} \label{a63} \\
&+& \frac{1}{2\widetilde{A}}{\rm
Tr}\left(\widehat{\Sigma}^{2}\right)+
\frac{1}{2\widetilde{B}}\left({\rm Tr} \ \widehat{\Sigma}\right)^{2}, \nonumber \\
{\mathcal{T}}= {\mathcal{T}}_{\rm tr}+{\mathcal{T}}_{\rm int} &=&
\frac{1}{2m}g^{ij}{p}_{i}{p}_{j}+ \frac{1}{2\widetilde{I}}g_{ik}
{\Sigma}^{i}{}_{j}
{\Sigma}^{k}{}_{l}g^{jl} \label{b63} \\
&+& \frac{1}{2\widetilde{A}}{\rm Tr}\left({\Sigma}^{2}\right)+
\frac{1}{2\widetilde{B}}\left({\rm Tr} \ {\Sigma}\right)^{2}. \nonumber
\end{eqnarray}
Let us observe that the metrical ($g$- and $\eta$-dependent) parts
of kinetic energies may be alternatively written down in terms of
the Cauchy and Green tensors:
\begin{eqnarray}
\frac{m}{2}C_{ij}v^{i}v^{j} +
\frac{I}{2}C_{ij}{\Omega}^{i}{}_{k}{\Omega}^{j}{}_{l}\widetilde{C}^{kl},
&\ &\frac{m}{2}G_{AB}\widehat{v}^{A}\widehat{v}^{B} +
\frac{I}{2}G_{AB}\widehat{\Omega}^{A}{}_{C}\widehat{\Omega}^{B}{}_{D}\widetilde{G}^{CD},
\nonumber \\
\frac{1}{2m}\widetilde{C}^{ij}p_{i}p_{j} +
\frac{1}{2\widetilde{I}}C_{ij}{\Sigma}^{i}{}_{k}{\Sigma}^{j}{}_{l}\widetilde{C}^{kl},
&\ & \frac{1}{2m}\widetilde{G}^{AB}\widehat{p}_{A}\widehat{p}_{B}
+
\frac{1}{2\widetilde{I}}G_{AB}\widehat{\Sigma}^{A}{}_{C}\widehat{\Sigma}^{B}{}_{D}\widetilde{G}^{CD}.
\nonumber
\end{eqnarray}
It is important that in a certain open rang of triples $(I,A,B)
\in {\mathbb{R}}^{3}$ the above kinetic energies are positively
definite and at the same time they have all geometrical and
analytical advantages of invariant geodetic system on the group
manifolds.

One can show that the spatially affine and materially metrical
model (\ref{a63}), or more precisely, its internal part, may be
expressed as follows:
\begin{equation}\label{c63}
 {\mathcal{T}}_{\rm int} = \frac{1}{2\alpha}{\rm
 Tr}\left(\widehat{\Sigma}^{2}\right)+ \frac{1}{2\beta}\left({\rm Tr} \
 \widehat{\Sigma}\right)^{2}+ \frac{1}{2\mu} \left\| V \right\|^{2},
\end{equation}
where
\begin{equation}\label{d64}
    \alpha = I+A, \qquad \beta = -\frac{1}{B}\left(I + A\right)\left(I + A +
    nB\right), \qquad \mu =\frac{1}{I}\left(I^{2} - A^{2}\right),
\end{equation}
and $\left\| V \right\|$ denotes the magnitude of the vorticity,
\[
\left\| V \right\|^{2}= - \frac{1}{2}{\rm Tr}\left(V^{2}\right).
\]
Denoting the $k$-th order Casimir invariant built of generators by
$C(k)$,
\[
C(k) = {\rm Tr}\left(\Sigma^{k}\right) = {\rm
Tr}\left(\widehat{\Sigma}^{k}\right),
\]
we can write simply:
\begin{equation}\label{a64}
 {\mathcal{T}}_{\rm int} = \frac{1}{2\alpha}C(2)+\frac{1}{2\beta}C(1)^{2}+\frac{1}{2\mu} \left\| V
 \right\|^{2}.
\end{equation}

Similarly, for the spatially metrical and materially affine model
(\ref{b63}) we have
\begin{equation}\label{b64}
 {\mathcal{T}}_{\rm int} = \frac{1}{2\alpha}C(2)+\frac{1}{2\beta}C(1)^{2}+\frac{1}{2\mu} \left\|
 S \right\|^{2},
\end{equation}
with the same as previously convention concerning the magnitude of
spin:
\[
\left\| S \right\|^{2} = - \frac{1}{2}{\rm Tr}\left(S^{2}\right).
\]
Let us note that $\left\| V \right\|$, $\left\| S \right\|$ are
simply second-order Casimir built of vorticity and spin.

For the model (\ref{a60}) affinely invariant both in the physical
and material space we have:
\begin{equation}\label{c64}
 {\mathcal{T}}^{0}_{\rm int} = \frac{1}{2A}C(2)+
 \frac{1}{2A\left(n+A/B\right)}C(1)^{2}.
\end{equation}
It is very convenient to separate dilatational and incompressible
motions, especially when affinely invariant kinetic energies
(metrics on $Q$) are used.

One can easily show that for the affine-affine model (\ref{a60})
we have
\begin{equation}\label{d65}
    T^{0}_{\rm int} = \frac{A}{2}{\rm Tr}\left(\nu^{2}\right) +
    \frac{n\left(A+nB\right)}{2}\dot{q}^{2}=T^{0}_{\rm sh}+T^{0}_{\rm dil}
\end{equation}
cf. (\ref{a47}). Performing the Legendre transformation we obtain
\begin{equation}\label{a65}
 {\mathcal{T}}^{0}_{\rm int} =\frac{1}{2A}{\rm
 Tr}\left(\sigma^{2}\right)+ \frac{1}{2n\left(A+nB\right)}p^{2}
 = {\mathcal{T}}^{0}_{\rm sh}+{\mathcal{T}}^{0}_{\rm dil}
\end{equation}
cf. (\ref{a48}), (\ref{b48}). Similarly, for the affine-metrical
and metrical-affine models one obtains respectively the following
expression:
\begin{eqnarray}
{\mathcal{T}}_{\rm int} &=&\frac{1}{2\left(I+A\right)} C_{{\rm
SL}(n)}(2)+ \frac{1}{2n\left(I+A+nB\right)}p^{2}
+\frac{I}{2\left(I^{2}-A^{2}\right)}\left\| V \right\|^{2},\qquad
\label{b65} \\
{\mathcal{T}}_{\rm int} &=&\frac{1}{2\left(I+A\right)} C_{{\rm
SL}(n)}(2)+ \frac{1}{2n\left(I+A+nB\right)}p^{2}
+\frac{I}{2\left(I^{2}-A^{2}\right)}\left\| S \right\|^{2},\qquad
\label{c65}
\end{eqnarray}
where $C_{{\rm SL}(n)}(k)$ are Casimir invariants built of
$\sigma$,
\[
C_{{\rm SL}(n)}(k)= {\rm Tr}\left(\sigma^{k}\right) = {\rm
Tr}\left(\widehat{\sigma}^{k}\right).
\]

Let us observe, the only difference is that concerning the last,
third term. And the both expressions reduce to (\ref{a65}) when we
substitute $I=0$. And conversely, they may be obtained from
(\ref{a65}) be replacing: $A \mapsto (I+A)$ and introducing the
mentioned terms,
\begin{eqnarray}
{\mathcal{T}}^{\rm aff-met}_{\rm int} = {\mathcal{T}}^{0}_{\rm
int}[A \mapsto I+A] + \frac{I}{2\left(I^{2}-A^{2}\right)}\left\| V
\right\|^{2}, \nonumber \\
{\mathcal{T}}^{\rm met-aff}_{\rm int} = {\mathcal{T}}^{0}_{\rm
int}[A \mapsto I+A] + \frac{I}{2\left(I^{2}-A^{2}\right)}\left\| S
\right\|^{2}. \nonumber
\end{eqnarray}

Our philosophy is to base the dynamics as for as possible on
geodetic affinely-invariant models. In particular, geodetic
affine-isometric and isometric-affine models are of special
interest. They are "as affine as possible" and at the same time
compatible with the positive definiteness demand. Nevertheless
some models with potentials are still of interest, and, for
non-affine models they are just unavoidable. So, we shall consider
also potential models
\[
H= {\mathcal{T}} + {V},
\]
where ${V}$ depends only on the configuration variables $(x,
\varphi)$. What concerns inertial properties we concentrated on
highly-symmetric models; in any case they are always spatialy- and
usually materially-isotropic (one can be general in formulation,
but no so much in effective analysis). It is natural to assume
that the potential energy ${V}$ is compatible with these
invariance properties of the kinetic term. So, ${V}$ is invariant
under internal spatial rotations if and only if it depends on
$\varphi$ through the Green tensor $G$. It is invariant under
material spatial rotations if and only if it depends on $\varphi$
through the Cauchy tensor.

And finally, ${V}$ is both spatially and materially isotropic in
internal degrees of freedom if and only if it depends on $\varphi$
only through the deformation invariants, parameterized, e.g., by
$q^{1}, \ldots , q^{n}$. There is a very important special case
when ${V}$ is invariant under the volume-preserving groups
SL$(V)$, SL$(U)$. This means that it depends on $\varphi$ through
the determinant $\det \varphi$. If we use the logarithmic scala of
deformation invariants, this means that ${V}$ is function of
$q=\left(q^{1}+ \cdots + q^{n}\right)/n$, the "centre of mass" of
logarithmic deformation invariants $q^{i}$, $i=1, \ldots, n$. In
kinetic energy models (\ref{a65}), (\ref{b65}), (\ref{c65})
dilatational and shear-rotational degrees of freedom
(incompressible motion) are mutually orthogonal; there is no
interaction between them. This suggests us to concentrate also on
adapted potentials where these degrees of freedom are explicitly
separated,
\begin{equation}\label{a67}
{V}\left(q^{1}, \ldots, q^{n}\right) = {V}_{\rm dil}(q)+ {V}_{\rm sh}\left(\ldots,
q^{i}-q^{j}, \ldots \right);
\end{equation}
the labels "dil" and "sh" refer respectively to "dilatation" and
"shear". The most natural scheme for ${V}_{\rm sh}$ is that of
"binary interactions" between deformation invariants
\begin{equation}\label{b67}
{V}_{\rm sh}= \sum_{i \neq j}f_{ij} \left(q^{i}-q^{j}\right).
\end{equation}
For isotropic models $H_{\rm int} = {\mathcal{T}}_{\rm int}+
{V}\left(q^{1}, \ldots, q^{n}\right)$ with ${\mathcal{T}}_{\rm
int}$ given by (\ref{a63}), (\ref{c63}), (\ref{b65}) the vorticity
$V$ is a constant of motion and the third term in (\ref{b65}) has
also the vanishing Poisson brackets with all terms of (\ref{b65}).
The structure of Poisson brackets and equations of motion
(\ref{a44}) implies that the evolution of variables
$\Sigma^{i}{}_{j}$, $q^{a}$, ruled by the above Hamiltonian
$H_{\rm int}$, is the same as one ruled by
\[
H^{0}_{\rm int}={\mathcal{T}}^{0}_{\rm int}[A \mapsto I+A]+
{V}\left(q^{1}, \ldots, q^{n}\right),
\]
where ${\mathcal{T}}^{0}_{\rm int}[A \mapsto I+A]$ is obtained
from (\ref{b60}) ${\mathcal{T}}^{0}_{\rm int}$ by substituting
$(I+A)$ instead $A$. The difference occurs only in degrees of
freedom ruled by SO$(V,g)$, SO$(U, \eta)$, i.e., in the time
evolution of quantities $L$, $R$ describing the orientation of
principal axes of deformation tensors $C$, $G$. If $V$ depends
only on the dilatational invariant $q$, than the total motion in
$Q$ is a direct product of two independent things: the geodetic
incompressible motion and the autonomous dynamics of the
$q$-variable. The deviator
\[
\sigma^{i}{}_{j} = \Sigma^{i}{}_{j}-
\frac{1}{n}\Sigma^{a}{}_{a}\delta^{i}{}_{j}
\]
is then a constant of motion. The general solution for geodetic
models based on ${\mathcal{T}}^{0}_{\rm int}$ (\ref{b60}) is
explicitly given by exponential mapping. Roughly speaking, it is
produced from initial conditions by one-parameter subgroups of
GL$(V)$, GL$(U)$. And it may be shown on the basis of the
properties of matrix exponents that for the incompressible
geodetic affine-affine model
\[
{T}^{\rm o \ sh}_{\rm int}= \frac{A}{2}{\rm
Tr}\left({\nu}^{2}\right), \qquad {\mathcal{T}}^{\rm o \ sh}_{\rm
int}=\frac{A}{2}{\rm Tr}\left({\sigma}^{2}\right)
\]
(with constrains $q=0$) the general solution contains an
open-subset of bounded (oscillating) motions and an open subset of
unbounded (escaping, dissociated) motions. When dilatations are
allowed, then for any Hamiltonian
\[
H = {\mathcal{T}}^{\rm o}_{\rm int}+V(q)
\]
with ${\mathcal{T}}^{\rm o}_{\rm int}$ given by (\ref{a65}) and
$V(q)$ stabilizing dilatations, there exists also an open
subfamily of bounded motions (and an open subfamily of unbounded
motions if $\sup{V}< \infty$). The same remains true for the
general geodetic affine-metrical model (\ref{a62}), (\ref{a63}),
(\ref{a64}) with incompressibility constraints $q=0$, and
similarly, without such constraints but with dilatations -
stabilizing potential $V(q)$.

The same arguments may be applied to dilatationnally stabilized
geodetic models in (\ref{b62}), (\ref{b63}), (\ref{b64}) invariant
under ${\rm O}(V,g) \times {\rm GL}(U)$ or purely geodetic
isochoric models with the symmetry group ${\rm O}(V,g) \times {\rm
SL}(U)$ (materially special-affine and spatially material models).
On the level of state variables $\widehat{\Sigma}^{A}{}_{B}$,
$q^{a}$ the time evolution is exactly identical with that based on
the affine-affine model of ${\mathcal{T}}_{\rm int}$ (again with
$A$ in (\ref{b63}) replaced by $I+A$).

Let us summarize the main message. Incompressible affine-affine,
affine-material and metrical-affine models may encode the dynamics
of elastic vibrations without any extra potential used, because
their general solutions contain open subset of bounded motions.
When no incompressibility constraints are imposed, the same may be
achieved by introducing some dilatations-stabilizing potential
${V}(q)$, e.g., some potential well oscillator $V(q)= (k/2)q^{2}$,
etc.

The bounded or unbounded character of motion has to do only with
the time evolution of $q^{a}$-variables, and from this point of
view the mentioned three models are essentially identical. The
difference appears only on to level of $L,R$ - degrees of freedom,
but these gyroscopic variables with compact topology cannot
influence the property of trajectories to be bounded or escaping.

To finish this classical description we describe everything in
terms of the two-polar decomposition. It is convenient to combine
the non-holonomic canonical momenta $\widehat{\rho}$,
$\widehat{\tau}$ in the following way:
\begin{equation}\label{c70}
    M:=-\widehat{\rho}-\widehat{\tau}, \qquad
    N:=\widehat{\rho}-\widehat{\tau}.
\end{equation}
This provides a partial diagonalization of the kinetic energy and
elimination of certain interference terms. Namely, after some
calculations one obtain the following expressions for Casimirs:
\begin{equation}\label{a70}
    C(2)= \sum_{a} p_{a}{}^{2}+ \frac{1}{16}\sum_{a,b}
    \frac{\left(M_{ab}\right)^{2}}{{\rm sh}^{2}\frac{q^{a}-q^{b}}{2}}
    - \frac{1}{16}\sum_{a,b}
    \frac{\left(N_{ab}\right)^{2}}{{\rm
    ch}^{2}\frac{q^{a}-q^{b}}{2}},
\end{equation}
and, obviously,
\begin{equation}\label{b70}
C(1)=p= \sum_{a} p_{a}.
\end{equation}

The first term in $C(2)$ may be suggestively decomposed into the
"relative" and the over-all ("centre-of-mass") parts:
\[
\frac{1}{2n} \sum_{a,b} \left(p_{a}-p_{b}\right)^{2}+
\frac{p^{2}}{n}.
\]
This enables one to separate the incompressible and purely
dilatational parts.

It is interesting that $C(2)$ and therefore the kinetic energy
itself has a characteristic lattice structure known from the
theory of one-dimensional many-body system. Here the logarithmic
deformation invariants $q^{a}$ are positions of $n$
indistinguishable fictitious "material points". Unlike in the
usual Sutherland, hyperbolic-Sutherland and Calogero-Moser
lattices where all binary coupling parameters were identical, now
the quantities $M^{i}{}_{j}$, $N^{i}{}_{j}$ are not only
non-identical, but also non-constant. Moreover, they are state
variables subject, together with other ones, to some closed system
of evolution equations (\ref{a44}), where $F$ runs over the
quantities $q^{i}$, $p_{i}$, $L$, $R$, $M^{i}{}_{j}$,
$N^{i}{}_{j}$.

Obviously, one should substitute to (\ref{a44}) the following
basic Poisson brackets
\begin{eqnarray}
\left\{q^{a},p_{b}\right\}&=&\delta^{a}{}_{b}, \nonumber \\
\left\{q^{a},M^{c}{}_{d}\right\}&=&
\left\{p_{a},M^{c}{}_{d}\right\}=
\left\{q^{a},N^{c}{}_{d}\right\}=
\left\{p_{a},N^{c}{}_{d}\right\}=0, \nonumber \\
\left\{M_{ab},M_{cd}\right\}&=& \left\{N_{ab},N_{cd}\right\}=
M_{cb}g_{ad} - M_{ad}g_{cb}+
M_{ac}g_{db}- M_{db}g_{ac}, \nonumber \\
\left\{M_{ab},N_{cd}\right\}&=& N_{cb}g_{ad} - N_{ad}g_{cb}+
N_{ac}g_{db}- N_{db}g_{ac}, \nonumber
\end{eqnarray}
where the shift of indices is meant in the $g$-sense. Obviously,
usually Cartesian orthonormal coordinates are used and then simply
$g_{ab}= \delta_{ab}$. The Poisson brackets for
$M=-\widehat{\rho}-\widehat{\tau}$,
$N=\widehat{\rho}-\widehat{\tau}$ follow from the following ones
for $\widehat{\rho}$, $\widehat{\tau}$:
\begin{eqnarray}
\left\{\widehat{\rho}_{ab},\widehat{\rho}_{cd}\right\}&=&
-\widehat{\rho}_{cb}g_{ad}+ \widehat{\rho}_{ad}g_{cb} -
\widehat{\rho}_{ac}g_{db} + \widehat{\rho}_{db}g_{ac}, \nonumber \\
\left\{\widehat{\tau}_{ab},\widehat{\tau}_{cd}\right\}&=&
-\widehat{\tau}_{cb}g_{ad}+ \widehat{\tau}_{ad}g_{cb} -
\widehat{\tau}_{ac}g_{db} + \widehat{\tau}_{db}g_{ac}, \nonumber \\
\left\{\widehat{\rho}_{ab},\widehat{\tau}_{cd}\right\}&=& 0.
\nonumber
\end{eqnarray}
And these brackets are based on the structure constants of
SO$(n,\mathbb{R})$, because $\widehat{\rho}$, $\widehat{\tau}$
(similarly like $S$, $-V$) are corresponding Hamiltonian
generators of SO$(n,\mathbb{R})$.

In affine kinetic energies (\ref{a65}), (\ref{b65}), (\ref{c65})
the second-order SL$(n,\mathbb{R})$-Casimir invariant has the
form:
\begin{equation}\label{a72}
C_{{\rm SL(n)}}= \frac{1}{2n} \sum_{a,b} \left(p_{a} -
p_{b}\right)^{2}+ \frac{1}{16}\sum_{a,b}
    \frac{\left(M_{ab}\right)^{2}}{{\rm sh}^{2}\frac{q^{a}-q^{b}}{2}}
    - \frac{1}{16}\sum_{a,b}
    \frac{\left(N_{ab}\right)^{2}}{{\rm
    ch}^{2}\frac{q^{a}-q^{b}}{2}}.
\end{equation}
It is seen from (\ref{a64}), (\ref{b64}), (\ref{c64}),
(\ref{a65}), (\ref{b65}), (\ref{c65}), (\ref{c70}), (\ref{a70}),
(\ref{b70}) that the $M$-term describes some effective centrifugal
repulsion of deformation invariants, whereas the $N$-term is a
model of "centrifugal attraction" between $q^{a}$-"particles".
This has nothing to do with any potential ${V}\left(q^{1}, \ldots,
q^{n}\right)$; this attraction is due only to the negative
contribution to the affine-affine kinetic energy (\ref{a65}). In
this way an apparently "embarrassing" turns out to be just
desirable.

The affine-metrical and metrical-affine models (\ref{b65}),
(\ref{c65}) may be respectively written as follows:
\begin{eqnarray}
{\mathcal{T}}^{\rm aff-met}_{\rm int} &=& {\mathcal{T}}^{\rm
aff-aff}_{\rm int} + \frac{1}{2\left(I^{2}-A^{2}\right)}\left\| V
\right\|^{2}, \label{a73} \\
{\mathcal{T}}^{\rm met-aff}_{\rm int} &=& {\mathcal{T}}^{\rm
aff-aff}_{\rm int} + \frac{1}{2\left(I^{2}-A^{2}\right)}\left\| S
\right\|^{2}, \label{b73}
\end{eqnarray}
where ${\mathcal{T}}^{\rm aff-aff}_{\rm int}$ here is just (\ref{a65}) but with $A$ replaced
by $(I+A)$. Let us repeat the explicit formula for ${\mathcal{T}}^{\rm aff-aff}_{\rm int}$ in
terms of the two-polar parametrization and the shear-dilatation splitting:
\begin{eqnarray}
{\mathcal{T}}^{\rm aff-aff}_{\rm int}&=& \frac{1}{4(I+A)n}
\sum_{a,b}\left(p_{a} - p_{b}\right)^{2} + \frac{1}{32(I+A)}
\sum_{a,b}\frac{\left(M_{ab}\right)^{2}}{{\rm
sh}^{2}\frac{q^{a}-q^{b}}{2}} \label{c73} \\
&-&\frac{1}{32(I+A)}\sum_{a,b}\frac{\left(N_{ab}\right)^{2}}{{\rm
ch}^{2}\frac{q^{a}-q^{b}}{2}} + \frac{p^{2}}{2n(I+A+nB)}.
\nonumber
\end{eqnarray}

It is interesting that for Hamiltonians of the form
\[
H={\mathcal{T}}+{V}\left(q^{1}, \ldots, q^{n}\right),
\]
with potentials depending on deformation invariants only, the
models (\ref{a73}), (\ref{b73}), (\ref{c73}) give exactly the same
evolution equations for the system of state variables:
\[
\left(\ldots, q^{a}, \ldots; \ldots, p_{a}, \ldots; \ldots,
M^{a}{}_{b}, \ldots; \ldots, N^{a}{}_{b}, \ldots\right).
\]
This follows from the basic Poisson brackets quoted above. The
only distinction between these three model appear on the level of
variables $L$, $R$, i.e., the principal axes of the Cauchy and
Green deformation tensors. These degrees of freedom have compact
topology, thus they do not influence anything concerning the
bounded or unbounded character of motion.

It is roughly seen from (\ref{a73}), (\ref{b73}), (\ref{c73}) and
may be rigorously shown that; the incompressible sector of our
state variables admits an open family of bounded motions and an
open family of unbounded ones even in the purely geodetic models
(without potential). And this is interesting because invariant
geodetic systems on Lie groups (SL$(n,\mathbb{R})$ this time) may
be successfully analyzed in terms of the exponential mapping and
special functions on groups. Obviously, the dilatational sector
violates these nice features. Without potential energy the
dilatational parameter $q$ moves uniformly in time and the total
motion is unbounded. The only bounded solutions $q=const$ are
exponentially unstable on the level of physical
$\varphi$-variables. Therefore, the "maximally geodetic"
affinely-invariant systems have the form:
\[
H={\mathcal{T}}+{V}(q),
\]
where ${\mathcal{T}}$ stabilizes dilatations. There is no
interaction between dilatational and shear-rotational degrees of
freedom. Dilatational parameter $q$ is subject to the
one-dimensional dynamics ruled by the Hamiltonian
\[
H_{\rm dil}= \frac{p^{2}}{2n(I+A+nB)}+ {V}(q).
\]
The same is true in a more general situation when the potential
energy depends also on the shear variables (non-geodetic models)
and has the explicitly separated form
\[
{V}\left(q^{1}, \ldots, q^{n}\right)= {V}_{\rm dil}(q)+ {V}_{\rm
sh}\left(\ldots, q^{i} - q^{j}, \ldots\right)
\]
cf (\ref{a67}), (\ref{b67}); usually the effective models of ${V}_{\rm sh}$ will have the
binary structure (\ref{a67}).

Finally, let us quote the two-polar representation of the
doubly-isotropic d'Alembert model:
\begin{equation}\label{a75}
{\mathcal{T}}_{\rm int} = \frac{1}{2I} \sum_{a}{P}_{a}{}^{2}+
\frac{1}{8I}\sum_{a,b}\frac{(M_{ab})^{2}}{\left(Q^{a}-Q^{b}\right)^{2}}
+\frac{1}{8I}\sum_{a,b}\frac{(N_{ab})^{2}}{\left(Q^{a}-Q^{b}\right)^{2}}.
\end{equation}

As mentioned earlier, on the purely geodetic level it would be
completely non-physical. Here it is seen explicitly that
${\mathcal{T}}_{\rm int}$ is purely repulsive on the level of
$Q$-variables. All realistic models should be based on some
potential term,
\[
H= {\mathcal{T}}_{\rm int}+ {V}\left(Q^{1}, \ldots, Q^{n}\right)
\]
The binary structure of ${\mathcal{T}}_{\rm int}$ resembles the
Calogero-Moser lattices. And in fact, the general scattering
solution of the Calogero-Moser chain is a subfamily of the general
solution of the geodetic model (\ref{a75}).

%koniec elizy ;)

\section{General ideas of quantization}

After all above classical preliminaries we can formulate the general ideas of quantization. We
practically restrict ourselves to models based on Riemannian structures in the configuration
space and on kinetic energies quadratic in velocities. Let us mention in connection with this
the idea of Capriz \cite{Cap89} about kinetic energies of more general type, i.e.,
non-quadratic ones. Such models in fact appear in relativistic problems and may be useful in
complicated problems of condensed matter theory, defects dynamics, etc. However, it may be
very difficult to use them in quantization problems, because they may need the use of
pseudo-differential operators; this may be hopelessly difficult in curved configuration
spaces. So, we remain within the traditional Schr\"odinger framework.

Let us assume that the classical problem is based (as above) on the Riemann structure
$\Gamma$, i.e., on the kinetic energy form
\[
T=\frac{1}{2}\Gamma_{\mu\nu}(q)\frac{dq^{\mu}}{dt}\frac{dq^{\nu}}{dt},
\]
or, if canonical language is used, on the kinetic Hamiltonian
\[
\mathcal{T}=\frac{1}{2}\Gamma^{\mu\nu}(q)p_{\mu}p_{\nu}.
\]
The Riemannian volume element is given by
\[
d\mu_{\Gamma}(q)=\sqrt{\left|\det\left[\Gamma_{\mu\nu}\right]\right|}dq^{1}\cdots dq^{f}.
\]

Quantum-mechanical formulation is based on the Hilbert space L$^{2}(Q,\mu_{\Gamma})$ of
$\mathbb{C}$-valued functions with the scalar product
\[
\langle\Psi_{1}|\Psi_{2}\rangle=\int \overline{\Psi}_{1}(q)\Psi_{2}(q)d\mu_{\Gamma}(q).
\]
Quantum operator of the kinetic energy is given by
\[
\mathbf{T}=-\frac{\hbar^{2}}{2}\Delta\left(\Gamma\right),
\]
where $\hbar$ is the "crossed" Planck constant and $\Delta\left(\Gamma\right)$ denotes the
Laplace-Bel-trami operator of $\Gamma$:
\begin{equation}\label{a77}
\Delta\left(\Gamma\right)=\frac{1}{\sqrt{|\Gamma|}}\sum_{\mu,\nu}\partial_{\mu}\left(
\sqrt{|\Gamma|}\Gamma^{\mu\nu}\partial_{\nu}\right)=\Gamma^{\mu\nu}\nabla_{\mu}\nabla_{\nu}.
\end{equation}
Obviously, $\nabla$ denotes the Levi-Civita covariant derivative induced by the
$\Gamma$-metric. This means that the quantum kinetic energy is obtained from the classical one
by the formal replacing of $p_{\mu}$ in $\mathcal{T}$ by the operator
$\mathbf{p}_{\mu}=(\hbar/i)\nabla_{\mu}$. Parallel transports preserve $\Gamma$ and
$\sqrt{|\Gamma|}$, thus, $\mathbf{p}_{\mu}$ is formally self-adjoint in
L$^{2}(Q,\mu_{\Gamma})$. When the classical problem is non-geodetic and based on some
potential $V(q^{1},\ldots,q^{f})$, i.e., on the Hamiltonian $H=\mathcal{T}+V$, then the
corresponding quantum Hamiltonian is given by
\[
\mathbf{H}=\mathbf{T}+\mathbf{V},
\]
where $\mathbf{V}$ denotes the operator multiplying wave functions by the potential $V$, i.e.,
$\mathbf{V}\Psi=V\Psi$; usually we do not distinguish them graphically. Velocity-dependent
generalized (magnetic) potentials are not considered here.

Strictly speaking, from the very principal point of view wave functions are not scalars but
scalar densities of weight $1/2$, and the squared moduli $\overline{\Psi}\Psi$ are scalar
densities of weight one. But in all realistic models, and we do not go outside this scope,
some Riemann structure is used and all tensor densities are factorized into tensors and
standard densities built of $\Gamma$. In particular, $1/2$-densities $\Psi$ describing pure
quantum states are factorizing as $\Psi=\Psi\sqrt[4]{|\Gamma|}$, where $\Psi$ are just the
aforementioned scalar wave functions.

Let us also mention that despite some current views, the one-valuedness of wave functions is
nota fundamental assumption of quantum mechanics. There are at least some situations where
multivalued amplitudes seem to be acceptable. First of all, it is so when the configuration
space $Q$ is multiply-connected and has a finite homotopy group. Then it is natural to define
the wave functions on the covering manifold $\overline{Q}$. They need not project onto $Q$ as
one-valued amplitudes but it seems natural to demand that, according to the statistical
interpretation, the squared moduli $\overline{\Psi}\Psi$ are uniquely projectable. This has to
do with the projective representation. And just such situations are interesting in our model,
where the configuration spaces of rigid and affinely-rigid bodies in dimensions $n\geq 3$ have
two-element homotopy groups. This point was stressed, e.g., in
\cite{ABB95,ABMB95,BBM92,Bar-Racz77,Pau39,Rei39}, where the possibility of doubly-valued wave
functions for quantized rigid body was pointed out.

Literally performed calculations of Laplace-Beltrami operators are usually very difficult and
the result is rather non-readable. It is much more convenient to use directly the operators
$\widehat{H}_{A}$, $H_{i}$, $\widehat{E}^{A}{}_{B}$, $E^{i}{}_{j}$ introduced formerly and the
classical expressions for kinetic Hamiltonians based on these quantities. Then we can define
easily quantum operators representing the corresponding physical quantities, e.g.,
$\mathbf{p}_{i}$, $\mathbf{p}^{A}{}_{i}$, $\widehat{\mathbf{\Sigma}}^{A}{}_{B}$,
$\mathbf{\Sigma}^{i}{}_{j}$.

Hilbert spaces may be constructed without calculating the complicate coordinate expressions
for the metric tensors $\Gamma_{\mu\nu}$ and their densities $\sqrt{|\Gamma|}$. Namely, our
configuration spaces may be in a sense identified with Lie groups (more precisely their group
spaces), therefore we can simply use Haar measures, which are explicitly known and given by
simple expressions. We are usually dealing with left- and right-invariant metrics $\Gamma$,
thus, the corresponding measures $\mu_{\Gamma}$ are also invariant, and just coincide with the
invariant Haar measures, because the latter ones are unique (modulo normalization).

First of all, let us observe that our configuration space $Q=M\times$ LI$(U,V)$ as an affine
space (with the translation space $V\times$ L$(U,V)$) is endowed with the natural Lebesgue
measure a unique up to normalization. Fixing metric tensors $g\in V^{\ast}\otimes V^{\ast}$,
$\eta\in U^{\ast}\otimes U^{\ast}$ and some adapted Cartesian coordinates (orthonormal with
respect to these tensors) $x^{i}$, $a^{K}$, $\varphi^{i}{}_{K}$, we can normalize $l$ as
follows:
\[
da(x,\varphi)=dx^{1}\cdots dx^{n}d\varphi^{1}{}_{1}\cdots d\varphi^{n}{}_{n}.
\]
When translational degrees of freedom are neglected, we use the usual Lebesgue measure $l$ on
LI$(U,V)$ as an open subset of the linear space L$(U,V)$:
\[
dl(\varphi)=d\varphi^{1}{}_{1}\cdots d\varphi^{n}{}_{n}.
\]
These measures are invariant under translations in the affine space $M\times$ LI$(U,V)$ and
under spatial and material isometries. They are, however, non-invariant under spatial and
material affine transformations. To achieve the affine invariance we must use the following
Haar measures $\alpha$, $\lambda$ on $Q=M\times$ LI$(U,V)$ and $Q_{\rm int}=$ LI$(U,V)$
induced from the affine group GAf$(n,\mathbb{R})\simeq$
GL$(n,\mathbb{R})\times_{s}\mathbb{R}^{n}$ and the linear group GL$(n,\mathbb{R})$:
\begin{eqnarray}
d\alpha(\varphi,x)&=&(\det\varphi)^{-n-1}da(x,\varphi)=(\det\varphi)^{-n-1}dx^{1}\cdots
dx^{n}d\varphi^{1}{}_{1}\cdots d\varphi^{n}{}_{n},\nonumber\\
d\lambda(\varphi)&=&(\det\varphi)^{-n}dl(\varphi)=(\det\varphi)^{-n}d\varphi^{1}{}_{1}\cdots
d\varphi^{n}{}_{n}.\nonumber
\end{eqnarray}
Expressing the measure $l$ in terms of the two-polar decomposition $\varphi=LDR^{-1}$ we
obtain
\[
d\lambda(\varphi)=d\lambda(L;q^{a};R)=\prod_{i\neq j}\left|{\rm
sh}\left(q^{i}-q^{j}\right)\right|dq^{1}\cdots dq^{n}d\mu(L)d\mu(R),
\]
where $\mu$ is the left- and right-invariant Haar measure on the
manifolds of linear isometries LIs$(\mathbb{R}^{n},\delta;V,g)$,
LIs$(\mathbb{R}^{n},\delta;U,\eta)$. Obviously, when LI$(U,V)$ is
identified with GL$(n,\mathbb{R})$ and the mentioned manifolds of
isometries are identified with the orthogonal group
SO$(n,\mathbb{R})$, then $\mu$ becomes simply the literally
understood Haar measure on SO$(n,\mathbb{R})$. As the manifolds
LIs and the group SO$(n,\mathbb{R})$ are compact, the measure
$\mu$ may, although need not, be normalized to unity (the manifold
volume equals the unity). In certain formulas it is convenient to
use the symbol
\begin{equation}\label{a81}
P_{\lambda}:=\prod_{i\neq j}\left|{\rm sh}\left(q^{i}-q^{j}\right)\right|,
\end{equation}
thus,
\[
d\lambda(\varphi)=P_{\lambda}dq^{1}\cdots dq^{n}d\mu(L)d\mu(R).
\]
One can also obtain the following convenient expression for the Lebesgue measure $l$:
\[
dl=P_{l}dQ^{1}\cdots dQ^{n}d\mu(L)d\mu(R),
\]
where
\begin{equation}\label{b81}
P_{l}=\prod_{i\neq j}\left(Q^{i}{}^{2}-Q^{j}{}^{2}\right)=\prod_{i\neq
j}\left(Q^{i}+Q^{j}\right)\left(Q^{i}-Q^{j}\right),
\end{equation}
and, as we remember,
\[
Q^{a}=\exp\left(q^{a}\right).
\]
The Haar measure on the internal configuration space of the isochoric (incompressible)
affinely-rigid body may be expressed in terms of the Dirac distribution
\[
d\lambda_{\rm SL}(\varphi)=P_{\lambda}\left(q^{1},\ldots,q^{n}\right)\delta\left(q^{1}+\cdots+
q^{n}\right)dq^{1}\cdots dq^{n}d\mu(L)d\mu(R).
\]

Our quantum-mechanical models will be based on Hilbert spaces L$^{2}(Q,a)$, L$^{2}(Q_{\rm int
},l)$, L$^{2}(Q,\alpha)$, and L$^{2}(Q_{\rm int},\lambda)$. Obviously, for affinely-invariant
models L$^{2}(Q,\alpha)$, L$^{2}(Q_{\rm int},\lambda)$ are more convenient. Similarly, for the
usual d'Alembert models L$^{2}(Q,a)$, L$^{2}(Q_{\rm int },l)$ are more natural. Nevertheless,
it is a matter of convenience; one should stress that both types of models may be formulated
in terms of any of these Hilbert spaces.

The spatial and material actions of GAf$(M)$ and GAf$(N)$ (\ref{b6}), (\ref{a6}) on the
configuration space $Q$ preserve the Haar measure $\alpha$. Similarly, (\ref{d6}), (\ref{c6})
preserve the Haar measure $\lambda$ on the internal configuration space $Q_{\rm int}$. On the
other hand, except isometries, they do not preserve the usual Lebesgue measures on affine
spaces, i.e., $a$, $l$. The latter ones are invariant, however, under the usual affine
translations given analytically by
\[
\left(\ldots,x^{i},\ldots;\ldots,\varphi^{i}{}_{A},\ldots\right)\mapsto
\left(\ldots,x^{i}+\xi^{i},\ldots;\ldots,\varphi^{i}{}_{A}+\xi^{i}{}_{A},\ldots\right)
\]
just the usual additive translations in $M\times$ L$(U,V)$ and L$(U,V)$ as affine spaces.
(Remark: on $Q$ and $Q_{\rm int}$, when L$(U,V)$ is replaced by its open subset LI$(U,V)$,
then these translations act only locally.) On the other hand, these additive translations in
general do not preserve the Haar measures $\alpha$, $\lambda$.

All the mentioned groups act argument-wise on wave functions. When they preserve the measure
on $Q$ or $Q_{\rm int}$, the resulting transformations of wave functions preserve the
corresponding L$^{2}$-spaces and are unitary, i.e., they preserve the scalar products too.

Let us quote explicitly some expressions, at least to fix the notation used later on.

For any $A\in$ GAf$(M)$ we define the operation $\mathbf{A}$ which transforms the wave
function $\Psi:$ AfI$(N,M)\rightarrow\mathbb{C}$ into such one that
\begin{equation}\label{a83}
\left(\mathbf{A}\Psi\right)(\Phi)=\Psi(A\circ\Phi).
\end{equation}
Similarly, for any $A\in$ GAf$(N)$ we define the operator $\mathbf{B}$ such that
\begin{equation}\label{c83}
\left(\mathbf{B}\Psi\right)(\Phi)=\Psi(\Phi\circ B).
\end{equation}
If translational degrees of freedom are neglected and we deal with wave functions $\Psi:$
LI$(U,V)\rightarrow\mathbb{C}$, then for any $A\in$ GL$(V)$, $B\in$ GL$(U)$ we define
\begin{eqnarray}
\left(\mathbf{A}\Psi\right)(\varphi)&=&\Psi(A\varphi),\label{b83}\\
\left(\mathbf{B}\Psi\right)(\varphi)&=&\Psi(\varphi B).
\end{eqnarray}
Obviously, for any $A$, $B$ the operators $\mathbf{A}$, $\mathbf{B}$ are unitary in
L$^{2}(Q,\alpha)$, L$^{2}(Q_{\rm int},\lambda)$, because the measures $\alpha$, $\lambda$ are
invariant under regular translations. Unlike this, they are not unitary in L$^{2}(Q,\alpha)$,
L$^{2}(Q_{\rm int},\lambda)$, unless $A$, $B$ are volume-preserving mappings, i.e., elements
of SAf$(M)$, SAf$(N)$, SL$(V)$, SL$(U)$ (more precisely, unimodularity is sufficient, i.e.,
det$L(A)=$ det$L(B)=\pm 1$). Obviously, the differential operators (vector fields) $H_{a}$,
$E^{i}{}_{j}$ defined in (\ref{a45}), (\ref{a43}) are generators of unitary groups defined in
(\ref{a83}), (\ref{b83}), therefore, they formally anti-self-adjoint in L$^{2}(Q,\alpha)$,
L$^{2}(Q_{\rm int},\lambda)$, or rather in the subspaces of smooth functions. Being
non-bounded (non-continuous) they are not anti-Hermitian in the rigorous mathematical sense.
However, they are so in rough terms used in physics. They possess anti-Hermitian extensions.
The following differential operators:
\[
{\bf P}_{a}=\frac{\hbar}{i}{\bf H}_{a}=\frac{\hbar}{i}\frac{\partial}{\partial x^{a}},\qquad
{\bf \Sigma}^{a}{}_{b}=\frac{\hbar}{i}{\bf E}^{a}{}_{b}=
\frac{\hbar}{i}\varphi^{a}{}_{K}\frac{\partial}{\partial \varphi^{b}{}_{K}}
\]
are formally Hermitian. They are respectively quantum linear momentum and hyperspin operators.
One can also introduce the operator of the total affine momentum (hypermomentum)
\[
{\bf J}^{a}{}_{b}=\mathbf{x}^{a}{\bf P}_{a}+{\bf \Sigma}^{a}{}_{b}={\bf
\Lambda}^{a}{}_{b}+{\bf \Sigma}^{a}{}_{b}.
\]
The ordering of non-commuting operators meant just as written above; it follows from their
geometric nature as group generators. Obviously, the coordinate operators are defined in the
usual way,
\[
\left(\mathbf{x}^{a}\Psi\right)(x,\varphi)=x^{a}\Psi(x,\varphi),\qquad
\left(\mathbf{\varphi}^{a}{}_{K}\Psi\right)(x,\varphi)=\varphi^{a}{}_{K}\Psi(x,\varphi).
\]
If we define quantum Poisson bracket in the usual way,
\[
\{{\bf A},{\bf B}\}:=\frac{1}{i\hbar}\left[{\bf A},{\bf B}\right]=\frac{1}{i\hbar}\left({\bf
A}{\bf B}-{\bf B}{\bf A}\right),
\]
then the above basic quantities satisfy the rules identical with the classical ones
(\ref{a43})-(\ref{a44}), (\ref{b45}).

The same concerns the co-moving representants based on differential operators
$\widehat{\mathbf{H}}_{A}$, $\widehat{\mathbf{E}}^{A}{}_{B}$. The corresponding formally
Hermitian operators
\[
{\bf \widehat{P}}_{A}=\frac{\hbar}{i}{\bf
\widehat{H}}_{A}=\frac{\hbar}{i}\varphi^{a}{}_{A}\frac{\partial}{\partial x^{a}},\qquad {\bf
\widehat{\Sigma}}^{A}{}_{B}=\frac{\hbar}{i}{\bf \widehat{E}}^{A}{}_{B}=
\frac{\hbar}{i}\varphi^{a}{}_{B}\frac{\partial}{\partial \varphi^{a}{}_{A}},
\]
and
\[
{\bf \widehat{J}}^{K}{}_{L}=\mathbf{a}^{K}{\bf \widehat{P}}_{L}+{\bf
\widehat{\Sigma}}^{K}{}_{L}={\bf \widehat{\Lambda}}^{K}{}_{L}+{\bf \widehat{\Sigma}}^{K}{}_{L}
\]
are quantum generators of GAf$(N)$, GL$(U)$.

Operators of angular momenta are given by the doubled $g$-skew-symmetric parts of affine
momenta:
\[
{\bf S}^{a}{}_{b}={\bf \Sigma}^{a}{}_{b}-g^{ac}g_{bd}{\bf \Sigma}^{d}{}_{c},\qquad {\bf
L}^{a}{}_{b}={\bf \Lambda}^{a}{}_{b}-g^{ac}g_{bd}{\bf \Lambda}^{d}{}_{c},
\]
and
\[
{\bf \bar{J}}^{a}{}_{b}={\bf J}^{a}{}_{b}-g^{ac}g_{bd}{\bf J}^{d}{}_{c}={\bf L}^{a}{}_{b}+{\bf
S}^{a}{}_{b}.
\]
They are respectively spin, orbital, and the total angular momentum. Similarly, the quantum
vorticity operator is given by
\[
{\bf V}^{A}{}_{B}={\bf \widehat{\Sigma}}^{A}{}_{B}-\eta^{AC}\eta_{BD}{\bf
\widehat{\Sigma}}^{D}{}_{C}.
\]
Canonical linear momentum conjugate to $\varphi^{i}{}_{A}$ is on the quantum level represented
by the operator
\[
{\bf P}^{A}{}_{i}=\frac{\hbar}{i}\frac{\partial}{\partial \varphi^{i}{}_{A}}.
\]
Important: it is not formally Hermitian in L$^{2}(Q,\alpha)$, L$^{2}(Q_{\rm int},\lambda)$.
Indeed, classically it generates additive translations in L$(U,V)$:
\[
\varphi^{i}{}_{A}\mapsto\varphi^{i}{}_{A}+\xi^{i}{}_{A}.
\]
And those do not preserve the Haar measures $\alpha$, $\lambda$. But they preserve the
Lebesgue measures $a$, $l$, therefore, ${\bf P}^{A}{}_{i}$ is formally Hermitian in
L$^{2}(Q,a)$ and L$^{2}(Q_{\rm int},l)$. Because of this the Hilbert spaces are more
convenient for describing quantization of models based on the usual d'Alembert principle.

However, in all kinds of models the "non-usual" Hilbert spaces may be also applied, simply the
definition of some operators must be modified. For example, affine models may be as well
quantized in L$^{2}(Q,a)$ and L$^{2}(Q_{\rm int},l)$, but the operators $\mathbf{A}$,
$\mathbf{B}$ in (\ref{a83}), (\ref{b83}) are to be replaced by $\mathbf{A}^{\prime}$,
$\mathbf{A}^{\prime}$ given by
\[
\mathbf{A}^{\prime}:={\rm det}L[A]^{-(n+1)/2}\mathbf{A},\qquad \mathbf{B}^{\prime}:={\rm
det}B^{-n/2}\mathbf{B}.
\]
More explicitly,
\begin{eqnarray}
\left(\mathbf{A}^{\prime}\Psi\right)(x,\varphi)&=&{\rm
det}L[A]^{-(n+1)/2}\Psi\left(A(x),L[A]\varphi\right),\nonumber\\
\left(\mathbf{B}^{\prime}\Psi\right)(x,\varphi)&=&{\rm det}B^{-n/2}\Psi\left(x,\varphi
B\right).\nonumber
\end{eqnarray}
Similarly, when considering only the action of $A\in$ GL$(V)$ on internal degrees of freedom,
we define
\[
\mathbf{A}^{\prime}:={\rm det}A^{-n/2}\mathbf{A},
\]
i.e.,
\[
\left(\mathbf{A}^{\prime}\Psi\right)(x,\varphi)={\rm det}A^{-n/2}\Psi\left(x,A\varphi\right).
\]
Due to the multiplicative terms, the above operators are unitary in Hilbert spaces based on
the Lebesgue measures. Their infinitesimal generators are then modified by additive correction
terms due to which they become formally Hermitian in L$^{2}(Q,a)$ and L$^{2}(Q_{\rm int},l)$.
For example, $\mathbf{\Sigma}^{a}{}_{b}$, $\mathbf{\Lambda}^{a}{}_{b}$, ${\bf
\widehat{\Sigma}}^{A}{}_{B}$ are respectively replaced by
\[
{}^{\prime}\mathbf{\Sigma}^{a}{}_{b}=\mathbf{\Sigma}^{a}{}_{b}+\frac{\hbar
n}{2i}\delta^{a}{}_{b},\qquad
{}^{\prime}\mathbf{\Lambda}^{a}{}_{b}=\mathbf{\Lambda}^{a}{}_{b}
+\frac{\hbar}{2i}\delta^{a}{}_{b},\qquad {}^{\prime}{\bf
\widehat{\Sigma}}^{A}{}_{B}={\bf
\widehat{\Sigma}}^{A}{}_{B}+\frac{\hbar n}{2i}\delta^{A}{}_{B}.
\]
It is easy to see that the linear momentum $\mathbf{P}_{a}$, spin $\mathbf{S}^{a}{}_{b}$,
vorticity ${\bf V}^{A}{}_{B}$ remain unchanged. The finite actions generated by them preserve
the Lebesgue measures.

Similarly, when quantizing the d'Alembert models with the use of non-usual for them
L$^{2}(Q,\alpha)$ and L$^{2}(Q_{\rm int},\lambda)$, we would have to modify ${\bf
P}^{A}{}_{i}=(\hbar/i)\partial/\partial\varphi^{i}{}_{A}$, but we shall not do this here.
Nevertheless, it must be stressed that quantizations in terms of "non-usual" Hilbert spaces
may be convenient when one is interested in comparison between various models.

As mentioned, the best tool when quantizing the d'Alembert model is offered by the geometry of
Hilbert spaces L$^{2}(Q,a)$ and L$^{2}(Q_{\rm int},l)$. Then the quantized version of thee
kinetic energy (\ref{a54}) is given by the operator
\begin{equation}\label{a88}
\mathbf{T}=\mathbf{T}_{\rm tr}+\mathbf{T}_{\rm
int}=\frac{1}{2m}g^{ij}\mathbf{p}_{i}\mathbf{p}_{j}
+\frac{1}{2}\widetilde{J}_{AB}\mathbf{P}^{A}{}_{i}\mathbf{P}^{B}{}_{j}g^{ij}.
\end{equation}
Explicitly, this is a kind of "Laplace operator" in the $n(n+1)$-dimensional Euclidean space:
\[
\mathbf{T}=-\frac{\hbar^{2}}{2m}g^{ij}\frac{\partial^{2}}{\partial x^{i}\partial x^{j}}
-\frac{\hbar^{2}}{2}\widetilde{J}_{AB}g^{ij}\frac{\partial^{2}}{\partial
\varphi^{i}{}_{A}\partial \varphi^{j}{}_{B}}.
\]

Unfortunately, geodetic (potential free) models are non-physical because they predict only
escaping, non-bounded classical motion and the purely continuous positive spectrum after
quantization (no bounded states). And for realistic potentials the variables
$(x^{i},\varphi^{i}{}_{A})$ are rather non-adequate. The classical expressions (\ref{a55}),
(\ref{a56}) are non-convenient for quantization because they suffer from the embarrassing
problem of the ordering of operators.

There are no such problems with models based on the affine invariance, i.e., (\ref{a58}),
(\ref{b58}). Let us remind that the first of them is affinely invariant in the physical space
and isometries-invariant in the material space. On the contrary, the other one is
isometries-invariant in the physical space and affinely invariant in the body. There are no
ordering problems and the quantum operators of the kinetic energy may be immediately obtained
via the simple replacement of the classical linear momentum and affine spin by the operators
just written down. Therefore, for the quantized versions of (\ref{a58}), (\ref{b58}) we
respectively obtain
\begin{eqnarray}
\mathbf{T}&=&\frac{1}{2m}\eta^{AB}{\bf \widehat{p}}_{A}{\bf \widehat{p}}_{B}+
\frac{1}{2}\widetilde{\mathcal{L}}^{B}{}_{A}{}^{D}{}_{C}{\bf \widehat{\Sigma}}^{A}{}_{B}{\bf
\widehat{\Sigma}}^{C}{}_{D}\nonumber\\
&=&-\frac{\hbar^{2}}{2m}\widetilde{C}^{ab}\frac{\partial^{2}}{\partial x^{a}\partial x^{b}}
-\frac{\hbar^{2}}{2}\widetilde{\mathcal{L}}^{B}{}_{A}{}^{D}{}_{C}\varphi^{k}{}_{B}
\frac{\partial}{\partial\varphi^{k}{}_{A}}\varphi^{l}{}_{D}
\frac{\partial}{\partial\varphi^{l}{}_{C}},\label{a89}\\
\mathbf{T}&=&\frac{1}{2m}g^{ij}{\bf p}_{i}{\bf p}_{j}+
\frac{1}{2}\widetilde{\mathcal{R}}^{j}{}_{i}{}^{l}{}_{k}{\bf \Sigma}^{i}{}_{j}{\bf
\Sigma}^{k}{}_{l}\nonumber\\
&=&-\frac{\hbar^{2}}{2m}g^{ij}\frac{\partial^{2}}{\partial x^{i}\partial x^{j}}
-\frac{\hbar^{2}}{2}\widetilde{\mathcal{R}}^{j}{}_{i}{}^{l}{}_{k}\varphi^{i}{}_{A}
\frac{\partial}{\partial\varphi^{j}{}_{A}}\varphi^{k}{}_{B}
\frac{\partial}{\partial\varphi^{l}{}_{B}},\label{b89}
\end{eqnarray}
where, as previously, $\widetilde{C}$ denotes the inverse Cauchy deformation tensor. The
ordering of operators $\varphi$, $\partial/\partial\varphi$ is essential, therefore, there
appear first-order differential operators, respectively,
\begin{eqnarray}
-\frac{\hbar^{2}}{2}\widetilde{\mathcal{L}}^{B}{}_{A}{}^{A}{}_{C}\varphi^{k}{}_{B}
\frac{\partial}{\partial\varphi^{k}{}_{C}}&=&-\frac{\hbar
i}{2}\widetilde{\mathcal{L}}^{B}{}_{A}{}^{A}{}_{C}{\bf \widehat{\Sigma}}^{C}{}_{B},\nonumber\\
-\frac{\hbar^{2}}{2}\widetilde{\mathcal{R}}^{j}{}_{i}{}^{l}{}_{j}\varphi^{i}{}_{A}
\frac{\partial}{\partial\varphi^{l}{}_{A}}&=&-\frac{\hbar
i}{2}\widetilde{\mathcal{R}}^{j}{}_{i}{}^{l}{}_{j}{\bf \Sigma}^{i}{}_{l}.\nonumber
\end{eqnarray}
The second-order terms are obvious:
\[
-\frac{\hbar^{2}}{2}\widetilde{\mathcal{L}}^{B}{}_{A}{}^{D}{}_{C}\varphi^{k}{}_{B}
\varphi^{l}{}_{D}\frac{\partial^{2}}{\partial\varphi^{k}{}_{A}\partial\varphi^{l}{}_{C}},\qquad
-\frac{\hbar^{2}}{2}\widetilde{\mathcal{R}}^{j}{}_{i}{}^{l}{}_{k}\varphi^{i}{}_{A}
\varphi^{k}{}_{B}\frac{\partial^{2}}{\partial\varphi^{j}{}_{A}\partial\varphi^{l}{}_{B}}.
\]
There "curved" structure is obvious. Geometrically this is due to the fact that the metric
tensors on $Q$ given by (\ref{a57}), (\ref{c58}) define there essentially Riemannian
structures with non-vanishing curvature tensors. Unlike this, the d'Alembert model is based on
the evidently flat, Euclidean geometry with the metric tensor (\ref{a55}). All this has to do
with strong nonlinearity encoded in geodetic terms of classical affine Hamiltonians. And this
strong nonlinearity follows from the "large" group of assumed symmetries.

As mentioned in the classical part, there are good reasons to concentrate the attention on
those metric tensors on $Q=M\times$ LI$(U,V)$ (those models of kinetic energy) which are:
\begin{enumerate}
\item affinely-invariant in the physical space and simultaneously isometry-inva\-riant
(homogeneous and isotropic) in the material space,\label{1}

\item and conversely, homogeneous and isotropic in the physical space (isometry-invariant) and
simultaneously affinely-invariant in the material space.\label{2}
\end{enumerate}
It is impossible to satisfy simultaneously both demands \ref{1} and \ref{2}. However, if
translational degrees of freedom are neglected, there exist metrics on $Q_{\rm int}=$
LI$(U,V)$ affinely (or rather centro-affinely) invariant both in the space and in the body.
They are always pseudo-Riemannian, i.e., have the non-definite hyperbolic signature.
Obviously, (\ref{a89}), (\ref{b89}) are formally Hermitian in L$^{2}(Q,\alpha)$. Just like
(\ref{a88}) is so in L$^{2}(Q,a)$.

The operators of translational kinetic energy $\mathbf{T}_{\rm tr}$ are exactly like in
general models (\ref{a89}), (\ref{b89}), so we concentrate on the internal parts
$\mathbf{T}_{\rm int}$. And they are just the very special cases of those in the general
formulas (\ref{a89}), (\ref{b89}). Due to their very peculiar features it is instructive to
quote them explicitly.

So, for internal degrees of freedom the quantized versions of (\ref{a63}), (\ref{b63}) are
obtained by the literal substitution of ${\bf \widehat{\Sigma}}$, $\mathbf{\Sigma}$ instead of
their classical counterparts, so, respectively,
\begin{eqnarray}
\mathbf{T}_{\rm int}&=&\frac{1}{2\widetilde{I}}\eta_{KL}\eta^{MN}{\bf
\widehat{\Sigma}}^{K}{}_{M}{\bf \widehat{\Sigma}}^{L}{}_{N}+\frac{1}{2\widetilde{A}}{\bf
\widehat{\Sigma}}^{K}{}_{L}{\bf \widehat{\Sigma}}^{L}{}_{K}+\frac{1}{2\widetilde{B}}{\bf
\widehat{\Sigma}}^{K}{}_{K}{\bf \widehat{\Sigma}}^{L}{}_{L},\label{a91}\\
\mathbf{T}_{\rm int}&=&\frac{1}{2\widetilde{I}}g_{ik}g^{jl}{\bf \Sigma}^{i}{}_{j}{\bf
\Sigma}^{k}{}_{l}+\frac{1}{2\widetilde{A}}{\bf \Sigma}^{k}{}_{l}{\bf
\Sigma}^{l}{}_{k}+\frac{1}{2\widetilde{B}}{\bf \Sigma}^{k}{}_{k}{\bf
\Sigma}^{l}{}_{l}\label{b91}
\end{eqnarray}
with the same meaning of inertial constants $\widetilde{I}$, $\widetilde{A}$, $\widetilde{B}$
like previously, (\ref{c62}). And again the second terms of both expressions are identical;
the same is true of the third ones. Let us write explicitly
\[
{\bf \widehat{\Sigma}}^{A}{}_{B}{\bf \widehat{\Sigma}}^{B}{}_{A}={\bf \Sigma}^{k}{}_{l}{\bf
\Sigma}^{l}{}_{k}=-\hbar^{2}\varphi^{k}{}_{B}\varphi^{l}{}_{A}
\frac{\partial^{2}}{\partial\varphi^{k}{}_{A}\partial\varphi^{l}{}_{B}}-in\hbar
\varphi^{k}{}_{A}\frac{\partial}{\partial\varphi^{k}{}_{A}}
\]
The geometric interpretation of the affine spin, usual spin, and vorticity as generators of
transformation groups implies that many quantum expressions involving them may be, as it was
just seen, directly obtained from classical formulas by simple substitution of appropriate
operators instead of the corresponding classical phase-space quantities, so that the difficult
ordering problems are avoided.

For example, the very convenient classical expressions (\ref{a64}), (\ref{b64}) remain valid
on the operator level as alternative expressions respectively for (\ref{a91}), (\ref{b91})
free of the inconvenient transposition term (one with $\widetilde{I}$):
\begin{equation}\label{a92}
\mathbf{T}_{\rm int}=\frac{1}{2\alpha}{\bf C}(2)+\frac{1}{2\beta}{\bf
C}(1)^{2}+\frac{1}{2\mu}\|{\bf V}\|^{2}
\end{equation}
(affine-metrical model), and
\begin{equation}\label{b92}
\mathbf{T}_{\rm int}=\frac{1}{2\alpha}{\bf C}(2)+\frac{1}{2\beta}{\bf
C}(1)^{2}+\frac{1}{2\mu}\|{\bf S}\|^{2}
\end{equation}
(metrical-affine model), where $\alpha$, $\beta$, $\mu$ are the same constants as (\ref{d64})
from classical formulas, and the operator Casimirs $\mathbf{C}(k)$, $\|{\bf V}\|^{2}$,
$\|\mathbf{S}\|^{2}$ are given by
\[
{\bf C}(1)={\bf \widehat{\Sigma}}^{A}{}_{A}={\bf \Sigma}^{a}{}_{a},\qquad {\bf C}(2)={\bf
\widehat{\Sigma}}^{A}{}_{B}{\bf \widehat{\Sigma}}^{B}{}_{A}={\bf \Sigma}^{a}{}_{b}{\bf
\Sigma}^{b}{}_{a},
\]
and analogously for $k>2$ (till $k=n$), and
\[
\|{\bf V}\|^{2}=-\frac{1}{2}{\bf V}^{A}{}_{B}{\bf V}^{B}{}_{A},\qquad \|{\bf
S}\|^{2}=-\frac{1}{2}{\bf S}^{a}{}_{b}{\bf S}^{b}{}_{a}.
\]

The more so the affine-affine model (\ref{c64}) retains its structure when quantized
(transformed to the operator form):
\begin{equation}\label{a93}
\mathbf{T}_{\rm int}=\frac{1}{2A}{\bf C}(2)+\frac{1}{2A(n+A/B)}{\bf C}(1)^{2}.
\end{equation}
And similarly, the splitting of the kinetic energy into incompressible and dilatational parts
survives smoothly the quantization procedure. Decomposition (\ref{a48}), (\ref{b48}) of the
affine spin into the spin-shear and dilatation parts has the following form:
\[
{\bf \Sigma}^{a}{}_{b}=\mathbf{s}^{a}{}_{b}+\frac{1}{n}{\bf p}\;\delta^{a}{}_{b},\qquad {\bf
\widehat{\Sigma}}^{A}{}_{B}=\mathbf{\widehat{s}}^{A}{}_{B}+\frac{1}{n}{\bf
p}\;\delta^{A}{}_{B},
\]
where the trace-less parts are given by
\[
\mathbf{s}^{a}{}_{b}={\bf \Sigma}^{a}{}_{b}-\frac{1}{n}{\bf
\Sigma}^{d}{}_{d}\delta^{a}{}_{b},\qquad \mathbf{\widehat{s}}^{A}{}_{B}={\bf
\widehat{\Sigma}}^{A}{}_{B}-\frac{1}{n}{\bf \widehat{\Sigma}}^{D}{}_{D}\delta^{A}{}_{B}.
\]
They are formally Hermitian operators generating the unitary actions of SL$(V)$, SL$(U)$ on
L$^{2}(Q_{\rm int},\lambda)$ in the sense of (\ref{a83}),
\begin{eqnarray}
A\in{\rm SL}(V)&:&\left(\mathbf{A}\Psi\right)(\varphi)=\Psi(A\varphi),\nonumber\\
B\in{\rm SL}(U)&:&\left(\mathbf{B}\Psi\right)(\varphi)=\Psi(\varphi B).\nonumber
\end{eqnarray}
Incidentally, they are also formally Hermitian in L$^{2}(Q_{\rm int},l)$, because the above
actions of SL$(V)$, SL$(U)$ are there unitary.

The operator ${\bf p}$ has the following form:
\[
{\bf p}={\bf \Sigma}^{a}{}_{a}={\bf
\widehat{\Sigma}}^{A}{}_{A}=\frac{\hbar}{i}\frac{\partial}{\partial q}.
\]
It is formally Hermitian in L$^{2}(Q_{\rm int},\lambda)$ (but not in L$^{2}(Q_{\rm int},l)$)
and generates the one-parameter unitary group of dilatations. This is the group induced by the
additive translations of logarithmic deformation invariants $q^{a}$, in particular, by the
additive translations of their "centre of mass" $q$. This is the group which acts as follows:
\begin{eqnarray}
{\bf \Lambda}=e^{\xi}\in\mathbb{R}^{+}&:& \left({\bf
\Lambda}\Psi\right)(\varphi)=\Psi(\lambda\varphi),\qquad q\mapsto q+\xi.\nonumber
\end{eqnarray}
Let us denote the second-order Casimir operator for SL$(n,\mathbb{R})$ by $\mathbf{C}_{{\rm
SL}(n)}(2)$:
\[
{\bf C}_{{\rm SL}(n)}(2)={\bf s}^{a}{}_{b}{\bf s}^{b}{}_{a}={\bf \widehat{s}}^{A}{}_{B}{\bf
\widehat{s}}^{B}{}_{A},
\]
and similarly for higher-order ones $\mathbf{C}_{{\rm SL}(n)}(k)$ (but of course
$\mathbf{C}_{{\rm SL}(n)}(1)=0$; for orthogonal groups all of the odd orders vanish).

On the quantized level the structure of affine-affine, affine-metrical, metrical-affine models
(respectively, (\ref{a65}), (\ref{b65}), (\ref{c65})) beautifully survives in operator
language. Namely, one obtains, respectively,
\begin{eqnarray}
\mathbf{T}^{0}_{\rm int}&=&\frac{1}{2A}{\bf C}_{{\rm SL}(n)}(2)
+\frac{1}{2n(A+nB)}{\bf p}^{2}=\mathbf{T}^{0}_{\rm sh}+\mathbf{T}^{0}_{\rm dil},\nonumber\\
\mathbf{T}_{\rm int}&=&\frac{1}{2(I+A)}{\bf C}_{{\rm SL}(n)}(2)+\frac{1}{2n(I+A+nB)}{\bf
p}^{2}+\frac{I}{2(I^{2}-A^{2})}\|{\bf V}\|^{2},\nonumber\\
\mathbf{T}_{\rm int}&=&\frac{1}{2(I+A)}{\bf C}_{{\rm SL}(n)}(2)+\frac{1}{2n(I+A+nB)}{\bf
p}^{2}+\frac{I}{2(I^{2}-A^{2})}\|{\bf S}\|^{2}.\nonumber
\end{eqnarray}
Obviously,
\[
\mathbf{p}^{2}=-\hbar^{2}\frac{\partial^{2}}{\partial q^{2}}.
\]
All these formulas are automatically obtained from the corresponding classical expressions
(\ref{a65}), (\ref{b65}), (\ref{c65}) by the formal substitution of operators instead of
phase-space quantities. It is so because one deals here with generators of the underlying
transformation groups, quantities of profound geometric interpretation.

Just as in the classical case the quantum unbounded dilatational motion should be stabilized
by some potential $V(q)$ if the model is to describe quantum elastic vibrations. The Hamilton
operator splits then into two independent mutually commuting parts
\[
\mathbf{H}=\mathbf{H}_{\rm sh}+\mathbf{H}_{\rm dil}.
\]
The same is true for more general doubly isotropic potentials separating explicitly the shape
and dilatation dynamics:
\[
V=V_{\rm sh}(\ldots,q^{i}-q^{j},\ldots)+V_{\rm dil}(q).
\]
We have then
\begin{eqnarray}
\mathbf{H}_{\rm sh}&=&\frac{1}{2A}{\bf C}_{{\rm SL}(n)}(2)+V_{\rm sh},\nonumber\\
\mathbf{H}_{\rm sh}&=&\frac{1}{2(I+A)}{\bf C}_{{\rm SL}(n)}(2)
+\frac{I}{2(I^{2}-A^{2})}\|{\bf V}\|^{2}+V_{\rm sh},\nonumber\\
\mathbf{H}_{\rm sh}&=&\frac{1}{2(I+A)}{\bf C}_{{\rm SL}(n)}(2)+\frac{I}{2(I^{2}-A^{2})}\|{\bf
S}\|^{2}+V_{\rm sh}\nonumber
\end{eqnarray}
respectively for the affine-affine, affine-metrical, and metrical-affine models. The first two
of them reduce to the third one when we put $I=0$.

And obviously
\[
\mathbf{H}_{\rm dil}=\frac{1}{2n(I+A+nB)}{\bf p}^{2}+V_{\rm dil}(q)=
-\frac{\hbar^{2}}{2n(I+A+nB)}\frac{\partial^{2}}{\partial q^{2}}+V_{\rm dil}(q).
\]
$V_{\rm dil}$ may be chosen as some qualitatively satisfactory phenomenological model, e.g.,
$V_{\rm dil}=(\kappa/2)q^{2}$, the finite or infinite potential well, etc.

When $V_{\rm sh}=0$, we are dealing with purely geodetic affinely-invariant Hamiltonians built
entirely of the group generators. In such situations one can expect solutions of the
eigenproblem based completely on some purely algebraic ladder procedure.

Finally, it is interesting to express everything in terms of the two-polar decomposition. And
now an unpleasant surprise, namely, the automatic replacing of classic quantities by seemingly
natural operators does not work any longer. Namely, although $p$ may be automatically
substituted by $\mathbf{p}=(\hbar/i)\partial/\partial q$, it is not the case with $p_{a}$,
they are not replace by $(\hbar/i)\partial/\partial q^{a}$ and $\sum_{a}p^{2}_{a}$ in
(\ref{a70}) is not "quantized" to
\[
-\hbar^{2}\Delta[q^{a}]=-\hbar^{2}\sum_{a}\frac{\partial^{2}}{\partial q^{a}{}^{2}}.
\]
The point is that the additive translations of logarithmic
deformation invariants are not geometrically fundamental
operations. So, whereas (\ref{a92}), (\ref{b92}), (\ref{a93}) are
automatically $(-\hbar^{2}/2)$-multiples of the corresponding
Laplace-Beltrami operators, there is no such automatism with the
two-polar expression of these operators. Fortunately, there are no
problems with the spin and vorticity operators
$\mathbf{S}^{i}{}_{j}$, ${\bf V}^{A}{}_{B}$ and with operators
$\mathbf{\widehat{r}}{}^{a}{}_{b}$,
$\mathbf{\widehat{t}}^{a}{}_{b}$ corresponding to the classical
quantities (\ref{c49}). The reason is again their
group-theoretical interpretation: spin and vorticity generate
respectively spatial and material rotations. And the operators
$\mathbf{\widehat{r}}^{a}{}_{b}$, $\mathbf{\widehat{t}}^{a}{}_{b}$
are their representations in terms of the principal axes of the
Cauchy and Green deformation tensors,
\[
\mathbf{\widehat{r}}^{a}{}_{b}=L^{a}{}_{i}L^{j}{}_{b}\mathbf{S}^{i}{}_{j},\qquad
\mathbf{\widehat{t}}^{a}{}_{b}=-R^{a}{}_{A}R^{B}{}_{b}{\bf
V}^{A}{}_{B};
\]
the ordering of operators just as written here. Just as in the
classical theory, $\mathbf{\widehat{r}}^{a}{}_{b}$,
$\mathbf{\widehat{t}}^{a}{}_{b}$ are generators (in the
quantum-Poisson-bracket sense) of the right action of
SO$(n,\mathbb{R})$ on the quantities $L:\mathbb{R}^{n}\rightarrow
V$, $R:\mathbb{R}^{n}\rightarrow U$
\[
L\mapsto LU,\qquad R\mapsto RU, \qquad U\in {\rm
SO}(n,\mathbb{R}).
\]
Just as $\mathbf{\widehat{r}}^{a}{}_{b}$,
$\mathbf{\widehat{t}}^{a}{}_{b}$, the operators
$\mathbf{S}^{i}{}_{j}$, ${\bf V}^{A}{}_{B}$ act only on
generalized coordinates $x^{\mu}$, $y^{\mu}$ parameterizing
respectively $L$ and $R$ (some Euler angles, rotation vectors,
first-kind canonical coordinates, and so on). Any of the mentioned
classical quantities $\mathbf{S}^{i}{}_{j}$, ${\bf V}^{A}{}_{B}$,
$\mathbf{\widehat{r}}^{a}{}_{b}$, $\mathbf{\widehat{t}}^{a}{}_{b}$
has the following form:
\[
f^{\mu}(x)p(x)_{\mu},\qquad g^{\mu}(y)p(y)_{\mu},
\]
where $p(x)_{\mu}$, $p(y)_{\mu}$ are canonical momenta conjugate respectively to $x^{\mu}$,
$y^{\mu}$. Due to the group-theoretical meaning of the mentioned quantities, the corresponding
quantum operators are given respectively by the operators
\[
\frac{\hbar}{i}f^{\mu}(x)\frac{\partial}{\partial x^{\mu}},\qquad
\frac{\hbar}{i}g^{\mu}(y)\frac{\partial}{\partial y^{\mu}};
\]
the ordering just as explicitly written.

In analogy to (\ref{c70}) we introduce the operators
\[
\mathbf{M}=-{\bf \widehat{r}}-{\bf \widehat{t}},\qquad \mathbf{N}={\bf \widehat{r}}-{\bf
\widehat{t}}.
\]
The kinetic energy operator for affine-affine model is then given as
\begin{eqnarray}
{\bf T}^{\rm aff-aff}_{\rm int}=&-&\frac{\hbar^{2}}{2A}{\bf
D}_{\lambda}+\frac{\hbar^{2}B}{2A(A+nB)}\frac{\partial^{2}}{\partial q^{2}}\nonumber\\
&+&\frac{1}{32A}\sum_{a,b}\frac{{\bf M}^{2}_{ab}}{{\rm
sh}^{2}\frac{q^{a}-q^{b}}{2}}-\frac{1}{32A}\sum_{a,b}\frac{{\bf N}^{2}_{ab}}{{\rm
ch}^{2}\frac{q^{a}-q^{b}}{2}},\label{a99}
\end{eqnarray}
where
\begin{eqnarray}\label{b99}
{\bf D}_{\lambda}=\frac{1}{P_{\lambda}}\sum_{a}\frac{\partial}{\partial
q^{a}}P_{\lambda}\frac{\partial}{\partial q^{a}}=\sum_{a}\frac{\partial^{2}}{\partial
q^{a}{}^{2}}+ \sum_{a}\frac{\partial \ln P_{\lambda}}{\partial q^{a}}\frac{\partial}{\partial
q^{a}}
\end{eqnarray}
and $P_{\lambda}$ is given by (\ref{a81}). The expression (\ref{a99}) exactly equals
$-(\hbar^{2}/2)\Delta(\Gamma^{0})$, where the Laplace-Beltrami operator $\Delta(\Gamma^{0})$
is built of the configuration metric $\Gamma^{0}$ (\ref{a60}), i.e., corresponds to the
classical expression (\ref{a60}), (\ref{b60}). It is seen that $\mathbf{D}_{\lambda}$ differs
from the $\mathbb{R}^{n}$-Laplace operator $\sum_{a}\partial^{2}/\partial q^{a}{}^{2}$ by some
first-order differential operator. This is just the mentioned breakdown of the naive classical
analogy between $p_{a}$ and $(\hbar/i)\partial/\partial q^{a}$. The reason of this breakdown
is that the additive translations
\[
q^{a}\mapsto q^{a}+u^{a}
\]
do not preserve the measures $\lambda$, $\alpha$. Because of this their argument-wise action
on wave functions is not unitary in L$^{2}(Q,\alpha)$, L$^{2}(Q_{\rm int},\lambda)$.
Incidentally, it is not unitary in L$^{2}(Q,a)$, L$^{2}(Q_{\rm int},l)$ either. And
infinitesimal generators $(\hbar/i)\partial/\partial q^{a}$, $(\hbar/i)\partial/\partial
Q^{a}$ are not formally self-adjoint in these Hilbert spaces.

The affine-metric and metric-affine models are respectively given by
\begin{eqnarray}
{\bf T}^{\rm aff-met}_{\rm int}=&-&\frac{\hbar^{2}}{2\alpha}{\bf
D}_{\lambda}-\frac{\hbar^{2}}{2\beta}\frac{\partial^{2}}{\partial q^{2}}\nonumber\\
&+&\frac{1}{32\alpha}\sum_{a,b}\frac{{\bf M}^{2}_{ab}}{{\rm
sh}^{2}\frac{q^{a}-q^{b}}{2}}-\frac{1}{32\alpha}\sum_{a,b}\frac{{\bf N}^{2}_{ab}}{{\rm
ch}^{2}\frac{q^{a}-q^{b}}{2}}+\frac{1}{2\mu}\|{\bf V}\|^{2},\nonumber\\
{\bf T}^{\rm met-aff}_{\rm int}=&-&\frac{\hbar^{2}}{2\alpha}{\bf
D}_{\lambda}-\frac{\hbar^{2}}{2\beta}\frac{\partial^{2}}{\partial q^{2}}\nonumber\\
&+& \frac{1}{32\alpha}\sum_{a,b}\frac{{\bf M}^{2}_{ab}}{{\rm
sh}^{2}\frac{q^{a}-q^{b}}{2}}-\frac{1}{32\alpha}\sum_{a,b}\frac{{\bf N}^{2}_{ab}}{{\rm
ch}^{2}\frac{q^{a}-q^{b}}{2}}+\frac{1}{2\mu}\|{\bf S}\|^{2}.\nonumber
\end{eqnarray}
Quite similarly, quantizing the doubly-isotropic d'Alembert model we obtain
\[
{\bf T}^{\rm d'A}_{\rm int}=-\frac{\hbar^{2}}{2I}{\bf D}_{l}+ \frac{1}{8I}\sum_{a,b}\frac{{\bf
M}^{2}_{ab}}{(Q^{a}-Q^{b})^{2}}+\frac{1}{8I}\sum_{a,b}\frac{{\bf
N}^{2}_{ab}}{(Q^{a}+Q^{b})^{2}},
\]
where
\begin{equation}\label{b101}
{\bf D}_{l}=\frac{1}{P_{l}}\sum_{a}\frac{\partial}{\partial
Q^{a}}P_{l}\frac{\partial}{\partial Q^{a}}=\sum_{a}\frac{\partial^{2}}{\partial Q^{a}{}^{2}}+
\sum_{a}\frac{\partial \ln P_{l}}{\partial Q^{a}}\frac{\partial}{\partial Q^{a}},
\end{equation}
and the weight factor $P_{l}$ is given by (\ref{b81}).

The ordering of non-commuting operators ${\bf D}_{\lambda}$, ${\bf D}_{l}$ is just as
explicitly written here. There are no other ordering problems because the operators ${\bf
M}_{ab}$, ${\bf N}_{ab}$ (equivalently ${\bf \widehat{r}}_{ab}$, ${\bf \widehat{t}}_{ab}$) do
commute with deformation invariants $q^{a}$, $Q^{a}$.

Let us observe that the first-order differential operators in (\ref{b99}), (\ref{b101}) may be
eliminated by introducing modified amplitudes $\varphi$ given respectively by
\begin{equation}\label{c101}
\varphi=\sqrt{P_{\lambda}}\Psi,\qquad \varphi=\sqrt{P_{l}}\Psi.
\end{equation}
Then the action of $\mathbf{D}$-operators on $\Psi$ is represented by the action of operators
$\widetilde{\mathbf{D}}$ given respectively by
\[
-\frac{\hbar^{2}}{2A}\widetilde{\mathbf{D}}_{\lambda}=
-\frac{\hbar^{2}}{2A}\sum_{a}\frac{\partial^{2}}{\partial
q^{a}{}^{2}}+\widetilde{V}_{\lambda},\qquad -\frac{\hbar^{2}}{2I}\widetilde{\mathbf{D}}_{l}=
-\frac{\hbar^{2}}{2I}\sum_{a}\frac{\partial^{2}}{\partial Q^{a}{}^{2}}+\widetilde{V}_{l},
\]
where $\widetilde{V}_{\lambda}$, $\widetilde{V}_{l}$ are auxiliary "artificial" potentials
\begin{eqnarray}
\widetilde{V}_{\lambda}&=&-\frac{\hbar}{2A}\frac{1}{P^{2}_{\lambda}}
+\frac{\hbar^{2}}{4A}\frac{1}{P_{\lambda}}\sum_{a}\left(\frac{\partial P_{\lambda}}{\partial
q^{a}}\right)^{2},\nonumber\\
\nonumber\widetilde{V}_{l}&=&-\frac{\hbar}{2I}\frac{1}{P^{2}_{l}}
+\frac{\hbar^{2}}{4I}\frac{1}{P_{l}}\sum_{a}\left(\frac{\partial P_{l}}{\partial
Q^{a}}\right)^{2}.
\end{eqnarray}
In other words
\[
{\bf D}\varphi=\sqrt{P}{\bf D}\Psi.
\]
All other terms of the kinetic energy operators commute with $\sqrt{P}$ interpreted as a
position-type operator. Obviously, the same concerns usual potential terms $V$. Therefore,
finally, the Hamilton operator $\mathbf{H}=\mathbf{T}+\mathbf{V}$ is represented in
$\varphi$-terms by $\widetilde{\mathbf{H}}$,
\[
\widetilde{\mathbf{H}}\varphi=\sqrt{P}\mathbf{H}\Psi,
\]
where analytically the action of $\widetilde{\mathbf{H}}$ differs from the action of
$\mathbf{H}$ in that the ${\bf D}$-operators are replaced by the usual
$\mathbb{R}^{n}$-Laplace operators and additional $\widetilde{V}$-potentials appear and are
combined with the "true" potentials $V$. The stationary Schr\"odinger equation, i.e.,
eigenequation
\[
\mathbf{H}\Psi=E\Psi
\]
is equivalent to
\[
\widetilde{\mathbf{H}}\varphi=E\varphi.
\]
This eigenproblem is meant in the Hilbert space based on the modified scalar product without
the weight factor $P$ in the integration element,
\[
(\varphi_{1}|\varphi_{2})=\int\overline{\varphi_{1}}\varphi_{2}dq^{1}\cdots
dq^{n}d\mu(L)d\mu(R)
\]
in affine models, and
\[
(\varphi_{1}|\varphi_{2})=\int\overline{\varphi_{1}}\varphi_{2}dQ^{1}\cdots
dQ^{n}d\mu(L)d\mu(R)
\]
in d'Alembert models. Obviously,
\[
(\varphi_{1}|\varphi_{2})=\langle\Psi_{1}|\Psi_{2}\rangle.
\]
There is no real simplification in replacing $\Psi$ by $\varphi$ because instead of the
complicated first-order differential operator the equally so complicated potential
$\widetilde{V}$ appears.

In geodetic problems and in problems with the doubly isotropic potentials
$V(q^{1},\ldots,q^{n})$, in particular, with the stabilizing dilatation potentials $V_{\rm
dil}(q)$, the above Schr\"odinger equations may be reduced to ones involving only coordinates
$q^{1},\ldots,q^{n}$, because the action of $\mathbf{H}$ on the $(L,R)$-dependence of wave
functions may be algebraized. This is based on the generalized Fourier analysis on the compact
group SO$(n,\mathbb{R})$.

To simplify the treatment we identify analytically $Q_{\rm int}$
with GL$^{+}(n,\mathbb{R})$ and use the matrix form of the
two-polar decomposition $\varphi=LDR^{-1}$. According to the
Peter-Weyl theorem, the wave functions may be expanded in
$(L,R)$-variables with respect to matrix elements of irreducible
unitary representations of SO$(n,\mathbb{R})$. Their expansion
coefficients are functions of deformation invariants $q^{a}$, or
equivalently, $Q^{a}$. Let $\Omega$ denote the set of irreducible
unitary representations of SO$(n,\mathbb{R})$ (more precisely, the
set of their equivalence classes). Obviously, due to the
compactness of the group SO$(n,\mathbb{R})$ these representations
are finite-dimensional; their dimensions will be denoted by
$N(\alpha)$. In the physical three-dimensional case $\Omega$ is
the set of all non-negative integers $s=0,1,2,\ldots$ and
$N(s)=2s+1$. If for some reasons we replace the rotation groups by
their universal coverings $\overline{{\rm SO}(n,\mathbb{R})}$ and
so admit half-integer angular momenta, then $\Omega$ is the set of
all non-negative half-integers and integers $s=0,1/2,1,3/2,\ldots$
and again $N(s)=2s+1$. Obviously, in the planar case
$\Omega=\mathbb{Z}$ is the set of all integers and $N(m)=1$ for
any $m\in\mathbb{Z}$ (Abelian group).

Let $\mathcal{D}^{\alpha}$ be $N(\alpha)\times N(\alpha)$ matrices of irreducible
representations. Then the mentioned expansion has the following form:
\begin{equation}\label{a105}
\Psi(\varphi)=\Psi(L,D,R)=\sum_{\alpha,\beta\in\Omega}
\sum_{m,n=1}^{N(\alpha)}\sum_{k,l=1}^{N(\beta)}
\mathcal{D}^{\alpha}{}_{mn}(L)f^{\alpha\beta}_{{}^{nk}_{ml}}
(D)\mathcal{D}^{\beta}{}_{kl}\left(R^{-1}\right).
\end{equation}
The non-uniqueness of the polar decomposition implies that the
deformation invariants $q^{1},\ldots,q^{n}$ ($Q^{1},\ldots,Q^{n}$)
are very complicated indistinguishable parastatistical "particles"
in $\mathbb{R}$. There is no place here to get into more details.
The point is that the reduced amplitudes
$f^{\alpha\beta}_{{}^{nk}_{ml}}$ as functions of
$q^{1},\ldots,q^{n}$ must satisfy certain conditions due to which
the resulting $\Psi$ as a function of $L$, $D$, and $R$ does not
distinguish triplets $(L,D,R)$ representing the same configuration
$\varphi$, i.e., $\Psi(L_{1},D_{1},R_{1})=\Psi(L_{2},D_{2},R_{2})$
if $L_{1}D_{1}R^{-1}_{1}=L_{2}D_{2}R^{-1}_{2}$. This is simply the
condition for $\Psi$ to be a one-valued function on the
configuration space $Q_{\rm int}$.

One can consider the matrix elements $\mathcal{D}^{\alpha}{}_{mk}$ as explicitly known. And in
fact, they are deeply investigated special functions on the orthogonal groups
SO$(n,\mathbb{R})$. In the physical case $n=3$ they are well-known functions
$\mathcal{D}^{j}{}_{mk}$ found by Wigner. Here $j=0,1,2,\ldots$ or, if we replace
SO$(3,\mathbb{R})$ by its universal covering SU$(2)$ (for the general $n$, SO$(n,\mathbb{R})$
is replaced by the group Spin$(n)$), $j=0,1/2,1,3/2,\ldots$. And according to the standard
convention $m=-j,-j+1,\ldots,j-1,j$, $k=-j,-j+1,\ldots,j-1,j$; for the fixed $j$, $m$, and $k$
have $(2j+1)$-element integer range with jumps by one both for the integer and half-integer
$j$.

The operators $\mathbf{S}^{i}{}_{j}$, ${\bf V}^{A}{}_{B}$, ${\bf \widehat{r}}^{a}{}_{b}$,
${\bf \widehat{t}}^{a}{}_{b}$ when acting on functions $\mathcal{D}^{\alpha}{}_{mk}$ may be
replaced by some standard algebraic operations. This enables one to reduce the Schr\"odinger
equation for the wave functions $\Psi$ depending on $n^{2}$ variables $\varphi^{i}{}_{A}$ to
some eigenproblems for the multi-component amplitudes $f^{\alpha\beta}$ depending only on the
$n$ deformation invariants $q^{a}$. Therefore, in a sense, the problem may be reduced to the
Cartan subgroup of diagonal matrices $\varphi$ (the maximal Abelian subgroup in
GL$(n,\mathbb{R})$).

In geodetic models and in models with doubly isotropic potentials (ones depending only on
deformation invariants; dilatations-stabilizing potentials $V(q)$ provide the simplest
example), the labels $m,n$ in (\ref{a105}) are good quantum numbers. The Hamilton operator
$\mathbf{H}$ commutes with the operators of spin and vorticity, i.e., $\mathbf{S}^{i}{}_{j}$,
$\mathbf{V}^{A}{}_{B}$. Also the representation labels $\alpha,\beta\in\Omega$ are good
quantum numbers. They are equivalent to the systems of eigenvalues of the Casimir invariants
built of $\mathbf{S}$, $\mathbf{V}$:
\begin{eqnarray}\label{a107}
{\bf C}(\mathbf{S},p)={\bf S}^{i}{}_{k}{\bf S}^{k}{}_{m}\cdots{\bf
S}^{r}{}_{z}{\bf S}^{z}{}_{i},\qquad {\bf
C}\left(\mathbf{V},p\right)={\bf V}^{A}{}_{K}{\bf
V}^{K}{}_{M}\cdots{\bf V}^{R}{}_{Z}{\bf V}^{Z}{}_{A}
\end{eqnarray}
($p$ factors). These eigenvalues will be denoted respectively by $C^{\alpha}(p)$,
$C^{\beta}(p)$. Obviously,
\[
{\bf C}(\mathbf{S},p)={\bf C}(\mathbf{\widehat{r}},p),\qquad {\bf
C}\left(\mathbf{V},p\right)={\bf
C}\left(\mathbf{\widehat{t}},p\right).
\]
The above Casimir invariants vanish trivially for the odd values of $p$; so in the physical
case $n=3$ there is only one possibility: ${\bf C}(\mathbf{S},2)$, ${\bf
C}\left(\mathbf{V},2\right)$. Due to the peculiarity of dimension three, where skew-symmetric
tensors may be identified with axial vectors, it is more convenient to use
\[
\|{\bf S}\|^{2}=-\frac{1}{2}{\bf S}^{a}{}_{b}{\bf S}^{b}{}_{a},\qquad \|{\bf
V}\|^{2}=-\frac{1}{2}{\bf V}^{A}{}_{B}{\bf V}^{B}{}_{A},
\]
i.e., $(-1/2)$-multiples of ${\bf C}(\mathbf{S},2)$, ${\bf C}\left(\mathbf{V},2\right)$. The
point is that for $n=3$
\[
\|{\bf S}\|^{2}={\bf S}^{2}_{1}+{\bf S}^{2}_{2}+{\bf S}^{2}_{3},\qquad \|{\bf V}\|^{2}={\bf
V}^{2}_{1}+{\bf V}^{2}_{2}+{\bf V}^{2}_{3},
\]
where
\[
{\bf S}_{a}=\frac{1}{2}\varepsilon_{ab}{}^{c}{\bf S}^{b}{}_{c},\qquad {\bf
V}_{A}=\frac{1}{2}\varepsilon_{AB}{}^{C}{\bf V}^{B}{}_{C}.
\]
The raising and lowering of indices is meant here in the sense of
orthonormal coordinates (Kronecker-delta trivial operation). The
same convention is used for ${\bf \widehat{r}}^{a}{}_{b}$, ${\bf
\widehat{t}}^{a}{}_{b}$, i.e.,
\[
{\bf \widehat{r}}_{a}=\frac{1}{2}\varepsilon_{ab}{}^{c}{\bf
\widehat{r}}^{b}{}_{c},\qquad {\bf
\widehat{t}}_{a}=\frac{1}{2}\varepsilon_{ab}{}^{c}{\bf
\widehat{t}}^{b}{}_{c}.
\]
Obviously,
\[
\|{\bf \widehat{r}}\|^{2}=\|{\bf S}\|^{2},\qquad \|{\bf
\widehat{t}}\|^{2}=\|{\bf V}\|^{2}.
\]
The corresponding eigenvalues are given by
\[
C(s,2)=\hbar^{2}s(s+1),\qquad C(j,2)=\hbar^{2}j(j+1),
\]
where $s,j$ are non-negative integers or non-negative integers and positive half-integers when
GL$^{+}(3,\mathbb{R})$, SL$(3,\mathbb{R})$ are replaced by their coverings $\overline{{\rm
GL}^{+}(3,\mathbb{R})}$, SU$(2)$.

It is convenient to use multi-component wave functions with values in the space of complex
$N(\alpha)\times N(\beta)$ matrices ($(2s+1)\times(2j+1)$ in the physical case $n=3$):
\begin{equation}\label{a109}
\Psi(\varphi)=\Psi^{\alpha\beta}(L,D,R)=
\mathcal{D}^{\alpha}(L)f^{\alpha\beta}(D)\mathcal{D}^{\beta}(R),
\end{equation}
where $f^{\alpha\beta}$ are complex $N(\alpha)\times N(\beta)$ matrices-reduced wave
amplitudes depending only on the deformation invariants. In the physical three-dimensional
case, when $\mathcal{D}^{\alpha}{}_{mn}$ are Wigner special functions
$\mathcal{D}^{s}{}_{mn}$, we, as usual, take $m,n$ running from $-s$ to $s$ and jumping by one
(also in the "spinorial" case when $s$ is half-integer). And then
\begin{eqnarray}
\|{\bf S}\|^{2}\Psi^{sj}=\hbar^{2}s(s+1)\Psi^{sj},&\qquad& \|{\bf
V}\|^{2}\Psi^{sj}=\hbar^{2}j(j+1)\Psi^{sj},\nonumber\\
{\bf S}_{3}\Psi^{sj}_{ml}=\hbar m\Psi^{sj}_{ml},&\qquad& {\bf
V}_{3}\Psi^{sj}_{ml}=\hbar l\Psi^{sj}_{ml}.\nonumber
\end{eqnarray}
And similarly, when the values $n,k$ in the superposition (\ref{a105}) are kept fixed and we
retain only the corresponding single term, for the resulting $\Psi$ we have
\[
{\bf \widehat{r}}_{3}\Psi^{sj}_{{}^{ml}_{nk}}=\hbar n\Psi^{sj}_{{}^{ml}_{nk}},\qquad {\bf
\widehat{t}}_{3}\Psi^{sj}_{{}^{ml}_{nk}}=\hbar k\Psi^{sj}_{{}^{ml}_{nk}}.
\]

Let us now describe in a few words the afore-mentioned algebraization procedure in the sector
of $(L,R)$-degrees of freedom. If a compact group counterpart of the usual Fourier-transform
algebraization, where $\partial/\partial x^{a}$ is represented by the point-wise
multiplication of the Fourier transform of $f(\bar{x})$, $\widehat{f}\left(\bar{k}\right)$ by
$ik^{a}$.

Let us introduce some auxiliary symbols.

The group SO$(n,\mathbb{R})$ may be parameterized by the first-kind canonical coordinates
$\omega$, namely,
\[
W(\omega)=\exp\left(\frac{1}{2}\omega^{a}{}_{b}E^{b}{}_{a}\right),
\]
where the basic matrices of the Lie algebra SO$(n,\mathbb{R})^{\prime}$ are given by
\[
\left(E^{b}{}_{a}\right)^{c}{}_{d}=\delta^{b}{}_{d}\delta^{c}{}_{a}- \delta^{bc}\delta_{ad},
\]
and the matrix $\omega$ is skew-symmetric in the "cosmetic" Kronecker sense. Therefore,
independent coordinates may be chosen as $\omega^{a}{}_{b}$, $a<b$, or conversely. However,
for the symmetry reasons it is more convenient to use the representation with the summation
extended over all possible $\omega^{a}{}_{b}$.

To be more "sophisticated", the groups SO$(V,g)$, SO$(U,\eta)$ are parameterized as follows:
\[
W(\omega)=\exp\left(\frac{1}{2}\omega^{i}{}_{j}E^{j}{}_{i}\right),\qquad
W(\omega)=\exp\left(\frac{1}{2}\widehat{\omega}^{A}{}_{B}\widehat{E}^{B}{}_{A}\right),
\]
where $E^{i}{}_{j}$, $\widehat{E}^{A}{}_{B}$ are basic matrices of Lie algebras
SO$(V,g)^{\prime}$, SO$(U,\eta)^{\prime}$ given by
\[
\left(E^{i}{}_{j}\right)^{k}{}_{l}=\delta^{i}{}_{l}\delta^{k}{}_{j}-g^{ik}g_{jl},\qquad
\left(\widehat{E}^{A}{}_{B}\right)^{C}{}_{D}=\delta^{A}{}_{D}\delta^{C}{}_{B}-
\eta^{AC}\eta_{BD}.
\]
The skew-symmetry of $\omega$ in the above exponential formulas is meant respectively as
follows:
\[
\omega^{a}{}_{b}=-g^{ac}g_{bd}\omega^{d}{}_{c},\qquad
\widehat{\omega}^{A}{}_{B}=-\eta^{AC}\eta_{BD}\widehat{\omega}^{D}{}_{C}.
\]
Now matrices of irreducible representations $\mathcal{D}^{\alpha}$ are given by
\[
\mathcal{D}^{\alpha}\left(L(l)\right)=\exp\left(\frac{1}{2}l^{a}{}_{b}
M^{\alpha}{}^{b}{}_{a}\right),\qquad
\mathcal{D}^{\alpha}\left(R(r)\right)=\exp\left(\frac{1}{2}r^{a}{}_{b}
M^{\alpha}{}^{b}{}_{a}\right),
\]
where $l$ and $r$ denote the $\omega$-parameters, respectively, for the $L$- and $R$-factors
of the two-polar decomposition. The anti-Hermitian matrices $M^{\alpha}$ will be expressed by
the Hermitian ones $S^{\alpha}$,
\[
S^{\alpha}{}^{a}{}_{b}=\frac{\hbar}{i}M^{\alpha}{}^{a}{}_{b}.
\]
The commutation rules for $M^{\alpha}{}^{a}{}_{b}$ are expressed through the structure
constants of SO$(n,\mathbb{R})$,
\[
[M^{s}{}_{ab},M^{s}{}_{cd}]=-g_{ad}M^{s}{}_{cb}+g_{cb}M^{s}{}_{ad}
-g_{bd}M^{s}{}_{ac}+g_{ac}M^{s}{}_{bd},
\]
and therefore
\[
\frac{1}{i\hbar}[S^{j}{}_{ab},S^{j}{}_{cd}]=g_{ad}S^{j}{}_{cb}-g_{cb}S^{j}{}_{ad}
+g_{bd}S^{j}{}_{ac}-g_{ac}S^{j}{}_{bd}.
\]
Indices here are shifted with the help of $g_{ab}$; as a rule we use orthonormal coordinates
when $g_{ab}=\delta_{ab}$.

In the physical three-dimensional case when we put
\[
S^{j}{}_{a}:=\frac{1}{2}\varepsilon_{a}{}^{bc}S^{j}{}_{bc},
\]
we obviously have
\[
\frac{1}{i\hbar}[S^{j}{}_{a},S^{j}{}_{b}]=\varepsilon_{ab}{}^{c}S^{j}{}_{c}.
\]
From the fact that $\mathcal{D}^{\alpha}$ are representations and
$(i/\hbar)\mathbf{S}^{k}{}_{l}$, $(i/\hbar)\mathbf{V}^{A}{}_{B}$,
$(i/\hbar)\mathbf{\widehat{r}}^{a}{}_{b}$, $(i/\hbar)\mathbf{\widehat{t}}^{a}{}_{b}$ are
infinitesimal generators of left and right orthogonal actions on the $(L,R)$-variables it
follows immediately that
\begin{eqnarray}
{\bf S}^{i}{}_{j}\Psi^{\alpha\beta}=S^{\alpha
i}{}_{j}\Psi^{\alpha\beta},&\qquad& {\bf
\widehat{r}}^{a}{}_{b}\Psi^{\alpha\beta}=\mathcal{D}^{\alpha}(L)S^{\alpha
a}{}_{b}f^{\alpha\beta}(D)\mathcal{D}^{\beta}\left(R^{-1}\right),\nonumber\\
{\bf V}^{A}{}_{B}\Psi^{\alpha\beta}=\Psi^{\alpha\beta}S^{\beta
A}{}_{B},&\qquad& {\bf
\widehat{t}}^{a}{}_{b}\Psi^{\alpha\beta}=\mathcal{D}^{\alpha}(L)f^{\alpha\beta}(D)S^{\beta
a}{}_{b}\mathcal{D}^{\beta}\left(R^{-1}\right).\nonumber
\end{eqnarray}
Therefore, spin and vorticity act on the wave amplitudes $\Psi^{\alpha\beta}$ as a whole, and
in a purely algebraic way. On the other hand, to describe in an algebraic way the action of
$\mathbf{\widehat{r}}^{a}{}_{b}$, $\mathbf{\widehat{t}}^{a}{}_{b}$, one must extract from
$\Psi^{\alpha\beta}$ the reduced amplitudes $f^{\alpha\beta}(q^{1},\ldots,q^{n})$. And it is
only this amplitude that is affected by the action of $\mathbf{\widehat{r}}^{a}{}_{b}$,
$\mathbf{\widehat{t}}^{a}{}_{b}$ according to the following rules:
\[
{\bf \widehat{r}}^{a}{}_{b}:\qquad f^{\alpha\beta}\mapsto
S^{\alpha}{}^{a}{}_{b}f^{\alpha\beta},\qquad {\bf
\widehat{t}}^{a}{}_{b}:\qquad f^{\alpha\beta}\mapsto
f^{\alpha\beta}S^{\beta}{}^{a}{}_{b}.
\]
It is very convenient to use the following notation:
\[
\overrightarrow{S^{\alpha}}{}^{a}{}_{b}f^{\alpha\beta}:=
S^{\alpha}{}^{a}{}_{b}f^{\alpha\beta},\qquad
\overleftarrow{S^{\beta}}{}^{a}{}_{b}f^{\alpha\beta}:= f^{\alpha\beta}S^{\beta}{}^{a}{}_{b}.
\]
As $\mathcal{D}^{\alpha}$ are irreducible, the matrices
\[
C(S^{\alpha},p):=S^{\alpha}{}^{a}{}_{b}S^{\alpha}{}^{b}{}_{c}\cdots
S^{\alpha}{}^{u}{}_{w}S^{\alpha}{}^{w}{}_{a}
\]
($p$ factors) are proportional to the $N(\alpha)\times N(\alpha)$ identity matrix,
\[
C(S^{\alpha},p)=C^{\alpha}(p)\mathbb{I}_{N(\alpha)},
\]
where $C^{\alpha}(p)$ are eigenvalues of (\ref{a107}).

In particular, in the physical case $n=3$ we have
\begin{eqnarray}
\|{\bf S}\|^{2}\Psi^{sj}=\|{\bf
\widehat{r}}\|^{2}\Psi^{sj}=\hbar^{2}s(s+1)\Psi^{sj},&\qquad&
{\bf S}_{a}\Psi^{sj}=S^{s}{}_{a}\Psi^{sj},\nonumber\\
\|{\bf V}\|^{2}\Psi^{sj}=\|{\bf
\widehat{t}}\|^{2}\Psi^{sj}=\hbar^{2}j(j+1)\Psi^{sj},&\qquad& {\bf
V}_{A}\Psi^{sj}=\Psi^{sj}S^{j}{}_{A},\nonumber
\end{eqnarray}
where $S^{s}{}_{a}$ are standard Wigner matrices of the angular momentum with the squared
magnitude $\hbar^{2}s(s+1)$. Multiplying them by $(i/\hbar)$ we obtain standard bases of
irreducible representations of the Lie algebra SO$(3,\mathbb{R})^{\prime}$. For the standard
Wigner representation the following is also true:
\[
{\bf S}_{3}\Psi^{sj}_{ml}=\hbar m\Psi^{sj}_{ml},\qquad {\bf V}_{3}\Psi^{sj}_{ml}=\hbar
l\Psi^{sj}_{ml}.
\]
Similarly, the action of $\mathbf{\widehat{r}}$, $\mathbf{\widehat{t}}$ operators is
represented by the following operations on the reduced amplitudes:
\[
{\bf \widehat{r}}_{a}:\qquad f^{sj}\mapsto
S^{s}{}_{a}f^{sj}=\overrightarrow{S^{s}}{}_{a}f^{sj},\qquad {\bf
\widehat{t}}_{a}:\qquad f^{sj}\mapsto
f^{sj}S^{j}{}_{a}=\overleftarrow{S^{j}}{}_{a}f^{sj}.
\]
In particular,
\[
{\bf \widehat{r}}_{3}:\qquad \left[f^{sj}_{ml}\right]\mapsto
\left[\hbar mf^{sj}_{ml}\right],\qquad {\bf
\widehat{t}}_{3}:\qquad \left[f^{sj}_{ml}\right]\mapsto
\left[\hbar lf^{sj}_{ml}\right].
\]
Using the well-known orthogonality relations for the matrix elements of irreducible unitary
representations $\mathcal{D}^{\alpha}{}_{mn}$ \cite{Ham62,Weyl31} we can rewrite the scalar
product in the following form:
\begin{equation}\label{a114}
\langle\Psi_{1}|\Psi_{2}\rangle=\sum_{\alpha,\beta\in\Omega}\frac{1}{N(\alpha)N(\beta)}
\int{\rm Tr}\left(f_{1}^{\alpha\beta
+}\left(q^{a}\right)f_{2}^{\alpha\beta}\left(q^{b}\right)\right) P_{\lambda}dq^{1}\cdots
dq^{n},
\end{equation}
where $P_{\lambda}$ is the weight factor given by (\ref{a81}), and the argument symbols like
$q^{a}$ are abbreviations for the system $(q^{1},\ldots,q^{n})$. The trace operation is meant
in the sense of matrix two-indices:
\[
{\rm Tr}\left(f_{1}^{\alpha\beta +}f_{2}^{\alpha\beta}\right)=
\sum^{N(\alpha)}_{n,m=1}\sum^{N(\beta)}_{k,l=1}\overline{f_{1}{\;}^{\alpha\beta}_{{}^{nk}_{ml}}}
f_{2}{\;}^{\alpha\beta}_{{}^{nk}_{ml}}.
\]
If no superposition over $m,l$ in (\ref{a105}) is performed and we use the matrix-valued wave
functions (\ref{a109}), the trace operation is meant in the usual sense.

Obviously, if we use the modified wave functions $\varphi$ (\ref{c101}), then the scalar
product expression is free of the weight factor $P_{\lambda}$,
\[
(\varphi_{1}|\varphi_{2})=\sum_{\alpha,\beta\in\Omega}\frac{1}{N(\alpha)N(\beta)} \int{\rm
Tr}\left(g_{1}^{\alpha\beta +}\left(q^{a}\right)g_{2}^{\alpha\beta}\left(q^{b}\right)\right)
dq^{1}\cdots dq^{n},
\]
where, obviously, $g=\sqrt{P_{\lambda}}f$. Quite analogous formulas are true for the
d'Alem\-bert models; simply $P_{\lambda}$ is replaced then by $P_{l}$ (\ref{b81}) and instead
$q^{a}$ we use their exponential functions $Q^{a}$,
\begin{eqnarray}
\langle\Psi_{1}|\Psi_{2}\rangle&=&\sum_{\alpha,\beta\in\Omega}\frac{1}{N(\alpha)N(\beta)}
\int{\rm Tr}\left(f_{1}^{\alpha\beta
+}\left(Q^{a}\right)f_{2}^{\alpha\beta}\left(Q^{b}\right)\right)
P_{l}dQ^{1}\cdots dQ^{n},\nonumber\\
(\varphi_{1}|\varphi_{2})&=&\sum_{\alpha,\beta\in\Omega}\frac{1}{N(\alpha)N(\beta)} \int{\rm
Tr}\left(g_{1}^{\alpha\beta +}\left(Q^{a}\right)g_{2}^{\alpha\beta}\left(Q^{b}\right)\right)
dQ^{1}\cdots dQ^{n}.\nonumber
\end{eqnarray}
Remark: it is implicit assumed in the above formulas that the Haar measure on the
$(L,R)$-manifolds is normalized to unity (the total "volume" of the corresponding manifolds
equals one).

We restrict ourselves to Hamiltonians of the form $\mathbf{H}=\mathbf{T}+\mathbf{V}$ with some
doubly-isotropic potentials $V(q^{1},\ldots,q^{n})$, in particular, with some
dilatation-stabilizing potentials $V(q)$ (affinely-invariant geodetic incompressible models).
The energy eigenproblem, i.e., stationary Schr\"odinger equation
\[
\mathbf{H}\Psi=E\Psi
\]
is equivalent to the infinite sequence of eigenequations for the reduced multi-component
amplitudes $f^{\alpha\beta}$:
\[
\mathbf{H}^{\alpha\beta}\Psi^{\alpha\beta}=E^{\alpha\beta}\Psi^{\alpha\beta}.
\]
The simultaneous spatial and material isotropy imply in the $N(\alpha)\times N(\beta)$-fold
degeneracy, i.e., for every component of the $N(\alpha)\times N(\beta)$-matrix amplitude
$f^{\alpha\beta}$ there exists an $N(\alpha)\times N(\beta)$-dimensional subspace of
solutions, just as seen from the symbol $f^{\alpha\beta}_{{}^{nk}_{ml}}$ used in (\ref{a105}).

The reduced Hamiltonians
\[
\mathbf{H}^{\alpha\beta}=\mathbf{T}^{\alpha\beta}+\mathbf{V}
\]
are $N(\alpha)\times N(\beta)$ matrices built of second-order differential operators.

For the affine-affine model of the kinetic energy we have
\begin{eqnarray}
{\bf T}^{\alpha\beta}f^{\alpha\beta}=&-&\frac{\hbar^{2}}{2A}{\bf
D}f^{\alpha\beta}+\frac{1}{32A}\sum_{a,b}\frac{\left(\overleftarrow{S^{\beta}}{}^{a}{}_{b}-
\overrightarrow{S^{\alpha}}{}^{a}{}_{b}\right)^{2}}{{\rm
sh}^{2}\frac{q^{a}-q^{b}}{2}}f^{\alpha\beta}\nonumber\\
&-&\frac{1}{32A}\sum_{a,b}\frac{\left(\overleftarrow{S^{\beta}}{}^{a}{}_{b}+
\overrightarrow{S^{\alpha}}{}^{a}{}_{b}\right)^{2}}{{\rm
ch}^{2}\frac{q^{a}-q^{b}}{2}}f^{\alpha\beta}
+\frac{\hbar^{2}B}{2A(A+nB)}\frac{\partial^{2}}{\partial q^{2}}f^{\alpha\beta}.\label{a117}
\end{eqnarray}
For the spatially metrical and materially affine model we obtain
\begin{eqnarray}
{\bf
T}^{\alpha\beta}f^{\alpha\beta}=&-&\frac{\hbar^{2}}{2\alpha}{\bf
D}f^{\alpha\beta}+\frac{1}{2\mu}C(\alpha,2)f^{\alpha\beta}+\frac{1}{32\alpha}\sum_{a,b}
\frac{\left(\overleftarrow{S^{\beta}}{}^{a}{}_{b}-
\overrightarrow{S^{\alpha}}{}^{a}{}_{b}\right)^{2}}{{\rm
sh}^{2}\frac{q^{a}-q^{b}}{2}}f^{\alpha\beta}\nonumber
\\
&-&\frac{1}{32\alpha}\sum_{a,b}\frac{\left(\overleftarrow{S^{\beta}}{}^{a}{}_{b}+
\overrightarrow{S^{\alpha}}{}^{a}{}_{b}\right)^{2}}{{\rm
ch}^{2}\frac{q^{a}-q^{b}}{2}}f^{\alpha\beta}
-\frac{\hbar^{2}}{2\beta}\frac{\partial^{2}}{\partial q^{2}}f^{\alpha\beta} ,\label{b117}
\end{eqnarray}
where $C(\alpha,2)$ is the $\alpha$-th eigenvalue of the rotational Casimir
$\|\mathbf{S}\|^{2}$, thus,
\[
-\frac{1}{2}S^{\alpha}{}^{i}{}_{j}S^{\alpha}{}^{j}{}_{i}=C(\alpha,2)\mathbb{I}_{N(\alpha)}.
\]
Obviously, for the physical dimension $n=3$, $f^{\alpha\beta}=f^{sj}$, we have
$C(s,2)=\hbar^{2}s(s+1)$. And similarly for the spatially affine and materially metrical model
we have
\begin{eqnarray}
{\bf
T}^{\alpha\beta}f^{\alpha\beta}=&-&\frac{\hbar^{2}}{2\alpha}{\bf
D}f^{\alpha\beta}+\frac{1}{2\mu}C(\beta,2)
f^{\alpha\beta}+\frac{1}{32\alpha}\sum_{a,b}
\frac{\left(\overleftarrow{S^{\beta}}{}^{a}{}_{b}-
\overrightarrow{S^{\alpha}}{}^{a}{}_{b}\right)^{2}}{{\rm
sh}^{2}\frac{q^{a}-q^{b}}{2}}f^{\alpha\beta}\nonumber\\
&-&\frac{1}{32\alpha}\sum_{a,b}\frac{\left(\overleftarrow{S^{\beta}}{}^{a}{}_{b}+
\overrightarrow{S^{\alpha}}{}^{a}{}_{b}\right)^{2}}{{\rm
ch}^{2}\frac{q^{a}-q^{b}}{2}}f^{\alpha\beta}
-\frac{\hbar^{2}}{2\beta}\frac{\partial^{2}}{\partial q^{2}}f^{\alpha\beta},\label{a118}
\end{eqnarray}
where $C(\beta,2)$ appears as the $\beta$-th eigenvalue of the vorticity Casimir
$\|\mathbf{V}\|^{2}$, and just as previously for $n=3$, $f^{\alpha\beta}=f^{sj}$, we have
$C(j,2)=\hbar^{2}j(j+1)$. It is so as if the doubly affine background ($\mathbf{T}$
affinely-invariant in the physical and material space) was responsible for some fundamental
part of the spectra, perturbated by some internal rotations of the body itself or of the
deformation axes. This perturbation and the resulting splitting of energy levels becomes
remarkable when $\mu$ is small, i.e., when the inertial constants $I$, $A$ differ slightly.
The suggestive terms
\[
\frac{\hbar^{2}}{2\mu}s(s+1),\qquad \frac{\hbar^{2}}{2\mu}j(j+1)
\]
as contributions to energy levels are very interesting and seem to be supported by
experimental data in various ranges of physical phenomena.

Finally, let us quote the corresponding form of $\mathbf{T}^{\alpha\beta}$ for the quantized
d'Alem\-bert model:
\begin{eqnarray}
{\bf T}^{\alpha\beta}f^{\alpha\beta}=&-&\frac{\hbar^{2}}{2I}{\bf
D}_{l}f^{\alpha\beta}+\frac{1}{8I}\sum_{a,b}
\frac{\left(\overleftarrow{S^{\beta}}{}^{a}{}_{b}-
\overrightarrow{S^{\alpha}}{}^{a}{}_{b}\right)^{2}}{(Q^{a}-Q^{b})^{2}}
f^{\alpha\beta}\nonumber\\
&+&\frac{1}{8I}\sum_{a,b}\frac{\left(\overleftarrow{S^{\beta}}{}^{a}{}_{b}+
\overrightarrow{S^{\alpha}}{}^{a}{}_{b}\right)^{2}}{(Q^{a}+Q^{b})^{2}}f^{\alpha\beta}.\nonumber
\end{eqnarray}
In this way the problem has been successfully reduced from $n^{2}$ internal degrees of freedom
(physically $9$, sometimes $4$) to the $n$ purely deformative degrees of freedom (physically
$3$, sometimes $2$). The price one pays for that is the use of multi-component wave functions
subject to the strange parastatistical conditions in the reduced $q^{a}$-variables. The
particular values of labels $\alpha,\beta$ and the corresponding matrices $S^{\alpha}{}_{ab}$,
$S^{\beta}{}_{ab}$ describe the influence of quantized rotational degrees of freedom on the
quantized dynamics of deformation invariants. It is interesting that on the classical level
there is no simple way to perform such a dynamical reduction to deformation invariants.

For any reduced problem with $\alpha,\beta$ labels the quantity
\[
\rho\left(q^{i}\right):={\rm Tr}\left(f_{1}^{\alpha\beta
+}\left(q^{a}\right)f_{2}^{\alpha\beta}\left(q^{b}\right)\right)P_{\lambda}\left(q^{c}\right)
\]
is the probability density for finding the object in the state of
deformation invariants $(q^{1},\ldots,q^{n})$. More precisely,
$\rho(q^{1},\ldots,q^{n})dq^{1}\cdots dq^{n}$ is the probability
that the values of deformation invariants will be detected in the
infinitesimal range $dq^{1}\cdots dq^{n}$ about the values
$(q^{1},\ldots,q^{n})$.

Similarly, performing the integration
\[
\rho\left(L,R\right)=\int\overline{\Psi}\left(L;q^{a};R\right)
\Psi\left(L;q^{a};R\right)P_{\lambda}\left(q\right)dq^{1}\cdots
dq^{n}
\]
one obtains the probability density for detecting the "gyroscopic" degrees of freedom $L,R$
(equivalently, the Cauchy and Green deformation tensors) in some range of the configuration
space. Obviously, this distribution is meant in the sense of the Haar measure $\mu$. The
integrals
\[
p^{\alpha\beta}_{{}^{nk}_{ml}}=
\int\overline{f{\;}^{\alpha\beta}_{{}^{nk}_{ml}}}\left(q^{a}\right)
f{\;}^{\alpha\beta}_{{}^{nk}_{ml}}\left(q^{b}\right)P_{\lambda}\left(q^{c}\right) dq^{1}\cdots
dq^{n}
\]
are probabilities of detecting the particular indicated values of angular momenta and
vorticities, $C(\alpha,2)$, $C(\beta,2)$, $\hbar m$, $\hbar l$, $\hbar n$, $\hbar k$. In the
physical three-dimensional case they are respectively $\hbar^{2}s(s+1)$, $\hbar^{2}j(j+1)$,
$\hbar m$, $\hbar l$, $\hbar n$, $\hbar k$, where $s,j$ are non-negative integers,
$m,n=-s,\ldots,s$, $k,l=-j,\ldots,j$, jumping by one. Particularly interesting are
\[
p^{\alpha\beta}_{ml}=\sum_{n,k}p^{\alpha\beta}_{{}^{nk}_{ml}}
\]
because they refer to the constants of motion $\mathbf{S}^{i}{}_{j}$, $\mathbf{V}^{A}{}_{B}$
and to "good" quantum numbers $\alpha,\beta,m,l$. Except the special case $n=2$,
$\widehat{\mathbf{r}}^{a}{}_{b}$, $\widehat{\mathbf{t}}^{a}{}_{b}$ are not constants of
motion, and $k,l$ are not "good" quantum numbers. Quite analogous statements are true for the
quantized d'Alembert model; the only formal difference is that the integration element in the
manifold of invariants is given by $P_{l}(Q^{1},\ldots,Q^{n})dQ^{1}\cdots dQ^{n}$.

Let us now write down the explicit formulas for the physical three-dimen\-sional case. For the
affine-affine model (\ref{a117}) we have now
\begin{eqnarray}
{\bf T}^{sj}_{\rm aff-aff}f^{sj}=&-&\frac{\hbar^{2}}{2A}{\bf
D}_{\lambda}f^{sj}+\frac{\hbar^{2}B}{2A(A+3B)}\frac{\partial^{2}}{\partial
q^{2}}f^{sj}\nonumber\\
&+&\frac{1}{16A}\sum_{a=1}^{3}\frac{(S^{s}{}_{a})^{2}f^{sj}-
2S^{s}{}_{a}f^{sj}S^{j}{}_{a}+f^{sj}(S^{j}{}_{a})^{2}}{{\rm sh}^{2}\frac{q^{b}-q^{c}}{2}}
\nonumber\\
&-&\frac{1}{16A}\sum_{a=1}^{3}\frac{(S^{s}{}_{a})^{2}f^{sj}+
2S^{s}{}_{a}f^{sj}S^{j}{}_{a}+f^{sj}(S^{j}{}_{a})^{2}}{{\rm
ch}^{2}\frac{q^{b}-q^{c}}{2}}.\label{a121}
\end{eqnarray}
where $S^{j}{}_{a}$ are the standard Wigner matrices for $j$-angular momentum, i.e.,
$\hbar^{2}j(j+1)$-magnitude, and for any $a$-th term of both summation we have obviously
$b\neq a$, $c\neq a$, $b\neq c$. Obviously, it does not matter in what an ordering $q^{b}$,
$q^{c}$ are written, because the denominators are sign-non-sensitive. $\mathbf{D}_{\lambda}$
is given by (\ref{b99}), where explicitly
\[
P_{\lambda}=\left|{\rm sh}\left(q^{2}-q^{3}\right){\rm sh}\left(q^{3}-q^{1}\right){\rm
sh}\left(q^{1}-q^{2}\right)\right|.
\]
And similarly, using the abbreviated form, we can write for the metrical-affine (\ref{b117})
and affine-metrical (\ref{a118}) models, respectively, as follows:
\begin{eqnarray}
{\bf T}^{sj}_{\rm met-aff}&=&{\bf T}^{sj}_{\rm aff-aff}\left[A\mapsto
I+A\right]+\frac{I}{2(I^{2}-A^{2})}\hbar^{2}s(s+1),\nonumber\\
{\bf T}^{sj}_{\rm aff-met}&=&{\bf T}^{sj}_{\rm aff-aff}\left[A\mapsto
I+A\right]+\frac{I}{2(I^{2}-A^{2})} \hbar^{2}j(j+1),\nonumber
\end{eqnarray}
where, obviously, ${\bf T}^{sj}_{\rm aff-aff}\left[A\mapsto I+A\right]$ is obtained from ${\bf
T}^{sj}_{\rm aff-aff}$ (\ref{a121}) simply by replacing $A$ with $\alpha=I+A$.

The doubly-isotropic d'Alembert model in three dimensions has the following form:
\begin{eqnarray}
{\bf T}^{sj}_{\rm d'A}f^{sj}=&-&\frac{\hbar^{2}}{2I}{\bf
D}_{l}f^{sj}
+\frac{1}{4I}\sum_{a=1}^{3}\frac{(S^{s}{}_{a})^{2}f^{sj}-
2S^{s}{}_{a}f^{sj}S^{j}{}_{a}+f^{sj}(S^{j}{}_{a})^{2}}{(Q^{b}-Q^{c})^{2}}
\nonumber\\
&+&\frac{1}{4I}\sum_{a=1}^{3}\frac{(S^{s}{}_{a})^{2}f^{sj}+
2S^{s}{}_{a}f^{sj}S^{j}{}_{a}+f^{sj}(S^{j}{}_{a})^{2}}{(Q^{b}+Q^{c})^{2}}\label{a122}
\end{eqnarray}
with the same convention as previously, ${\bf D}_{l}$ given by (\ref{b101}), and explicitly
\[
P_{l}=\left|\left(\left(Q^{2}\right)^{2}-\left(Q^{3}\right)^{2}\right)
\left(\left(Q^{3}\right)^{2}-\left(Q^{1}\right)^{2}\right)
\left(\left(Q^{1}\right)^{2}-\left(Q^{2}\right)^{2}\right)\right|.
\]

Let us mention that in principle half-integer angular momentum of extended objects may be
formally introduced by replacing the group GL$(3,\mathbb{R})$ by its universal covering
$\overline{{\rm GL}(3,\mathbb{R})}$. There are some indications that the physical usefulness
of such models is not excluded. Formally, the procedure is as follows. In (\ref{a105})
specialized to $n=3$ we replace the group SO$(3,\mathbb{R})$ by its covering SU$(2)$ and write
the following expression involving the known Wigner matrices $\mathcal{D}^{s}{}_{mn}$:
\begin{equation}\label{a123}
\Psi(u,D,v)=\sum_{s,j}\sum^{s}_{m,n=-s}\sum^{j}_{k,l=-j}
\mathcal{D}^{s}{}_{mn}(u)f^{sj}_{{}^{nk}_{ml}}(D)\mathcal{D}^{j}{}_{kl}\left(v^{-1}\right),
\end{equation}
where $u,v\in$ SU$(2)$, $D\in$ Diag$(3,\mathbb{R})\subset$ GL$(3,\mathbb{R})$, and both the
integer and half-integer values of $s,j$ are admissible. However, if the function of triples
$(u,D,v)$ is to represent a function on $\overline{{\rm GL}^{+}(3,\mathbb{R})}$, then the
values of $s,j$ in the above series must have the same "halfness", i.e., either both $s,j$ in
(\ref{a123}) are integers or both are non-integers. And no superposition between elements of
these two function spaces is admitted (a kind of superselection rule). The point is that for
such "halfness-mixing" superpositions the squared modulus $\overline{\Psi}\Psi$ would be
two-valued from the point of view of SO$(3,\mathbb{R})$. This would be violation of the
probabilistic interpretation of $\Psi$ in GL$^{+}(3,\mathbb{R})$. If there is no mixing, then
in the case of superposing over half-integer $s,j$ in (\ref{a123}) the resulting $\Psi$ is
two-valued on GL$^{+}(3,\mathbb{R})$, i.e., it does not project from $\overline{{\rm
GL}^{+}(3,\mathbb{R})}$ to GL$^{+}(3,\mathbb{R})$ but $\overline{\Psi}\Psi$ does project,
i.e., it is single-valued in GL$^{+}(3,\mathbb{R})$.

The simplest possible situation in (\ref{a121}), (\ref{a122}) is $s=j=0$, i.e., purely scalar
amplitude $f^{00}$. Then $\mathbf{T}^{00}$ reduces respectively to
\[
{\bf T}^{00}=-\frac{\hbar^{2}}{2A}{\bf
D}_{\lambda}+\frac{\hbar^{2}B}{2A(A+3B)}\frac{\partial^{2}}{\partial q^{2}},\qquad {\bf
T}^{00}=-\frac{\hbar^{2}}{2I}{\bf D}_{l},
\]
i.e., there is no direct contribution from internal degrees of freedom.

If we admit half-integers, then the next simple situation is $s=j=1/2$. Then
$S^{1/2}{}_{a}=(\hbar/2)\sigma_{a}$, where $\sigma_{a}$ are Pauli matrices. Therefore,
$\left(S^{1/2}{}_{a}\right)^{2}=(\hbar^{2}/4)\mathbb{I}_{2}$.

Finally, let us briefly describe the two-dimensional situation, i.e., "Flatland" \cite{Abb84},
$n=2$. Obviously, it may have some direct physical applications when we deal with flat
molecules or other structural elements. But besides, the two-dimensional models shed some
light on the general situation and enable one to make it more comprehensible and lucid.
Indeed, let us observe that the expressions (\ref{a72}) and (\ref{c73}) (without the last
$p^{2}$-term) are superpositions of two-dimensional clusters corresponding to all possible
$\mathbb{R}^{2}$-subspaces in $\mathbb{R}^{n}$. Obviously, these terms in general are
non-disjoint and for $n=3$ they simply cannot be disjoint (all two-dimensional linear
subspaces in $\mathbb{R}^{3}$ have intersections of dimension higher than null; if different,
they always intersect along one-dimensional linear subspaces).

There are some very exceptional features of the dimension $n=2$. They are very peculiar, in a
sense pathological. But nevertheless the resulting simplifications generate some ideas and
hypotheses concerning the general dimension. Of course, later on they must be verified on the
independent basis. Let us begin with the classical description.

The one-dimensional group of planar rotations SO$(2,\mathbb{R})$ is Abelian, therefore,
$\widehat{\rho}=\rho=S$, $\widehat{\tau}=\tau=-V$. In doubly-isotropic models $S$ and $V$ are
constants of motion and so are $\widehat{\rho}$, $\widehat{\tau}$, $M$, $N$ if $n=2$. It is
not the case for $n>2$, where, as always in isotropic models, $S$, $V$ are constants of motion
but $\widehat{\rho}$, $\widehat{\tau}$ do not equal $S$, $-V$ and are non-constant. But it is
exactly the use of $\widehat{\rho}$, $\widehat{\tau}$ and their combinations $M$, $N$ that
simplifies the problem and leads to a partial separation of variables. In two-dimensional
space these things coincide and the problem may be effectively reduced to the dynamics of
two-deformation invariants both on the classical and quantum level. The two-polar
decomposition $\varphi=LDR^{-1}$ will be parameterized in a standard way; using the matrix
language we have
\begin{eqnarray}
L&=&\left[\begin{tabular}{cc}
  $\cos\alpha$ & $-\sin\alpha$ \\
  $\sin\alpha$ & $\cos\alpha$
\end{tabular}\right],\qquad
R=\left[\begin{tabular}{cc}
  $\cos\beta$ & $-\sin\beta$ \\
  $\sin\beta$ & $\cos\beta$
\end{tabular}\right],\nonumber\\
D&=&\left[\begin{tabular}{cc}
  $Q^{1}$ & $0$ \\
  $0$ & $Q^{2}$
\end{tabular}\right]=\left[\begin{tabular}{cc}
  $\exp q^{1}$ & $0$ \\
  $0$ & $\exp q^{2}$
\end{tabular}\right].\nonumber
\end{eqnarray}
To separate the dilatational and incompressible motion we introduce new variables:
\[
q=\frac{1}{2}\left(q^{1}+q^{2}\right),\qquad x=q^{2}-q^{1}.
\]
Their conjugate momenta are given by
\[
p=p_{1}+p_{2},\qquad p_{x}=\frac{1}{2}\left(p_{2}-p_{1}\right).
\]
Angular velocities are given by the following matrices:
\begin{eqnarray}
\chi=\frac{dL}{dt}L^{-1}=L^{-1}\frac{dL}{dt}=\widehat{\chi}&=&
\frac{d\alpha}{dt}\left[\begin{tabular}{cc}
 $0$ & $-1$ \\
 $1$ & $0$
\end{tabular}\right],\nonumber\\
\vartheta=\frac{dR}{dt}R^{-1}=R^{-1}\frac{dR}{dt}=\widehat{\vartheta}&=&
\frac{d\beta}{dt}\left[\begin{tabular}{cc}
 $0$ & $-1$ \\
 $1$ & $0$
\end{tabular}\right].\nonumber
\end{eqnarray}
Spin and vorticity essentially coincide with canonical conjugate
momenta $p_{\alpha}$, $p_{\beta}$, i.e.,
\[
S=\rho=\widehat{\rho}=p_{\alpha}\left[\begin{tabular}{cc}
 $0$ & $1$ \\
 $-1$ & $0$
\end{tabular}\right],\qquad V=-\tau=-\widehat{\tau}=
p_{\beta}\left[\begin{tabular}{cc}
 $0$ & $1$ \\
 $-1$ & $0$
\end{tabular}\right].
\]
With this convention the pairing between velocities and momenta has the form:
\[
p_{\alpha}\frac{d\alpha}{dt}=\frac{1}{2}{\rm Tr}\left(S\chi\right),\qquad
p_{\beta}\frac{d\beta}{dt}=\frac{1}{2}{\rm Tr}\left(V\vartheta\right).
\]
The diagonalizing quantities $M=-\widehat{\rho}-\widehat{\tau}$,
$N=\widehat{\rho}-\widehat{\tau}$ are also expressed by matrices
\[
M=\textsf{m}\left[\begin{tabular}{cc}
 $0$ & $1$ \\
 $-1$ & $0$
\end{tabular}\right],\qquad
N=\textsf{n}\left[\begin{tabular}{cc}
 $0$ & $1$ \\
 $-1$ & $0$
\end{tabular}\right],
\]
where
\[
\textsf{m}=p_{\beta}-p_{\alpha},\qquad
\textsf{n}=p_{\beta}+p_{\alpha}.
\]
In some formulas it is convenient to use modified variables
\[
\gamma=\frac{1}{2}(\beta-\alpha),\qquad
\delta=\frac{1}{2}(\beta+\alpha).
\]
Their conjugate momenta just coincide with the above $\textsf{m}$,
$\textsf{n}$, i.e.,
\[
p_{\gamma}=\textsf{m},\qquad p_{\delta}=\textsf{n}.
\]
The magnitudes of $S$, $V$ have the form:
\[
\|S\|=|p_{\alpha}|=\frac{1}{2}|\textsf{n}-\textsf{m}|,\qquad
\|V\|=|p_{\beta}|=\frac{1}{2}|\textsf{n}+\textsf{m}|.
\]
As mentioned, $p_{\alpha}$, $p_{\beta}$, $\textsf{m}$,
$\textsf{n}$ are constants of motion because in doubly isotropic
models $\alpha$, $\beta$ are cyclic variables. The corresponding
affine-affine, metrical-affine, and affine-metrical kinetic
energies of internal degrees of freedom are respectively given by
\begin{eqnarray}
\mathcal{T}^{\rm aff-aff}&=&\frac{1}{2A}\left(p^{2}_{1}+p^{2}_{2}\right)
-\frac{B}{2A(A+2B)}p^{2}\nonumber\\
&+&\frac{1}{16A}\frac{\textsf{m}^{2}}{{\rm
sh}^{2}\frac{q^{2}-q^{1}}{2}}-\frac{1}{16A}\frac{\textsf{n}^{2}}{{\rm
ch}^{2}\frac{q^{2}-q^{1}}{2}},\nonumber\\
\mathcal{T}^{\rm met-aff}&=&\mathcal{T}^{\rm aff-aff}\left[A\mapsto I+A\right]
+\frac{I}{8(I^{2}-A^{2})}(\textsf{n}-\textsf{m})^{2},\nonumber\\
\mathcal{T}^{\rm aff-met}&=&\mathcal{T}^{\rm
aff-aff}\left[A\mapsto
I+A\right]+\frac{I}{8(I^{2}-A^{2})}(\textsf{n}+\textsf{m})^{2},\nonumber
\end{eqnarray}
where, as usual, $\mathcal{T}^{\rm aff-aff}\left[A\mapsto
I+A\right]$ denotes $\mathcal{T}^{\rm aff-aff}$ with $A$ replaced
by $I+A$.

Separating dilatational and incompressible motion we obtain respectively the following
expressions:
\begin{eqnarray}
\mathcal{T}^{\rm aff-aff}_{\rm
int}&=&\frac{p^{2}}{4(A+2B)}+\frac{p^{2}_{x}}{A}+
\frac{(p_{\alpha}-p_{\beta})^{2}}{16A{\rm
sh}^{2}\frac{x}{2}}-\frac{(p_{\alpha}+p_{\beta})^{2}}{16A{\rm
ch}^{2}\frac{x}{2}},\nonumber\\
\mathcal{T}^{\rm met-aff}_{\rm int}&=&\mathcal{T}^{\rm
aff-aff}\left[A\mapsto
I+A\right]+\frac{Ip^{2}_{\alpha}}{I^{2}-A^{2}},
\nonumber\\
\mathcal{T}^{\rm aff-met}_{\rm int}&=&\mathcal{T}^{\rm
aff-aff}\left[A\mapsto
I+A\right]+\frac{Ip^{2}_{\beta}}{I^{2}-A^{2}}.\nonumber
\end{eqnarray}

Canonical momenta $p_{\alpha}$, $p_{\beta}$, or equivalently,
$\textsf{m}$, $\textsf{n}$, are constants of motion and their
Poisson brackets with the variables $q$, $x$, $p$, $p_{x}$.
Therefore, if we are interested only in the evolution of variables
$q$, $x$ but not in that of $\alpha$, $\beta$, we can simply
replace $p_{\alpha}$, $p_{\beta}$, $\textsf{m}$, $\textsf{n}$ in
the above expressions by constants characterizing a given family
of solutions. The are effective coupling constants for the
interaction between deformation invariants $q^{1}$, $q^{2}$. The
sh$^{-2}(x/2)$-term controlled by $\textsf{m}$ is always repulsive
and singular at the coincidence $x=0$ (non-deformed shape but
dilatation admitted). The ch$^{-2}(x/2)$-term controlled by
$\textsf{n}$ is attractive and finite at $x=0$. At large
"distances" of deformation invariants, $|x|\rightarrow\infty$,
attraction prevails if and only if $|\textsf{m}|>|\textsf{n}|$,
i.e., if $p_{\alpha}$, $p_{\beta}$ have the same signs,
$p_{\alpha}p_{\beta}>0$. If $|\textsf{m}|<|\textsf{n}|$, i.e.,
$p_{\alpha}p_{\beta}<0$, then the time evolution of $x$ is
unbounded. This is just the very special ($n=2$) example of that
was said formerly, namely that in the incompressible and
affinely-invariant geodetic regime there exists an open family of
bounded motions ("elastic vibrations") and an open family of
unbounded motions ("dissociation", decay). If the total
deformative motion is to be bounded, then some
dilatations-stabilizing potential $V(q)$ must be included into
Hamiltonian. But even if there is no $x$-dependent potential, our
affine geodetic model in the non-compact configuration space of
incompressible motion may encode bounded elastic vibrations. The
same is true for $n>2$, however the situation is more complicated
then because $M_{ab}$, $N_{ab}$ are not constants of motion and
also undergo some vibrations.

Analogous statements are true on the quantum level. The Haar
measure in our coordinates is given by
\[
d\lambda\left(\alpha;q,x;\beta\right)=\left|{\rm
sh}x\right|dqdxd\alpha d\beta,
\]
its weight factor equals $P_{\lambda}=\left|{\rm sh}x\right|$. The
wave functions $\Psi$ are expanded in the double Fourier series:
\[
\Psi\left(\alpha;q,x;\beta\right)=
\sum_{m,n\in\mathbb{Z}}f^{mn}(q,x)e^{im\alpha}e^{in\beta}.
\]
This is obviously the Peter-Weyl theorem specialized to the
two-dimensional torus group T$^{2}$. Our integers
$m,n\in\mathbb{Z}$ are just the labels $\alpha$, $\beta$ from the
general theory.

For the affine-affine model the reduced operator of the kinetic
energy is given by
\begin{eqnarray}
{\bf T}^{mn}_{\rm aff-aff}f^{mn}=&-&\frac{\hbar^{2}}{A}{\bf
D}_{\lambda}f^{mn}-\frac{\hbar^{2}}{4(A+2B)}
\frac{\partial^{2}f^{mn}}{\partial q^{2}}\nonumber\\
&+&\frac{\hbar^{2}(n-m)^{2}}{16A{\rm
sh}^{2}\frac{x}{2}}f^{mn}-\frac{\hbar^{2}(n+m)^{2}}{16A{\rm
ch}^{2}\frac{x}{2}}f^{mn},\nonumber
\end{eqnarray}
where now $\mathbf{D}_{\lambda}$ is expressed as follows:
\[
{\bf D}_{\lambda}=\frac{1}{|{\rm sh}x|}\frac{\partial}{\partial
x}\left(|{\rm sh}x|\frac{\partial}{\partial x}\right).
\]
Similarly for the metric-affine and affine-metric models we obtain
respectively
\begin{eqnarray}
{\bf T}^{mn}_{\rm met-aff}&=&{\bf T}^{mn}_{\rm
aff-aff}\left[A\mapsto I+A\right]
+\frac{I\hbar^{2}m^{2}}{I^{2}-A^{2}},\nonumber\\
{\bf T}^{mn}_{\rm aff-met}&=&{\bf T}^{mn}_{\rm
aff-aff}\left[A\mapsto
I+A\right]+\frac{I\hbar^{2}n^{2}}{I^{2}-A^{2}},\nonumber
\end{eqnarray}
To avoid the purely continuous spectrum one must include into
Hamiltonian at least some dilatations-stabilizing potential
$V(q)$. The problems is then (as usual) separable in
$(x,q)$-variables. And just as on the classical level, for
affinely-invariant incompressible dynamics, i.e., for the
$x$-sector of the above operators, there exists discrete spectrum
if $|n+m|>|n-m|$, i.e., if $mn>0$. This is the quantum bounded
motion.

In three-dimensional problems the above condition will be replaced
by some more complicated one between quantum numbers labelling the
reduced amplitudes $f$.

Let us observe that in more general, not necessarily geodetic,
problems in two dimensions with explicitly separable potentials
$V(q,x)=V_{\rm dil}(q)+V_{\rm sh}(x)$, the Schr\"odinger equation
\[
\mathbf{H}^{mn}f^{mn}=Ef^{mn},
\]
where $\mathbf{H}^{mn}=\mathbf{T}^{mn}+V_{\rm dil}(q)+V_{\rm
sh}(x)$ with any of the above $\mathbf{T}^{mn}$, splits into two
one-dimensional Schr\"odinger equations. The reduced wave function
is sought in the form
\[
f^{mn}(q,x)=\varphi^{mn}(x)\chi(q),
\]
where $\varphi^{mn}$, $\chi$ satisfy the following eigenequations:
\[
\mathbf{H}^{mn}_{\rm sh}\varphi^{mn}=E_{\rm sh}\varphi^{mn},\qquad
\mathbf{H}_{\rm
dil}\chi=-\frac{\hbar^{2}}{4(A+2B)}\frac{d^{2}\chi}{dq^{2}}+V_{\rm
dil}\chi=E_{\rm dil}\chi.
\]
Here the shear-rotational Hamiltion operator $\mathbf{H}^{mn}_{\rm
sh}$ is given by
\[
{\bf H}^{mn}_{\rm sh-aff-aff}=-\frac{\hbar^{2}}{A}{\bf
D}_{\lambda}+\frac{\hbar^{2}(n-m)^{2}}{16A{\rm
sh}^{2}\frac{x}{2}}-\frac{\hbar^{2}(n+m)^{2}}{16A{\rm
ch}^{2}\frac{x}{2}}+V_{\rm sh}(x)
\]
in the affine-affine model, and by
\[
{\bf H}^{mn}_{\rm sh-met-aff}={\bf H}^{mn}_{\rm
sh-aff-aff}+\frac{I\hbar^{2}m^{2}}{I^{2}-A^{2}},\qquad {\bf
H}^{mn}_{\rm sh-aff-met}={\bf H}^{mn}_{\rm
sh-aff-aff}+\frac{I\hbar^{2}n^{2}}{I^{2}-A^{2}}
\]
respectively in the metric-affine and affine-metric models.
Obviously, the total energy equals $E=E_{\rm sh}+E_{\rm dil}$. It
is seen that the main point of the analysis is the affine-affine
model because with fixed $m$, $n$ the other ones differ form it by
($m,n$-dependent) $c$-numbers.

Just as in the classical model, for $|n+m|>|n-m|$, i.e., $nm>0$,
the "centrifugal" term
\[
V_{\rm cfg}:=\frac{\hbar^{2}(n-m)^{2}}{16A{\rm
sh}^{2}\frac{x}{2}}-\frac{\hbar^{2}(n+m)^{2}}{16A{\rm
ch}^{2}\frac{x}{2}}
\]
is singular repulsive at $x=0$ and finite-attractive for
$|x|\rightarrow\infty$. And then even for the purely geodetic
incompressible model ($V_{\rm sh}=0$) there exist bounded states
and discrete energy spectrum for $E_{\rm sh}$. For $nm<0$ the
energy spectrum is continuous (scattering states).

Finally, let us quote the corresponding formulas for the quantized
d'Alembert model. Obviously, using the same notation as above we
have the following expression for the classical kinetic
Hamiltonian:
\[
\mathcal{T}^{\rm d.A}_{\rm
int}=\frac{1}{2I}\left(P^{2}_{1}+P^{2}_{2}\right)
+\frac{1}{4I}\frac{\textsf{m}^{2}}{\left(Q^{1}-Q^{2}\right)^{2}}+
\frac{1}{4I}\frac{\textsf{n}^{2}}{\left(Q^{1}+Q^{2}\right)^{2}}.
\]
With fixed values of $\textsf{m}$, $\textsf{n}$ the problem
reduces again to the dynamics of deformation invariants $Q^{1}$,
$Q^{2}$. In the two-polar coordinates the Lebesgue measure element
is given by
\[
dl\left(\alpha;Q^{1},Q^{2};\beta\right)=P_{l}\left(Q^{1},Q^{2}\right)
dQ^{1}dQ^{2}d\alpha d\beta,
\]
where
\[
P_{l}=\left|\left(Q^{1}\right)^{2}-\left(Q^{2}\right)^{2}\right|=
\left|\left(Q^{1}+Q^{2}\right)\left(Q^{1}-Q^{2}\right)\right|.
\]
The reduced amplitudes $f^{mn}$ satisfy the eigenequations
\begin{equation}\label{a134}
{\bf H}^{mn}f^{mn}={\bf T}^{mn}f^{mn}+
V\left(Q^{1},Q^{2}\right)f^{mn}=E^{mn}f^{mn}
\end{equation}
with
\[
{\bf T}^{mn}f^{mn}=-\frac{\hbar^{2}}{2I}{\bf
D}_{l}f^{mn}+\frac{\hbar^{2}m^{2}}{4I\left(Q^{1}-Q^{2}\right)^{2}}f^{mn}
+\frac{\hbar^{2}n^{2}}{4I\left(Q^{1}+Q^{2}\right)^{2}}f^{mn},
\]
where
\[
{\bf D}_{l}=\frac{1}{P_{l}}\frac{\partial}{\partial
Q^{1}}\left(P_{l}\frac{\partial}{\partial
Q^{1}}\right)+\frac{1}{P_{l}}\frac{\partial}{\partial
Q^{2}}\left(P_{l}\frac{\partial}{\partial Q^{2}}\right).
\]
The coordinates $Q^{1}$, $Q^{2}$ are very badly non-separable even
in the very kinetic energy expression. There are however other
coordinates on the plane of deformation invariants, much better
from this point of view. The simplest ones are coordinates
$Q^{+}$, $Q^{-}$ obtained from $Q^{1}$, $Q^{2}$ by the rotation by
the angle $\pi/4$,
\[
Q^{+}:=\frac{1}{\sqrt{2}}\left(Q^{1}+Q^{2}\right),\qquad
Q^{-}:=\frac{1}{\sqrt{2}}\left(Q^{1}-Q^{2}\right),
\]
where $Q^{+}$ and $Q^{-}$ may be expressed in terms of polar and
elliptic coordinates respectively as
\[
Q^{+}=r\cos\varphi,\qquad Q^{-}=r\sin\varphi,
\]
and
\[
Q^{+}={\rm ch}\rho \cos \lambda,\qquad Q^{-}={\rm sh}\rho \sin
\lambda.
\]
In all these variables the Hamilton-Jacobi and Schr\"odinger equations without potential are
separable. Obviously, the geodetic d'Alembert model $T=(I/2){\rm
Tr}(\dot{\varphi}^{T}\dot{\varphi})$ is completely non-physical. However, the coordinate
systems $(Q^{+},Q^{-})$, $(r,\varphi)$, $(\rho,\lambda)$ enable one to find a class of
potentials which are physically realistic and at the same time both the Hamilton-Jacobi and
Schr\"odinger equations are separable for the corresponding Hamiltonians.

The reduced Schr\"odinger eigenproblem (\ref{a134}) with doubly
isotropic potentials is separable if
\[
V\left(Q^{1},Q^{2}\right)=V_{+}(Q^{+})+V_{-}(Q^{-}).
\]
Namely, we have then ${\bf H}^{mn}={\bf H}^{mn}_{+}+{\bf
H}^{mn}_{-}$, where
\begin{eqnarray}
{\bf
H}^{mn}_{+}&=&-\frac{\hbar^{2}}{2I}\left(\frac{\partial^{2}}{\partial
Q^{+}{}^{2}}+\frac{1}{Q^{+}}\frac{\partial}{\partial
Q^{+}}\right)f^{mn}_{+}
+\left(\frac{\hbar^{2}(m-n)^{2}}{8IQ^{+}{}^{2}}+V_{+}\right)f^{mn}_{+},\nonumber\\
{\bf
H}^{mn}_{-}&=&-\frac{\hbar^{2}}{2I}\left(\frac{\partial^{2}}{\partial
Q^{-}{}^{2}}+\frac{1}{Q^{-}}\frac{\partial}{\partial
Q^{-}}\right)f^{mn}_{-}
+\left(\frac{\hbar^{2}(m+n)^{2}}{8IQ^{-}{}^{2}}+V_{-}\right)f^{mn}_{-},\nonumber
\end{eqnarray}
where
\[
{\bf H}^{mn}_{+}f^{mn}_{+}=E^{mn}_{+}f^{mn}_{+},\qquad {\bf
H}^{mn}_{-}f^{mn}_{-}=E^{mn}_{-}f^{mn}_{-},
\]
and
\[
f^{mn}(Q^{1},Q^{2})=f^{mn}_{+}(Q^{+})f^{mn}_{-}(Q^{-}),\qquad
E^{mn}=E^{mn}_{+}+E^{mn}_{-}.
\]
Obviously, the volume element is given by
\[
dl\left(\alpha;Q^{+},Q^{-};\beta\right)=2\left|Q^{+}\right|\left|Q^{-}\right|
dQ^{+}dQ^{-}d\alpha d\beta.
\]
The doubly isotropic models separable in coordinates $(r,\varphi)$
are based on potentials of the form
\[
V(r,\varphi)=V_{r}(r)+\frac{1}{r^{2}}V_{\varphi}(\varphi).
\]
The wave functions are factorized as follows:
\[
\Psi=e^{im\alpha}e^{in\beta}f^{mn}(r,\varphi)=
e^{im\alpha}e^{in\beta}R^{mn}(r)\Phi^{mn}(\varphi)
\]
and then of course
\[
\mathbf{S}\Psi=\mathbf{\widehat{r}}\Psi=\hbar m\Psi,\qquad
\mathbf{V}\Psi=-\mathbf{\widehat{t}}\Psi=\hbar n\Psi.
\]
The reduced Hamiltonian has the form:
\[
\mathbf{H}^{mn}=\mathbf{H}^{mn}_{r}+\frac{1}{r}\mathbf{H}^{mn}_{\varphi},
\]
where
\[
\mathbf{H}^{mn}_{r}=-\frac{\hbar^{2}}{2I}\left(\frac{\partial^{2}}{\partial
r^{2}}+\frac{3}{r}\frac{\partial}{\partial r}\right)+V_{r}
\]
is as a matter of fact independent of $m,n$; unlike this,
$\mathbf{H}^{mn}_{\varphi}$ depends explicitly on $m,n$:
\[
\mathbf{H}^{mn}_{\varphi}=-\frac{\hbar^{2}}{2I}\left(\frac{\partial^{2}}{\partial
\varphi^{2}}+2{\rm ctg}(2\varphi)\frac{\partial}{\partial
\varphi}\right)+\frac{\hbar^{2}}{2I}
\frac{m^{2}+2mn\cos(2\varphi)+n^{2}}{\sin^{2}(2\varphi)}+V_{\varphi}.
\]
The functions $\Phi^{mn}$, $R^{mn}$ satisfy eigenequations
\begin{eqnarray}
\mathbf{H}^{mn}_{\varphi}\Phi^{mn}&=&E^{mn}_{\varphi}\Phi^{mn},\label{a136}\\
\left(\mathbf{H}^{mn}_{r}+\frac{1}{r}E^{mn}_{\varphi}\right)R^{mn}&=&ER^{mn}.\label{b136}
\end{eqnarray}
The $\Phi$-equation (\ref{a136}) is to be solved as first. Then
the resulting quantized values of $E^{mn}_{\varphi}$, labelled by
an additional quantum number $k$ are to be substituted to
(\ref{b136}) and because of this the labels $m,n$ appear in $R$
although the radial operator $\mathbf{H}^{mn}_{r}$, in spite of
the used notation, is independent of $m,n$. There are also two
additional quantum numbers in $R$, namely $k$ itself appearing
through $E^{mnk}_{\varphi}$ and the proper radial quantum number
$\mu$, thus, $E$ obtained from (\ref{b136}) will be denoted by
$E^{mnk\mu}$.

Let us quote some very interesting model qualitatively compatible
with standard demands of the macroscopic nonlinear elasticity,
\[
V=\frac{2\kappa}{r^{2}\cos(2\varphi)}+\frac{\kappa}{2}r^{2}=
\kappa\left(\frac{1}{D_{1}D_{2}}+\frac{D_{1}^{2}+D^{2}_{2}}{2}\right).
\]
In the natural state of elastic equilibrium $r=0$, $\varphi=0$. We
do not quote more complicated and rather non-useful
one-dimensional equations for the elliptic coordinates $\rho$,
$\lambda$. Let us only mention the general shape of separable
doubly-isotropic potentials
\[
V(\rho,\lambda)=\frac{V_{\rho}(\rho)}{2\left({\rm
ch}^{2}\rho-\cos^{2}\lambda\right)}+\frac{V_{\lambda}(\lambda)}{2\left({\rm
ch}^{2}\rho-\cos^{2}\lambda\right)}.
\]
Finally, we quote a three-parameter family of doubly-isotropic
potentials for which both the classical and quantum problems are
simultaneously separable in all the aforementioned coordinate
systems:
\begin{eqnarray}
V&=&\frac{A}{Q^{+}{}^{2}}+\frac{B}{Q^{-}{}^{2}}
+C\left(Q^{+}{}^{2}+Q^{-}{}^{2}\right)\nonumber\\
&=&\frac{2A}{\left(Q^{1}+Q^{2}\right)^{2}}+\frac{2B}{\left(Q^{1}-Q^{2}\right)^{2}}
+C\left(\left(Q^{1}\right)^{2}+\left(Q^{2}\right)^{2}\right).\nonumber
\end{eqnarray}
Here $A,B,C$ are arbitrary constants. It is well-known that the simultaneous separability of
Hamilton-Jacobi and Schr\"odinger equations in a few coordinate systems has to do with
degeneracy and hidden symmetries.

Two-dimensional models are interesting not only from the philosophical point of view of the
"Flatland" geometry. They may be practically useful in the theory of surfaces of structured
bodies and in the dynamics of elongated molecules or other structure elements.

\subsection*{Acknowledgements}

The paper presented here as a contribution to this book has a long
story. It was initiated during my stay in Piza in 2001 at the
Istituto Nazionale di Alta Matematica "Francesco Severi",
Universit\`{a} di Piza, with professor Gianfranco Capriz and owes
very much to my discussions with him and with professor Carmine
Trimarco. In a sense, it should be interpreted as a document of
our common work. Later on, during the visit of professor Paolo
Mariano at our Institute of Fundamental Technological Research in
Warsaw, we discussed the material contained here and I must say
that this intellectual interaction influenced in a very deep way
its final shape. Certain parts were prepared during my stay in
Berlin in 2004 at the Institute of Theoretical Physics, Berlin
Technical University, with professors K.~E.~Hellwig and H.~H.~von
Borzeszkowski, and discussions with my German colleagues were very
essential for me.

There is, perhaps unfortunately, no intellectual work without
financial support. I was blessed in this sense by the help of
Istituto Nazionale di Alta Matematica "Francesco Severi", Gruppo
Nazionale per la Fizica Matematica Firenze in Piza and by
Alexander von Humboldt Stiftung in Berlin.

I am really very grateful to all mentioned professors and
Institutions.

\end{document}